\documentclass[preprint,aps,floatfix,nofootinbib,superscriptaddress]{revtex4-1}

\usepackage{amsmath, amssymb, amsfonts, amsthm, latexsym, epsfig, mathrsfs, xcolor, bbm, subfig, caption}

\usepackage[inline]{enumitem}

\usepackage{setspace}
\usepackage[marginal, multiple]{footmisc}

\usepackage[utf8]{inputenc}

\usepackage[colorlinks, allcolors=blue!70!black, linktocpage]{hyperref}

\setlist[enumerate]{noitemsep, label=(\arabic*), ref=(\arabic*)}

\numberwithin{equation}{section}

\usepackage{cleveref}

\crefname{equation}{Eq.}{Eqs.}
\creflabelformat{equation}{#2#1#3}

\crefname{section}{Sec.}{Sec.}
\crefname{appendix}{Appendix}{Appendices}
\crefname{figure}{Fig.}{Figs.}

\crefname{definition}{Def.}{Defs.}
\crefname{prop}{Prop.}{Props.}
\crefname{lemma}{Lemma}{Lemmas}
\crefname{corollary}{Cor.}{Cors.}
\crefname{thm}{Theorem}{Theorems}
\crefname{remark}{Remark}{Remarks}

\def\deltaslash{\delta\hspace{-0.50em}\slash\hspace{-0.05em} }

\makeatletter
\def\p@subsection{}
\def\p@subsubsection{}
\makeatother

\let\OLDfootnote\footnote
\renewcommand\footnote[1]{%
        \setlength{\footnotesep}{0.55\baselineskip}%
        {\footnotesize \OLDfootnote{#1}}%
}

\let\OLDthebibliography\thebibliography
\renewcommand\thebibliography[1]{%
        \setstretch{1.079} 
        \OLDthebibliography{#1}%
        \small %
        \setlength{\itemsep}{0.2\baselineskip} 
}

\usepackage{array,multirow}
\newcolumntype{L}[1]{>{\raggedright\let\newline\\\arraybackslash\hspace{0pt}}m{#1}}
\newcolumntype{C}[1]{>{\centering\let\newline\\\arraybackslash\hspace{0pt}}m{#1}}
\newcolumntype{R}[1]{>{\raggedleft\let\newline\\\arraybackslash\hspace{0pt}}m{#1}}

\newcommand{\scri}{\mathscr{I}}

\newcommand{\lie}{\pounds}

\newcommand{\mc}{\mathcal}

\newcommand{\ms}{\mathscr}
\newcommand{\mf}{\mathfrak}

\newcommand{\twosphere}{{S^2}}
\newcommand{\diff}{{\rm diff}}
\newcommand{\df}[1]{\boldsymbol{#1}}

\newcommand{\hateq}{\mathrel{\mathop {\widehat=} }} 

\newcommand{\lb}{\left}
\newcommand{\rb}{\right}

\renewcommand{\bar}{\overline}

\newcommand{\goesto}{\to}

\newcount\hh
\newcount\mm
\mm=\time
\hh=\time
\divide\hh by 60
\divide\mm by 60
\multiply\mm by 60
\mm=-\mm
\advance\mm by \time
\def\hhmm{\number\hh:\ifnum\mm<10{}0\fi\number\mm}

\def\be{\begin{equation}}
\def\ee{\end{equation}}
\def\pullback{\Pi}
\def\pushback{\Upsilon}

\begin{document}
\count\footins = 1000
\setstretch{1.2}

\title{Symmetries and charges of general relativity at null boundaries}

\author{Venkatesa Chandrasekaran}\email{ven\_chandrasekaran@berkeley.edu}
\affiliation{Center for Theoretical Physics and Department of Physics, University of California, Berkeley, CA 94720}

\author{\'Eanna \'E. Flanagan}\email{eef3@cornell.edu}
\affiliation{Department of Physics, Cornell University, Ithaca, NY 14853}
\affiliation{Cornell Laboratory for Accelerator-based Sciences and Education (CLASSE), Cornell University, Ithaca, NY 14853}

\author{Kartik Prabhu}\email{kartikprabhu@cornell.edu}
\affiliation{Cornell Laboratory for Accelerator-based Sciences and Education (CLASSE), Cornell University, Ithaca, NY 14853}


\begin{abstract}
We study general relativity at a null boundary using the covariant
phase space formalism. We define a covariant phase
space and compute the algebra of symmetries at the null
boundary by considering the boundary-preserving diffeomorphisms that
preserve this phase space. This algebra is
the semi-direct sum of diffeomorphisms on the two sphere and
a nonabelian algebra of supertranslations that has some similarities
to supertranslations at null infinity. By using the
general prescription developed by Wald and Zoupas, we derive the
localized charges of this algebra at cross sections of the null
surface as well as the associated fluxes.
Our analysis is covariant and
applies to general non-stationary null surfaces.
We also derive the global charges that generate the symmetries
for event horizons, and show that these obey the same algebra as the linearized diffeomorphisms, without any central extension.
Our results show that
supertranslations play an important role not just at null infinity but
at all null boundaries, including non-stationary event horizons.
They should facilitate further investigations of
whether horizon symmetries and conservation laws in black hole spacetimes
play a role in the information loss problem, as suggested by
Hawking, Perry, and Strominger.

\end{abstract}

\maketitle
\tableofcontents

\section{Introduction}\label{sec:intro}

It is well known that gauge transformations of a diffeomorphism
invariant theory can become genuine symmetries of the theory at
boundaries of the spacetime. In general relativity, diffeomorphisms of
asymptotically flat spacetimes that preserve the fall-off conditions for
the metric near null infinity yield the standard BMS group
\cite{Bondi:1962px, Sachs:1962wk, Sachs:1962zza}. Similarly, in QED
there exists an infinite set of symmetries at null infinity comprised
of large gauge transformations
\cite{He:2014cra,Kapec:2015ena}. Associated to the various symmetries
are global conserved charges which act as generators of the symmetries
\cite{WZ,Strominger:2017zoo}.  There are in addition localized charges
such as Bondi mass which quantify the amount of charge in subregions
of the spacetime boundary, which can be calculated using a variety of
formalisms \cite{WZ, Ashtekar:1981bq, Dray:1984rfa}.

More recently, it has been found that stationary black holes also
possess an infinite number of symmetries beyond the usual horizon
Killing symmetries
\cite{Donnay:2016ejv,Donnay:2015abr,Eling:2016xlx,Cai:2016idg,
  Hawking:2015qqa, Hawking:2016sgy,
  Hawking:2016msc,Carlip:2017xne,Blau:2015nee,Penna:2017bdn,Grumiller:2018scv} (see
\cite{Koga:2001vq} for older work on this topic, and
\cite{Mao:2016pwq} for the electromagnetic case).
The new symmetries are
diffeomorphisms which preserve the near horizon geometry under
specific gauge conditions, and a subclass of them are similar to
the supertranslations at null infinity. These horizon supertranslations
give rise to contributions to the global charges associated with
supertranslations, in addition to the contribution from null infinity.
In \cite{Hawking:2016sgy,
  Hawking:2016msc, Hawking:2015qqa} it was suggested that this
enlarged group of horizon symmetries and its associated charges and
conservation laws play a role in how information is released as a
black hole evaporates, and may lead to a resolution of the information
loss paradox (see also \cite{Penna:2015gza,Strominger:2014pwa}). At the least, a complete analysis of
supertranslation conservation laws in black hole spacetimes cannot be
undertaken without first knowing what the supertranslation charges and
fluxes are on general, non-stationary event horizons. It is therefore
of considerable interest to gain a deeper, more unified understanding
of such symmetries and charges.

A natural question is whether supertranslations are symmetries
of general relativity at \emph{any} null surface, with stationary horizons
and null infinity being special cases. This would give null boundaries in general relativity
quite a rich structure from the phase space point of view, and put
supertranslations on far more general footing.
As one of the main results of this paper, we systematically calculate
the group and algebra of symmetries of general relativity at a null boundary at
a finite location in spacetime, and show that this is indeed the
case. We do so using covariant
phase space methods, which clarifies the geometric meaning of the symmetries.
The symmetry group is the semidirect product of the group of
diffeomorphisms of the base space (typically the two-sphere) with
a nonabelian group of supertranslations, which contains
angle-dependent displacements of affine parameter as well as
angle-dependent rescalings of affine parameter\footnote{Our symmetry
group does not coincide exactly with any of the several different groups
in Refs.\ \cite{Penna:2017bdn,Koga:2001vq,Cai:2016idg,Eling:2016xlx,Donnay:2016ejv,Donnay:2015abr}, since
we preserve a particular geometric structure on the null surface
which defines our field configuration space,
and other authors preserve other quantities such as the near horizon geometry.}.
The results apply to nonstationary black hole horizons as well as
cosmological horizons.

We next turn to the charges and conservation laws associated with
these symmetries.  We distinguish between {\it global charges} and
associated global conservation laws -- the independence of integrals
over Cauchy surfaces $\Sigma$ of the choice of Cauchy surface -- and
{\it localized charges} and localized conservation laws, which involve
integrals over hypersurfaces $\Sigma$ that are not Cauchy surfaces.
For the global charges, we compute explicitly the contribution to the
charges from integrals over event horizons.  The complete
charges and complete formulation of the conservation laws requires an
understanding of how the symmetries of the event horizon mesh with
asymptotic symmetries at null infinity.  This has been worked
out in some special cases \cite{Hawking:2016sgy,Hawking:2016msc}, but
the general case is a subject
for future investigations.

Localized charges, for example the Bondi mass at cross sections of future
null infinity, are associated with localized conservation laws that
express the difference between the charges at two successive cross
sections with the integral of a flux over the intervening region of the boundary.
These charges are not generators of symmetries on phase space.  Wald
and Zoupas \cite{WZ} give a general prescription for computing such charges,
by starting with the integral of a symplectic current that defines the
variation of the global charge, and restricting the domain of
integration to a hypersurface which is not a Cauchy surface, in order
to attempt to obtain the charge contained within some of the degrees
of freedom of the theory.  This quantity is not in general a total
variation and so cannot be integrated up in phase space to obtain the
charge.  Wald and Zoupas give a prescription for adding a correction
term that overcomes this obstacle, thus allowing the definition of
finite charges.  Their prescription gives the conventional answers for
localized charges and fluxes at null infinity \cite{WZ}.

In this paper we describe how to adapt the prescription to a finite
null surface, and calculate the charges and fluxes of the symmetry
algebra at the surface. In particular, we obtain simple
expressions for the supertranslation charges and fluxes. The result
applies to a very general class of null surfaces including, most
importantly, non-stationary event horizons. The fluxes manifestly
satisfy the property that they vanish on stationary solutions at the
null surface, as one would desire if the charges are to be physically
meaningful.

An interesting question is the physical interpretation of the localized charges at the null
surface.  At null infinity, such an interpretation of supertranslation
charges is provided by the memory effect.  The supertranslation that
relates two different stationary regions (or vacua) can be measured as
a gravitational wave memory \cite{Strominger:2014pwa, Hollands:2016oma}.
Outgoing radiation can be though of as causing a transition from one
vacuum to another.  A similar situation likely occurs at a black hole
horizon, when accretion of radiation causes a transition from one
state to a supertranslated state, with the supertranslation being measurable
by near-horizon observers as a memory effect.  While some
aspects of this memory have been uncovered \cite{Hawking:2016msc}
there are still open questions.

Aside from the above motivations, which are centered around black
holes, an understanding of the gravitational symmetry algebra at a
null surface is important in and of itself: null surfaces play a
crucial role in information theoretic constraints and dynamics within
field theory and semi-classical gravity
\cite{Bousso:1999xy,Wall:2011hj,Casini:2017roe}, in holographic
settings and action formulations \cite{Brown:2015lvg,Lehner:2016vdi,
  Hopfmuller:2016scf}, in derivations of the generalized second law \cite{Wall:2011hj},
and even in quantum gravity
\cite{Wieland:2017zkf, Wieland:2017cmf}. The covariant phase space
formalism for spacetimes with boundary is also important in studying
the contribution of edge modes to entanglement entropy in gauge
theories and gravity \cite{Donnelly:2016auv,Speranza:2017gxd}. As such, a complete
description of the symmetries and charges of general non-stationary
solutions at null surfaces could provide further insight into
gravity, just as it did at null infinity.

Our work is complementary to the recent derivation of Hopfmuller and Friedel of
boundary currents for arbitrary null surfaces and associated local conservation laws,
for arbitrary vector fields tangent to the null surface \cite{Hopfmuller:2018fni}.
Earlier treatments of the symplectic structure of general relativity
on null surfaces and in 2+2 formulations can be found in Refs.\
\cite{Brady:1995na,Epp:1995uc,Reisenberger:2007pq,Parattu:2015gga}.

The paper is organized as follows. Section \ref{sec:cov-phase} reviews
the covariant phase space formulation of boundary symmetries and conserved charges
of diffeomorphism covariant theories, and Sec.\
\ref{sec:null-geometry} establishes our conventions for describing the
local geometry of null surfaces.  In Sec.\ \ref{sec:universal} we
define a universal intrinsic structure for null
hypersurfaces, and derive its invariance group and algebra.  Section
\ref{sec:cpsnb} defines a covariant phase space for general relativity
with a null boundary, and shows the associated symmetry algebra of linearized diffeomorphisms is the
same as that of the universal intrinsic structure.  The global and
localized charges associated with these symmetries are discussed
in Sec.\ \ref{sec:wzcharges}, and global conservation laws in Sec.\ \ref{sec:globalcons}.
Section \ref{sec:central} shows that for event horizons, the algebra of global
charges under Dirac brackets coincides with the algebra of linearized diffeomorphisms under Lie brackets.
Section \ref{sec:discussion} discusses other applications to black holes and concludes.

\subsection{Notation and conventions}

We use the sign convention $(-,+,+,+)$ throughout. We use the
following conventions for tensor indices:
\begin{itemize}
\item Tensors on the spacetime $M$ will be denoted by lowercase Roman abstract indices
$a$, $b$, $c$ etc. from the first half of the alphabet.
\item Tensors on the null surface ${\cal N}$ will be denoted by lowercase Roman
abstract indices $i$, $j$, $k$ etc. from the second half of the alphabet.
\item Tensors built on the vector space of covectors $w_i$ orthogonal
  to the normal $\ell^i$ at a point on ${\cal N}$ will be denoted by
uppercase Roman abstract indices $A$, $B$, $C$ etc.
\end{itemize}
Boldface quantities like $\df \omega$ will denote differential forms.
In Sec.\ \ref{sec:cov-phase} we will work in $d$ spacetime dimensions,
but in the remainder of the paper we will specialize to 4 spacetime dimensions.

\section{Review of the covariant phase space formalism}
\label{sec:cov-phase}

In this section we review the generally covariant phase space framework
for describing symmetries in a diffeomorphism covariant theory on a
manifold $M$ with boundary $\partial M$
\cite{1987thyg.book..676C,ASHTEKAR1991417,WZ,LW,W-noether-entropy,Khavkine:2014kya,2017NuPhB.924..312G}.
We mostly follow the notations and terminology of Wald and Zoupas
\cite{WZ}, with one or two exceptions noted below.
The framework is very general and can be applied to arbitrary theories
and boundary conditions.  It was applied to vacuum general relativity
at null infinity in Ref.\ \cite{WZ}, and will be applied to vacuum
general relativity at finite null boundaries in later sections of this paper.

A summary of the properties of the various charges and conservation laws reviewed in this section is given in
Table \ref{tab:summary}.

\begin{table*}[t]
\centering
\scriptsize
\begin{tabular}{|C{2.5cm}|C{3.2cm}|C{3.2cm}|C{3.2cm}|C{3.2cm}|}
\hline
Property&Noether charge&
Boundary symmetry ``charge variation''
& Localized (Wald-Zoupas) boundary symmetry charge &
Global symmetry generator charge\\
\hline
\hline
Symbol & $Q_\xi$ &
$\deltaslash \mc Q_{\xi,j}$\ \ \ \ \ {}\footnote{The slash on the
  variation symbol is included here as a reminder
that this is a one form on phase space which need not be exact, the corresponding charge may not exist.} &
$\mc Q_\xi^{\rm loc}$
 & $\mc Q_\xi$
\\
\cline{1-5}
Defining equations & (\ref{eq:noether-current}), (\ref{eq:J-Q-C}), (\ref{noether2}) & (\ref{Hamiltonian Definition1})
 & (\ref{localcharge1}), (\ref{Thetadef1}), (\ref{cc11}), (\ref{stat-req})
 & (\ref{Hamiltonian Definition}) with $\Sigma$ a Cauchy surface
\\
\cline{1-5}
Always well defined? & Yes & Yes
 & Requires the existence of presymplectic potential $\df \Theta$
 satisfying certain properties
 & Yes (assuming validity of conjecture of Sec.\ \ref{sec:WZ-charge})
\\
\cline{1-5}
Interpretation as generator of symmetry? & No & No
 & No
 & Yes
\\
\cline{1-5}
Depends on? & Field configuration
$\phi$, $(d-2)$-surface ${\cal S}$, boundary symmetry $\xi^a$ at
${\cal S}$
& Field configuration
$\phi$, field variation $\delta \phi$, $(d-2)$-surface ${\cal S}$,
boundary symmetry $\xi^a$ at
${\cal S}$
 & Field configuration
$\phi$, $(d-2)$-surface ${\cal S}$, boundary symmetry $\xi^a$ at
${\cal S}$
 & Field configuration
$\phi$, global boundary symmetry $\xi^a$ (assuming global conservation laws valid)
\\
\cline{1-5}
Nature of associated conservation law
 & Conserved Noether current (\ref{eq:noether-current}) on spacetime &
Conserved presymplectic current (\ref{eq:omega-Q}) on spacetime
 &
Exact $(d-1)$-form (\ref{localcharge1}), (\ref{eq:WZ-flux}) on component of boundary
 & Conjectured law is that integral of symplectic current
 (\ref{eq:omega-Q}) over Cauchy surface $\Sigma$ and then in phase
 space independent of $\Sigma$ (Sec.\ \ref{sec:globalcons}). Established in some special cases
\\
\hline\hline
\end{tabular}

\caption{
A summary of the properties of the various charges and conservation laws reviewed in this section.}
\label{tab:summary}
\end{table*}

\normalsize

\subsection{Definitions of field configuration space and covariant
  phase space}

We consider a \(d\)-dimensional manifold $M$ with boundary $\partial
M$, on which we want to define a theory of some dynamical fields
$\phi$, tensors\footnote{One can also include dynamical
  fields that are gauge-covariant fields defined on a principal
  bundle over \(M\) \cite{KP-first-law}.} on $M$ (we suppress tensor
indices on $\phi$).  In the following sections of the paper we will specialize
to vacuum general relativity for which $\phi = g_{ab}$.
The boundary of $M$ can
consist of a number of different components ${\cal B}_j$,
\be
\partial M = \cup_j \ {\cal B}_j.
\ee
The boundary components can either be at a finite location, as for a
black hole horizon, or can be asymptotic boundaries.  In the latter
case the manifold $M$ will be the unphysical spacetime of the
conformal completion framework.

\begin{figure}
\subfloat[]{\includegraphics[width=0.3\textwidth]{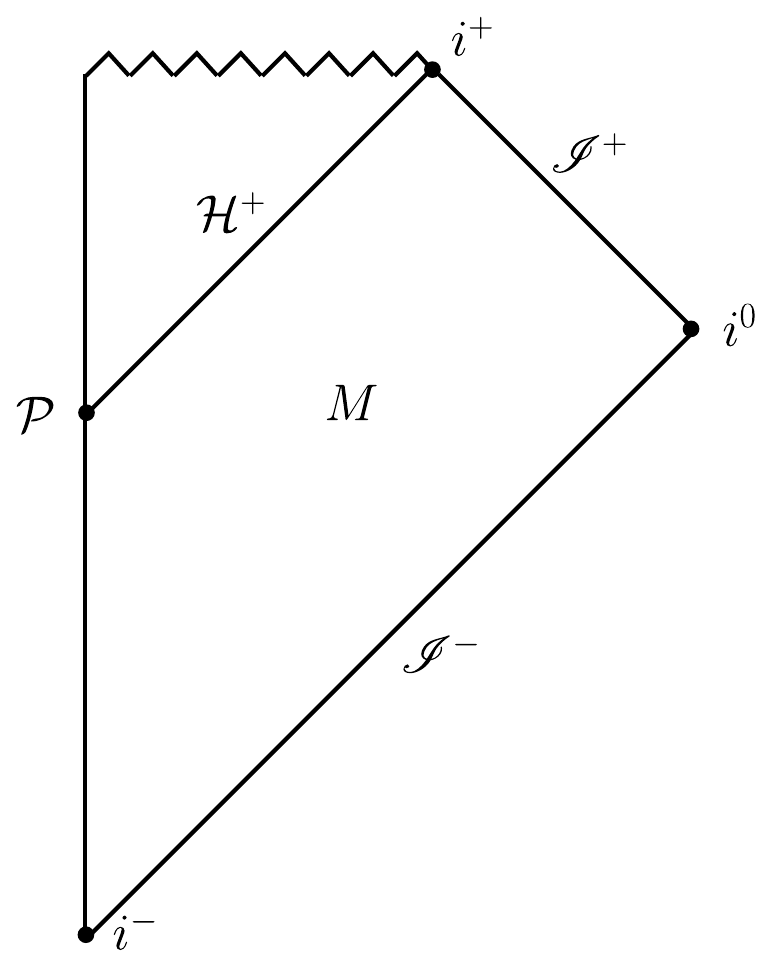}}
\hspace{0.1\textwidth}
\subfloat[]{\includegraphics[width=0.5\textwidth]{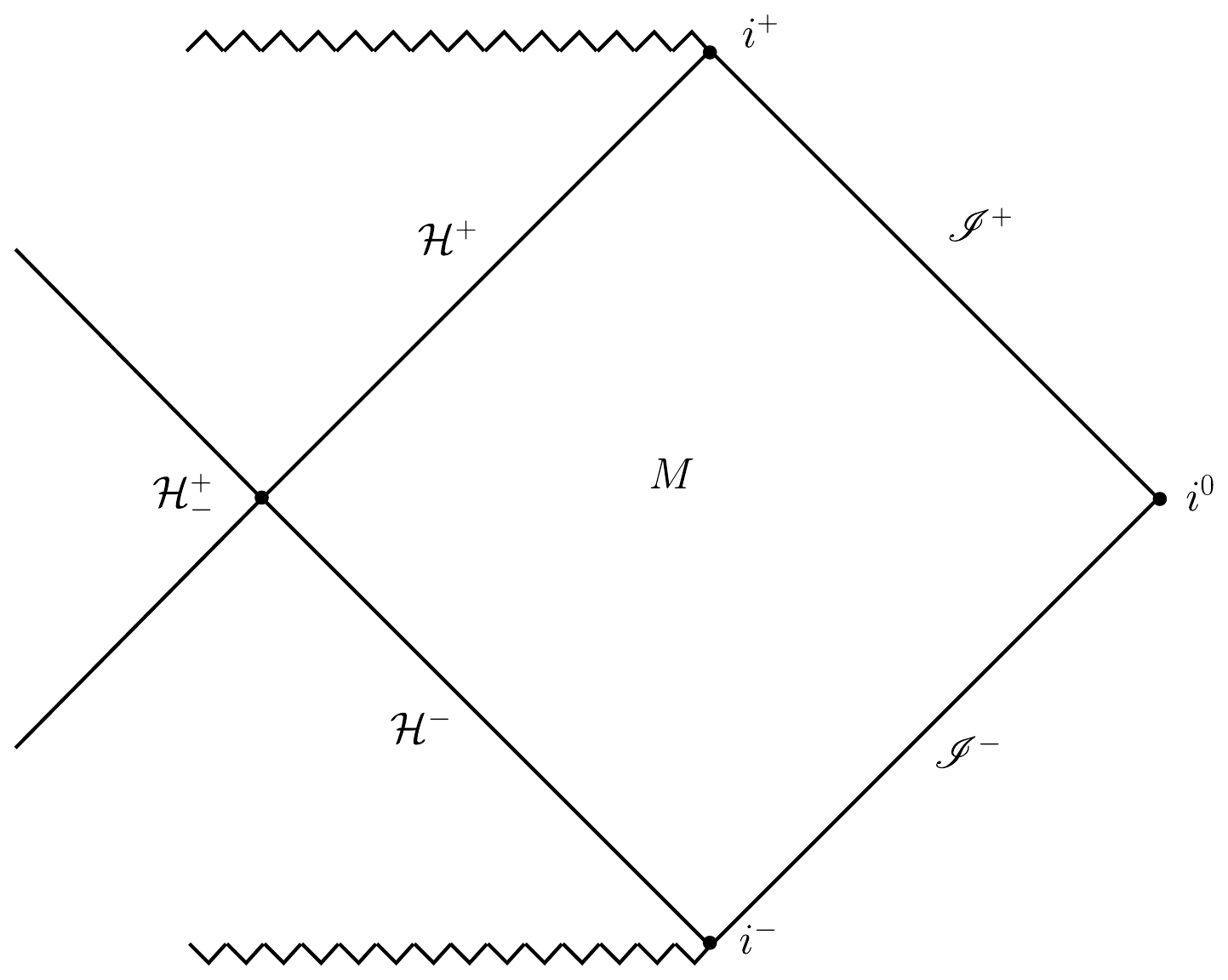}}
\caption{An illustration of two situations we will consider for the spacetime \(M\).
(a) $M$ is taken to be the domain of outer communications of a black hole formed in a gravitational collapse, with boundary elements $\scri^-$, $\scri^+$ and ${\cal H}^+$.
(b) $M$ is taken to be the domain of outer communications of an eternal black hole, with boundary elements $\scri^-$, $\scri^+$, ${\cal H}^-$ and ${\cal H}^+$.}
\label{fig:spacetimes}
\end{figure}

Two prototypical examples of setups we will want to consider are shown
in Figure \ref{fig:spacetimes}.  In the first, the manifold $M$ is the
domain of outer communications of a black hole formed in a
gravitational collapse, and the boundary elements are future null
infinity $\scri^+$, past null infinity $\scri^-$, and the future
event horizon ${\cal H}^+$.  In the second, the manifold is the domain
of outer communications of an eternal black hole, and the boundary
elements contain in addition the past event horizon ${\cal H}^-$.
We will also be concerned with the boundaries ${\cal H}^+_\pm$,
$\scri^+_\pm$ etc of these boundary elements, where ${\cal H}^+_+$ ($\scri^+_+$) is to be
interpreted as the limit of cuts ${\cal S}$ of ${\cal H}^+$ ($\scri^+$) in
the limit as ${\cal S}$ approaches future timelike infinity $i^+$,
${\cal H}^+_-$ is the bifurcation two-sphere in the second case, and
$\scri^+_-$ is the limit of cuts tending to spatial infinity $i^0$.

A crucial role in the formalism is the definition of a field
configuration space $\ms F$ of fields $\phi$ on $M$.  The fields are required
to be smooth on ${\cal M}$ and to obey suitable boundary conditions at each boundary component
${\cal B}_j$ and at their intersections.  A key goal of this paper is to determine appropriate boundary
conditions for vacuum general relativity, for a boundary component which is a general null surface
${\cal N}$ at a finite location in spacetime.  These boundary
conditions should allow the computation of symmetries and charges.
Boundary conditions that achieve this
are specified in
Sec.\ \ref{sec:cps} below.

\subsection{Definitions of currents}

We next review how conserved currents associated with spacetime
symmetries are obtained from the Lagrangian \cite{WZ}.
We assume that the dynamics of the theory is obtained from a
\(d\)-form Lagrangian
\be\label{Lagrangian}
\df L = \df L (\phi)
\ee
which depends locally and covariantly on the fields $\phi$. Such a
Lagrangian is independent of any ``background fields''.
Under a field variation $\phi \to \phi + \delta \phi$ the variation of
the Lagrangian can always be written as
\be
\label{Lagrangian Variation}
\delta \df L = \df E(\phi) \cdot\delta \phi + d\df\theta(\phi,\delta \phi),
\ee
where the tensor-valued \(d\)-form $\df E(\phi)$ represents the
equations of motion and $\cdot$ represents contraction over any suppressed tensor
indices.  The \((d-1)\)-form $\df\theta(\phi,\delta \phi)$ is the
presymplectic potential, which is locally and
covariantly constructed out of \(\phi\) and \(\delta \phi\) and
finitely many of their derivatives.  The subspace of \(\ms F\)
satisfying the equations of motion \(\df E = 0\) forms the
covariant phase space \(\bar{\ms F}\) of the theory.

Given two independent field variations $\delta_1 \phi$ and $\delta_2
\phi$ we define the presymplectic current
\begin{align}
\df{\omega}(\phi, \delta_1 \phi, \delta_2 \phi) = \delta_1 \df{\theta}(\phi, \delta_2 \phi) - \delta_2 \df{\theta}(\phi, \delta_1 \phi). \label{presymplectic current}
\end{align}
If $\phi$ satisfies the equations of motion and $\delta_1 \phi$ and
$\delta_2 \phi$ satisfy the linearized equations of motion, then
the presymplectic current is conserved,
\be
d \df \omega=0.
\label{cc}
\ee
We also define, for any vector field $\xi^a$ on spacetime, the Noether current \((d-1)\)-form $\df{j}_{\xi}$ by
\be\label{eq:noether-current}
\df{j}_{\xi} = \df{\theta}(\phi,\lie_\xi \phi) - i_{\xi}\df{L},
\ee
where $i_\xi$ denotes contraction of the vector field with the differential form on the first index.
It follows from Eqs.\ (\ref{Lagrangian Variation}) and (\ref{eq:noether-current}) that $d \df j_\xi=0$ on shell.
For any local and covariant theory it can be shown that the Noether current can always be written in the form (see \cite{IW-euclidean,SeiW})
\be\label{eq:J-Q-C}
\df j_{\xi} = d\df Q_\xi + \xi^a\df C_a,
\ee
where $\df Q_\xi(\phi)$ is the Noether charge \((d-2)\)-form and $\df C_a(\phi)$ are the constraints which vanish when the equations of motion hold. Taking a variation of the Noether current (\ref{eq:noether-current}) and using Eqs.\ (\ref{Lagrangian Variation}), (\ref{presymplectic current}) and (\ref{eq:J-Q-C}) we get for on-shell perturbations
\be\label{eq:omega-Q}
        \df\omega(\phi, \delta \phi, \lie_\xi \phi) =  d[ \delta \df{Q}_\xi - i_{\xi} \df{\theta}(\phi,\delta \phi) ].
\ee

\subsection{Definition of presymplectic form on covariant
  phase space}
\label{sec:presymform}

We next define the quantity
\begin{align}
\label{Omegadef}
\Omega_{\Sigma}(\phi, \delta_1 \phi, \delta_2 \phi) = \int_{\Sigma}\df{\omega}(\phi, \delta_1 \phi, \delta_2 \phi),
\end{align}
where $\Sigma$ is any hypersurface embedded in $M$.  We would like to use the definition (\ref{Omegadef})
specialized to a Cauchy surface $\Sigma$ to define the presymplectic
form\footnote{The presymplectic form $\Omega_\Sigma$ is usually
  degenerate.  One can
factor the configuration space $\ms F$ by the orbits of the degeneracy subspaces of
$\Omega_\Sigma$ to obtain a phase space $\Gamma$ on which there exists
a nondegenerate symplectic form \cite{LW}.  However this will not be
needed in what follows.}
 of the theory, a two-form on the covariant phase space $\bar{\ms
 F}$. There are a number of properties that we would like
 $\Omega_\Sigma$ to satisfy, some of which inform and restrict the definition of
 field configuration space $\ms F$.  These properties are:

\begin{itemize}

\item {\it Invariance under gauge transformations:} One might expect
  that $\Omega_\Sigma(\phi,\delta_1 \phi, \delta_2 \phi)$ should be
  invariant under independent linearized diffeomorphisms acting on
  $\delta_1 \phi$ and $\delta_2 \phi$.  This would require that
  $\Omega_\Sigma(\phi, \delta \phi, \lie_\xi \phi)=0$ for any $\phi
  \in \bar{ \ms F}$ and for any vector fields $\xi^a$ and variations
  $\delta \phi$ for which $\delta \phi$ and $\lie_\xi \phi$ are
  tangent to $\bar{\ms F}$.  However, this is not true in general.
  Instead, from
  Eqs.\ (\ref{eq:omega-Q}) and (\ref{Omegadef}) we have that, on
  shell, for a Cauchy surface $\Sigma$,
\be
\Omega_\Sigma( \phi, \delta \phi, \lie_\xi \phi) =
  \int_{\partial \Sigma} \delta \df{Q}_\xi(\phi) - i_{\xi} \df{\theta}(\phi,\delta \phi),
\label{Omegadef1}
\ee
where \(\partial\Sigma\) is the boundary of $\Sigma$, a $d-2$-surface in
$\partial M$.  This quantity vanishes for vector fields whose support lies in the interior of $M$, but not
in general for vector fields which are nonzero on the boundary
$\partial M$.  As is well known,
the fact that these diffeomorphisms do not correspond to degeneracy
directions of the presymplectic form reflects the fact that the
corresponding degrees of freedom are physical and not
gauge\footnote{One can choose to restore full diffeomorphism
  invariance by performing the Stueckelberg trick
and introducing new physical degrees of freedom on the boundary,
so-called edge modes \cite{Donnelly:2016auv,Speranza:2017gxd,2017NuPhB.924..312G}.}.

\item  {\it Finiteness at asymptotic boundaries:}  The definition
  (\ref{Omegadef}) is invariant under local deformations of the hypersurface
  $\Sigma$ when on-shell, from Eq.\ (\ref{cc}).  We would like the
  presymplectic form (\ref{Omegadef}) to have a well defined limit as
  $\Sigma$ approaches $\scri^+$ or $\scri^-$, which will be true if
  the presymplectic current $\df \omega$ has a well defined limit on
  those boundaries.  Boundary conditions at $\scri^+$ and $\scri^-$
  that are sufficient to ensure this are given by Wald and Zoupas
  \cite{WZ} (see their footnote 16).  These boundary conditions
  supplement the standard definition of asymptotic flatness at null
  infinity \cite{Wald-book} by specializing the gauge\footnote{Here by
    gauge we mean both diffeomorphism freedom and choice of conformal
    factor.}, and are necessary for $\df \omega$ to have a finite
  limit.  In the context of null boundaries at finite locations discussed in
this paper, we will also for convenience specialize the gauge at the
boundary (see Sec.\ \ref{sec:cps} below).  There is a tension between
gauge specializations at the boundary and the fact that some
of the diffeomorphism degrees of freedom on the boundary become
physical: one does not want to restrict physical degrees of freedom in
the definition of the field configuration space $\ms F$.
A general strategy for dealing with this tension is discussed in Sec.\
\ref{sec:cps} below.

\item {\it Independence of choice of Cauchy surface:}
In order for $\Omega_\Sigma$ to define a presymplectic form on the
covariant phase space $\bar{\ms F}$, one would like it to be
independent of the choice of Cauchy surface $\Sigma$.
While the integral (\ref{Omegadef}) is invariant under local
deformations of the hypersurface $\Sigma$, when one takes a limit to
the boundary of spacetime there can nonzero contributions to the
limiting integral from ``corners'' of the spacetime where boundary
elements intersect, such as spatial infinity $i^0$.  One would like to
specialize the definition of the
field configuration space $\ms F$ to eliminate such contributions.
This issue is closely related to the question of the validity of the
global conservation laws discussed in Sec.\ \ref{sec:globalcons} below.

\end{itemize}

\subsection{Global charges that generate boundary symmetries}
\label{sec:globalcharges}

We now turn to a discussion of spacetime symmetries, which we will
also call boundary symmetries since only the action of the symmetry
near the boundary $\partial M$ of spacetime will be important \cite{WZ}.
Infinitesimal diffeomorphisms are parametrized by vector fields
$\xi^a$ on $M$, under which fields transform as $\phi \to \phi + \delta
\phi$, where
\be
\delta \phi = \lie_\xi \phi.
\label{spacetimesymmetry}
\ee
Fix attention on one component ${\cal B}_j$ of the boundary $\partial
M$.  We denote by $G_j$ the set of smooth vector fields \(\xi^a\) on
\(M\) such that the
diffeomorphism generated by \(\xi^a\) preserves the boundary
\(\partial M\), and such that for any solution $\phi \in \bar{\ms F}$,
the transformed solution $\phi + \lie_\xi \phi$
satisfies any boundary conditions at ${\cal B}_j$
imposed on fields in \(\ms F\), to
linear order in $\xi^a$.
We will call such a vector field a \emph{representative of an
  infinitesimal boundary symmetry at} ${\cal B}_j$.
We also define $G$ to be the set of smooth vector fields whose
diffeomorphisms preserve $\partial M$ and map $\bar {\ms F}$ to $\bar
{\ms F}$ under pullback, which we call \emph{representatives of infinitesimal
boundary symmetries} \footnote{The set $G$ will generally be a proper
subset of $\cap_j G_j$, because of boundary conditions imposed at
intersections of boundary elements in the definition of $\ms F$
(for example continuity at a bifurcation two-sphere in an eternal
black hole spacetime).  See Sec.\ \ref{sec:globalcons} below for further discussion.}.

Consider now a representative of an infinitesimal boundary symmetry
$\xi^a$.
We would like to construct a charge ${\cal Q}_\xi$, a function on ${\ms F}$, which generates the
boundary symmetry (\ref{spacetimesymmetry}).  This means that ${\cal
  Q}_\xi$ should satisfy \cite{WZ}
\be\label{Hamiltonian presymplectic}
\delta {\cal Q}_{\xi} = \Omega_{\Sigma}(\phi, \delta \phi, \lie_{\xi}
\phi) = \int_{\Sigma} \df{\omega}(\phi, \delta \phi, \lie_\xi \phi)
\ee
for all $\phi \in \bar{\ms F}$ and for all $\delta \phi$, $\lie_\xi
\phi$ tangent to ${\ms F}$, where $\Sigma$ is a Cauchy surface.
The charge
${\cal Q}_\xi$ can be interpreted as a Hamiltonian\footnote{Here we
  depart slightly from the terminology used by Wald and Zoupas
  \cite{WZ}, who call all such charges Hamiltonians and denote them by
  $H_\xi$.  The definition
  of Wald and Zoupas -- their Eq.\ (8) -- is also more general since they do not impose
  that $\Sigma$ be a Cauchy surface.  We will return to this
  generalization in Sec.\ \ref{sec:WZ-charge} below.} in the special case when $\xi$ is a timelike vector field.
We call the charges (\ref{Hamiltonian presymplectic}) {\it global
  charges} since they are obtained by an integral over a complete
Cauchy surface and so involve all the degrees of freedom in the
theory, in contrast to the localized charges discussed in Sec.\
\ref{sec:WZ-charge} below.

We next discuss the conditions under which the boundary symmetry
generator charge ${\cal Q}_\xi$
will exist.  Since Eq.\ (\ref{Hamiltonian presymplectic}) is attempting to define an exact
one-form on field configuration space, the right hand side should be a
closed one-form. It follows from Eq.\ (\ref{eq:omega-Q}) that the variation of the charge is a
surface term on-shell:
\be\label{Hamiltonian Definition}
\delta {\cal Q}_{\xi} =  \int_{\partial \Sigma} \delta \df{Q}_\xi - i_{\xi} \df{\theta}(\phi,\delta \phi).
\ee
If the boundary $\partial \Sigma$ consists of a number of disconnected components ${\cal S}_j$, then
$\delta {\cal Q}_\xi = \sum_j \delta {\cal Q}_{\xi,j}$ where
\be
\delta {\cal Q}_{\xi,j} =  \int_{{\cal S}_j} \delta \df{Q}_\xi - i_{\xi} \df{\theta}(\phi,\delta \phi).
\label{Hamiltonian Definition1}
\ee
Taking a second variation and using the definition (\ref{presymplectic current}) of the
presymplectic current gives \cite{WZ}
\be\label{Hamiltonian Condition}\begin{split}
        0  = (\delta_1 \delta_2 - \delta_2 \delta_1) {\cal Q}_{\xi}
                 = - \int_{\partial\Sigma} i_\xi \df \omega(\phi,\delta_1\phi, \delta_2 \phi).
\end{split}\ee
The quantity (\ref{Hamiltonian Condition}) must vanish for all
$\delta_1 \phi$ and $\delta_2 \phi$ tangent to $\bar {\ms F}$ in order
for the charge ${\cal Q}_\xi$ to exist.  When it does vanish\footnote{Note that in general
  the second
  term in Eq.\ (\ref{Hamiltonian Definition}) can give a nonvanishing contribution, so that
  the charge differs from the Noether charge, even when the
  obstruction (\ref{Hamiltonian Condition}) vanishes.  This occurs for example for ADM
  charges at spatial infinity \cite{IW-noether-entropy}.}, the
definition (\ref{Hamiltonian presymplectic}) determines the charge on
$\bar{\ms F}$ up to constants of integration on phase space, which can be specified by
demanding that the charge vanish on a reference solution on each connected component of $\bar{\ms F}$ \cite{WZ}.
This prescription is discussed in more detail in the more general context
of localized charges in Sec.\ \ref{sec:WZ-charge} below.

In all cases that we are aware of, the condition (\ref{Hamiltonian Condition}) is satisfied
whenever $\Sigma$ is taken to be a Cauchy surface, as here.  While we
are not aware of a general proof, there is a physical argument
indicating that the condition should be satisfied: a non-vanishing
pullback of the symplectic current to $\partial \Sigma$ in
(\ref{Hamiltonian Condition}) reflects an interaction between degrees
of freedom that have been included in the integral (\ref{Hamiltonian presymplectic}) and those
that have been excluded, and Cauchy surfaces include all of the
degrees of freedom.  Some examples of cases where the condition
(\ref{Hamiltonian Condition}) is satisfied include:
\begin{itemize}
\item Spacetimes in general relativity that are asymptotically
  flat at spatial infinity $i^0$ and vacuum in a neighborhood of $i^0$,
  and spacelike Cauchy surfaces $\Sigma$
  that extend to $i^0$.  In this case the presymplectic
  current extends continuously to the boundary but has vanishing
  pullback there \cite{WZ}.

\item Asymptotically flat spacetimes in vacuum general relativity with
  no horizons, with $\Sigma$ taken to be future null infinity
  $\scri^+$, with certain fall off conditions on the News tensor.
  Consider the integrand in the obstruction (\ref{Hamiltonian
    Condition}),
in the limit where the cut ${\cal S}$ of $\scri^+$
approaches $\scri^+_+$ or $\scri^+_-$, i.e., $i^+$ or $i^0$.  Denoting
affine parameter by $u$, the integrand
is given by Eq.\ (72) of \cite{WZ} and scales like a symmetry
generator $\sim u$, times a shear tensor $\sim u^0$, times a News
tensor.  Hence if the News tensor decays faster than $1/|u|$ as $|u|
\to \infty$ the
result vanishes:
\be
\int_{\scri^+_\pm} i_\xi \df \omega =0.
\ee
In the Christodoulou-Klainerman class of spacetimes \cite{Christodoulou:1993uv}
the News decays like $|u|^{-3/2}$.

\item In the previous example, if the spacetime contains in addition a
  future event horizon ${\cal H}^+$, then the Cauchy surface can be
  taken to be ${\cal H}^+ \cup \scri^+$ and the integral (\ref{Hamiltonian presymplectic}) will
  contain contributions from both ${\cal H}^+$ and $\scri^+$:
\be
\delta {\cal Q}_\xi = \int_{{\cal H}^+} {\df \omega}(\phi,\delta \phi,
\lie_\xi \phi) + \int_{\scri^+} {\df \omega}(\phi,\delta \phi,
\lie_\xi \phi).
\label{split}
\ee
Here the first term will depend only on the limiting form of the
symmetry $\xi^a$ near ${\cal H}^+$, and the second term only on the
limiting form near $\scri^+$.  The integrability analysis described
above can be applied to each of these terms separately.
  In Appendix \ref{sec:bh} we show that
the condition (\ref{Hamiltonian Condition}) is satisfied
 for the integral over ${\cal H}^+$ under certain conditions (as well as for the
  integral over $\scri^+$).
\end{itemize}

To summarize this discussion, the definition (\ref{Hamiltonian presymplectic}) should be
sufficient to compute global charges ${\cal Q}_\xi$ that generate
boundary symmetries when $\Sigma$ is a Cauchy surface.  See the review article
by Strominger \cite{Strominger:2017zoo} for several specific calculations of charges of this type.
In Sec.\ \ref{sec:ham} below we will compute explicitly the
contribution to such charges from boundary elements that are null
surfaces at a finite location in spacetime, and in
Sec.\ \ref{sec:globalcons} we will discuss global conservation laws
that are satisfied by global charges ${\cal Q}_\xi$.

\subsection{Boundary symmetry algebras of linearized diffeomorphisms}
\label{sec:algebra}

We next discuss the symmetry algebras associated with each component
${\cal B}_j$ of the boundary $\partial M$ of spacetime.  These are
obtained
from the set $G_j$ of representatives of infinitesimal boundary
symmetries at ${\cal B}_j$ by
modding out the trivial representatives whose charges (\ref{Hamiltonian Definition}) vanish \cite{WZ}.
Specifically, we define an equivalence relation on representatives $\xi^a$ by
\be\label{eq:boundary-symm-equiv}
        \xi^a \sim \xi'^a \text{\ \ \  if \ \ \ } \xi^a \hateq \xi'^a
        \text{ and }
\int_{\cal S}( \delta {\df Q}_\xi - i_\xi \df \theta) = \int_{\cal S}( \delta {\df Q}_{\xi'} - i_{\xi'} \df \theta).
\ee
Here the notation $\hateq$ means equal when evaluated on ${\cal B}_j$,
and the integrals must
coincide for all $\phi \in \bar{\ms F}$ and $\delta \phi$ tangent to
$\bar {\ms F}$ and for all cross sections ${\cal S}$ of ${\cal B}_j$.
We define the symmetry algebra
\be
\mf g_j = G_j / \sim,
\ee
which for example gives the BMS algebra at null infinity \cite{WZ}.
In Sec.\ \ref{sec:symgroup} below we will derive the corresponding symmetry
algebra for a null surface at a finite location.

We similarly define the global symmetry algebra $\mf g = G / \sim$,
where now the equivalence relation is defined by imposing Eq.\
(\ref{eq:boundary-symm-equiv}) at all cross sections ${\cal S}$ of all
boundary components ${\cal B}_j$.  In general $\mf g$ will be a proper
subalgebra of the direct sum algebra
\be
\bigoplus_j \mf g_j,
\ee
because of boundary conditions imposed at the intersections of
boundary components in the definition of $\ms F$, cf.\ the discussions
in Sec.\ \ref{sec:globalcharges} above and \ref{sec:globalcons} below.

\subsection{Localized (Wald-Zoupas) charges, fluxes and conservation laws}
\label{sec:WZ-charge}

We now turn to a discussion of a different type of charge which we call
{\it localized charges}, whose physical interpretation is roughly the amount
of charge in a subset of the degrees of freedom of the theory.
Studies of this type of charge
have a long history in general relativity.
For example, there have been many attempts made to define the
total mass in a finite region of space, using various notions of
quasilocal mass \cite{Szabados2009}, but no natural and generally
accepted definition has emerged.
On the other hand, as is well known, the total amount of
4-momentum\footnote{Or more generally any BMS charge.} radiated through any
finite region of future null infinity is uniquely defined
\cite{Ashtekar:1981bq,Dray:1984rfa}.
Wald and Zoupas \cite{WZ} give a very general prescription for
defining localized charges of this type at a boundary of spacetime, for any
diffeomorphism invariant theory and for a large class of boundary
conditions. They show that their general prescription gives the conventional
results \cite{Ashtekar:1981bq,Dray:1984rfa} for BMS charges at null infinity.  In this subsection we
review and specialize slightly their general construction, and in
Sec.\ \ref{sec:wz} below we apply it to compute localized charges at a
spacetime boundary consisting of a null surface at a finite location.

One trivial kind of localization was already encountered in Sec.\
\ref{sec:globalcharges} above.  In the example (\ref{split}), the charge variation
$\delta \mc Q_\xi$ was expressed as a sum of an integral over the future
event horizon ${\cal H}^+$ and an integral over future null infinity
$\scri^+$, each of which individually satisfies the integrability
condition (\ref{Hamiltonian Condition}).  Here we want to go further and
consider charges localized to subregions of boundary components.

Consider a region $\Delta {\cal B}_j$ of a boundary ${\cal B}_j$ whose
boundary consists of two crosssections ${\cal S}$ and ${\cal S}'$,
and a representative $\xi^a$ of an infinitesimal boundary symmetry at ${\cal B}_j$.
Given a solution $\phi \in \bar{\ms F}$,
we would like to define an exact 3-form $d \df {\mc Q}_\xi^{\rm loc}$
on ${\cal B}_j$ for which the charge in the region $\Delta
{\cal B}_j$ is
\be
\int_{\Delta {\cal B}_j} d \df {\mc Q}_\xi^{\rm loc}  =
{\mc
  Q}_\xi^{\rm loc}({\cal S}') - {\mc
  Q}_\xi^{\rm loc}({\cal S}),
\label{qlocdef}
\ee
where
\be
  {\mc Q}_\xi^{\rm loc}({\cal S}) = \int_{\cal S} \df {\mc Q}_\xi^{\rm
  loc}
\label{localcharge}
\ee
is the charge at crosssection ${\cal S}$.
We will call the quantity (\ref{localcharge}) a localized or
Wald-Zoupas charge.
The prototypical example of a quantity like
this is the Bondi mass at a cross section ${\cal S}$ of $\scri^+$, which is the
total mass of the spacetime minus the mass radiated up to ${\cal S}$.
In Sec.\ \ref{sec:wz} we will define a similar quantity at cuts of a null
boundary, which for a future event horizon will be the total charge at
the bifurcation twosphere of the black hole (if any) plus the total charge accreted by
the black hole up to the cut ${\cal S}$\footnote{Our
  orientation convention is such that (\ref{qlocdef}) is valid at $\scri^+$
  when ${\cal S}$ is to the future of ${\cal S}'$, while at a future
  event horizon ${\cal H}^+$ it is valid when ${\cal S}'$ is to the future of
  ${\cal S}$.}.

In the limit  $\Delta {\cal B}_j \to {\cal B}_j$, the quantity
(\ref{qlocdef}) should reduce to the contribution from ${\cal B}_j$ to
the global charge $\mc Q_\xi$.  A natural candidate prescription for
defining a $d-2$-form
$\df {\mc Q}_\xi^{\rm loc}$ that would achieve this is given by taking
$\Sigma = \Delta {\cal B}_j$ in the definition (\ref{Hamiltonian
  presymplectic}), or, from Eqs.\ (\ref{Hamiltonian Definition}) and
(\ref{qlocdef}),
\be
\delta \df {\mc Q}_\xi^{\rm loc} = \delta \df Q_\xi - i_\xi \df
\theta.
\label{candidate}
\ee
However, the corresponding charge (\ref{localcharge}) will generally
not exist because of the obstruction (\ref{Hamiltonian Condition}).
One would like to modify the right hand side of Eq.\ (\ref{candidate})
in such a way as to remove this obstruction, without changing the
integral on the left hand side of (\ref{qlocdef}) in the limit $\Delta {\cal B}_j \to {\cal B}_j$.
One would also like to find a natural prescription for this
modification that yields unique charges.
One could then interpret Eq.\ (\ref{qlocdef}) as a localized
conservation law, which equates a flux through a region of ${\cal
  B}_j$ with the difference between the charges at the two
crosssections.  (A distinct kind of global conservation law involving
global charges $\mc Q_\xi$ is discussed in Sec.\ \ref{sec:globalcons} below.)

Wald and Zoupas \cite{WZ} suggested a prescription of this kind that gives unique
answers under certain conditions, which can be summarized as follows
(we omit some subtleties related to taking the limit to asymptotic
boundaries that will not be relevant for our application):
\begin{enumerate}[label=\arabic*.]
\item Compute the pullback $\df \omega(\bar{\phi},\bar{\delta_1
  \phi},\bar{\delta_2 \phi})$ to the boundary component ${\cal B}_j$ of
  the presymplectic current $\df \omega(\phi, \delta_1\phi,
  \delta_2\phi)$.  Here the barred fields are the dynamical fields on
  the boundary induced by the solution $\phi \in \bar{\ms F}$ and linearized solutions
  $\delta_1 \phi, \delta_2 \phi$ tangent to $\bar{\ms F}$, obtained by
  taking pullbacks of these fields (and possibly their derivatives) to
  the boundary.

\item Choose a presymplectic potential $\df \Theta(\bar{\phi},\bar{
    \delta \phi})$ on ${\cal B}_j$ for the pullback $\df\omega$, that is, a $d-1$-form
    which satisfies
\be
{\df \omega}(\bar{\phi}, \bar{\delta_1 \phi}, \bar{\delta_2 \phi}) =
\delta_1 \df \Theta(\bar{\phi}, \bar{\delta_2 \phi})
-\delta_2 \df \Theta(\bar{\phi}, \bar{\delta_1 \phi}).
\label{Thetadef1}
\ee
We require that the dependence of $\df \Theta$ on the dynamical fields
on the boundary, as well as the dependence on fields
in any universal background structure on ${\cal B}_j$ inherent in the
definition of the field configuration space $\ms F$, be local and
covariant\footnote{\label{localcovariant}What this means is as follows. The presympletic potential $\df \Theta$
depends on a field configuration $\phi$, its variation $\delta \phi$, a universal background structure on ${\cal B}_j$ which we denote by $\mf p$, and on the boundary ${\cal B}_j$: $\df \Theta = \df \Theta(\phi,\delta \phi, \mf p, {\cal B}_j)$.  Locality and covariance requires that for any diffeomorphism $\psi: M \to M$,
$$
\psi_* \df \Theta(\phi,\delta \phi, \mf p, {\cal B}_j) = \df \Theta(\psi_* \phi,\psi_* \delta \phi, \psi_* \mf p, \psi^{-1}({\cal B}_j)),
$$
where $\psi_*$ is the pullback.
If we specialize to diffeomorphisms which preserve the boundary, $\psi^{-1}({\cal B}_j) = {\cal B}_j$, and the universal background structure on the boundary, $\psi_* \mf p = \mf p$, then
$\psi_* \df \Theta(\phi,\delta \phi, \mf p, {\cal B}_j) = \df \Theta(\psi_* \phi,\psi_* \delta \phi, \mf p, {\cal B}_j)$.}.  (See Secs.\ \ref{sec:universal} and \ref{sec:cpsnb} for more details on universal
background structures.)
\item Add the term $i_\xi \df \Theta$ to the right hand side of Eq.\
  (\ref{candidate}), thus giving from Eq.\ (\ref{localcharge}) the following
  formula for the variation of the localized charge:
\be
  \delta {\mc Q}_\xi^{\rm loc}({\cal S}) = \int_{\cal S} \delta \df
  {\mc Q}_\xi^{\rm loc} = \int_{\cal S} \delta \df
  Q_\xi - i_\xi \df \theta + i_\xi \df \Theta.
\label{localcharge1}
\ee

\item Now repeating the computation that led to Eq.\ (\ref{Hamiltonian
    Condition}) shows that the obstruction now vanishes.  The
definition (\ref{localcharge1}) therefore determines the charge ${\mc
  Q}_\xi^{\rm loc}({\cal S})$ on
$\bar{\ms F}$ up to constants of integration on phase space, which can be specified by
demanding that the charges vanish on a reference solution\footnote{And
on all solutions related to $\phi_0$ by linearized diffeomorphisms.
See Appendix \ref{app:background} for further discussion of this point.} $\phi_0$ on
each connected component of $\bar{\ms F}$,
\be
\left. {\mc Q}_\xi^{\rm loc}({\cal S})\right|_{\phi = \phi_0} = 0,
\label{cc11}
\ee
for all symmetry
representatives $\xi^a$ and cuts ${\cal S}$ \cite{WZ}.

\item In order to reduce the non-uniqueness in the boundary
  presymplectic potential $\df \Theta$, we
  impose the requirement that
\be
\df \Theta(\bar{ \phi}, \bar {\delta \phi}) =0
\label{stat-req}
\ee
for all $\bar {\delta \phi}$ whenever ${\bar \phi}$ is
stationary\footnote{\label{statdef}By ``stationary at ${\cal B}_j$''
  we mean that
there exists a representative $\tau^a$
  of an infinitesimal boundary symmetry at ${\cal B}_j$ which
is timelike and satisfies the Killing equation on ${\cal B}_j$ and to
first order in deviations off ${\cal B}_j$.  This is a weaker notion
than used in \cite{WZ}.}
at
${\cal B}_j$.  We also impose that the reference solution $\phi_0$ be
stationary at ${\cal B}_j$.

\end{enumerate}

The motivation for the fifth requirement is as follows \cite{WZ}.  It is natural
on physical grounds to demand that the flux $d \df {\mc Q}_\xi^{\rm
  loc}$ vanish for solutions which are stationary at the boundary
${\cal B}_j$.
Taking the exterior derivative of the integrand in Eq.\
(\ref{localcharge1}) and using Eq.\ (\ref{eq:omega-Q}) and the fact
that $d$ and $\delta$ commute we get
\begin{eqnarray}
        \delta d \df {\mc Q}_\xi^{\rm loc} &=& \df\omega(\phi;\delta
        \phi, \lie_\xi \phi) + d \lb[ i_\xi \df\Theta(\phi; \delta \phi)
        \rb] = \df\omega(\phi;\delta \phi, \lie_\xi \phi) + \lie_\xi
        \df\Theta(\phi; \delta \phi) \nonumber \\
&=& \delta \df\Theta(\phi;\lie_\xi \phi).
\end{eqnarray}
To integrate this on \(\bar{\ms F}\), note that $\df {\mc Q}_\xi^{\rm
  loc}$ must vanish identically on $\phi_0$ by Eq.\ (\ref{cc11}), while
$\df \Theta(\phi_0,\delta \phi)$ vanishes by Eq.\ (\ref{stat-req}).  Thus we
obtain
\be
d \df{\mc Q}_{\xi}^{\rm loc} = \df{\Theta}(\phi; \lie_\xi \phi),
\label{eq:WZ-flux}
\ee
and so the flux vanishes identically on stationary solutions as
desired, by Eq.\ (\ref{stat-req}).

A useful method of parameterizing choices of $\df \Theta$ that automatically satisfy
all the requirements apart from the stationary requirement (\ref{stat-req}) is
\be
\df \Theta = \df \theta  - \delta \df \alpha,
\label{Generalized Potential Prescription}
\ee
where the first term on the right hand side is the pullback of the
presymplectic potential $\df \theta$, and $\df \alpha$ is some $d-1$-form
on ${\cal B}_j$ constructed from $\bar{\phi}$.
Inserting this into Eq.\ (\ref{localcharge1}), integrating in the covariant
phase space $\bar {\ms F}$ and using Eq.\ (\ref{cc11}) now gives
\be
\mc Q_\xi^{\rm loc}({\cal S}) = \int_{\cal S}  \df Q_\xi -
  i_\xi \df \alpha,
\ee
if the right hand side vanishes on the reference solution $\phi =
\phi_0$. In \cref{sec:wzcharges} we will show that at a null boundary
for vacuum general relativity one can choose \(\df\alpha\) so that
$\df{\Theta}$ satisfies the criteria outlined above, with the
definition of stationary of footnote \ref{statdef} replaced by the
weaker notion of shear free and expansion free.

Finally, the global charges $\mc Q_\xi$ discussed in Sec.\ \ref{sec:globalcharges}
above can often be written in terms of the localized charges $\mc Q_\xi^{\rm
  loc}({\cal S})$ discussed here, specialized to specific cross sections ${\cal S}$:
\be
\mc Q_\xi = \sum_j \mc Q_\xi^{\rm loc}({\cal S}_j) = \sum_j \int_{{\cal S}_j} \, \df {\mc Q}_\xi^{\rm loc},
\label{locglobal}
\ee
where the boundary $\partial \Sigma$ of a Cauchy surface $\Sigma$ is a union
$\partial \Sigma = \cup_j {\cal S}_j$ of disconnected components ${\cal S}_j$.
The relation (\ref{locglobal}) will hold when the correction term $i_\xi \df
\Theta$ in the definition (\ref{localcharge1}) of the localized charge vanishes on
$\partial \Sigma$, from the definition (\ref{Hamiltonian Definition}),
if the same reference solution is used for the localized and global charges.
We expect the correction term $i_\xi \df \Theta$ to generically vanish
on $\partial \Sigma$ when $\Sigma$ is a Cauchy surface.  Some examples
where this occurs are:
\begin{itemize}
\item At future null infinity $\scri^+$,
the correction term $i_\xi \df \Theta$ is proportional to the
generator $\xi^a$ times the News tensor (Eq.\ (73) of
  \cite{WZ}).  Letting $u$ denote an affine parameter along $\scri^+$,
  the generator scales as $\sim |u|$ as $u \to \pm \infty$, and so if the
  News tensor decays faster than $1/|u|$, the contributions from the
  boundaries $\scri_\pm^+$ of $\scri^+$ will vanish [cf.\ the
  discussion before Eq.\ (\ref{split}) above].

\item For a future event horizon ${\cal H}^+$, we show in Appendix \ref{sec:bh}
  that the contribution to the correction term from the future
  boundary ${\cal H}_+^+$ (the limit to $i^+$) of the horizon vanishes, if the shear obeys a
  suitable decay condition near ${\cal H}^+_+$.  We also show that the contribution from
  a bifurcation two-sphere ${\cal H}^+_-$ vanishes.

\end{itemize}

Explicit expressions for $\mc Q_\xi^{\rm loc}({\cal S})$ for cross
sections ${\cal S}$ of future null infinity $\scri^+$ are given in
Eqs.\ (92) and (98) of Wald and Zoupas \cite{WZ}, and specialized to
Bondi coordinates in Eq.\ (3.5) of Ref.\ \cite{Flanagan:2015pxa}.  For
cross sections of an arbitrary null surface, our result for $\mc Q_\xi^{\rm loc}({\cal S})$
is given in Eq.\ (\ref{GenHam}) below.

\subsection{Potential ambiguities in global and localized charges}
\label{sec:ambiguities}

We next discuss some ambiguities that can arise in the definitions
and constructions outlined above of global and localized charges
\cite{WZ,Jacobson:1993vj,2017NuPhB.924..312G}.  Wald and Zoupas show that
these ambiguities can be resolved
in vacuum general relativity at future null infinity.  We will
similarly argue that they can be resolved at null boundaries at finite
locations.  However, they may be significant for other theories or at other
types of boundary.

First, the definition (\ref{Lagrangian Variation}) of the presymplectic
potential $\df \theta$
determines it up to a closed
form. Since we require that \(\df\theta\) be local and covariant this
closed form is also exact \cite{W-closed}.  The corresponding
ambiguities are
\begin{subequations}
\label{amb2}
\begin{eqnarray}
\label{amb2aa}
\df \theta(\phi,\delta \phi) &\to& \df \theta(\phi,\delta \phi) +
d \df Y(\phi,\delta \phi), \\
\df \omega(\phi,\delta_1 \phi,\delta_2 \phi) &\to& \df
\omega(\phi,\delta_1 \phi,\delta_2 \phi) + d \left[
 \delta_1 \df  Y(\phi,\delta_2 \phi) -  \delta_2 \df  Y(\phi,\delta_1 \phi) \right]
\end{eqnarray}
\end{subequations}
for some $(d-2)$-form $\df Y$.
These give rise to the following transformations of the presymplectic
potential $\df \Theta$ and of the localized charge $\mc Q_\xi^{\rm
  loc}({\cal S})$:
\begin{subequations}
\label{amb2a}
\begin{eqnarray}
\df \Theta(\phi,\delta \phi) &\to& \df \Theta(\phi,\delta \phi) +
d \df Y(\phi,\delta \phi), \\
\mc Q_\xi^{\rm  loc}({\cal S})
&\to& \mc Q_\xi^{\rm  loc}({\cal S}) + \int_{\cal S} \df Y(\phi,
\lie_\xi \phi).
\end{eqnarray}
\end{subequations}
One can demand that the maximum number of derivatives of
the fields $\phi$ or their variations $\delta \phi$ in the $(d-2)$-form
$\df Y$ be two less then the number of derivatives appearing in the
Lagrangian.  This requirement is in some sense natural,
since otherwise the number of derivatives in $\df \theta$ from Eq.\ (\ref{amb2aa}) exceeds
what one would naively expect from Eq.\ (\ref{Lagrangian Variation}).
In Sec.\ \ref{sec:noether} below we argue that this requirement
eliminates the ambiguity (\ref{amb2}) for vacuum
general relativity.

Second, the definition (\ref{Thetadef1}) of the presymplectic potential $\df
\Theta$ determines it only up a transformation of the form
\be
\df \Theta(\phi,\delta \phi) \to \df \Theta(\phi,\delta \phi) + \delta
\df W(\phi),
\ee
where $\df W$ is constructed locally and covariantly from the field
$\phi$ and from any universal background structure on ${\cal B}_j$.
The localized charge transforms under this ambiguity as
\be
\mc Q_\xi^{\rm loc}({\cal S}) \to \mc Q_\xi^{\rm loc}({\cal S}) +
\int_{\cal S} i_\xi \df W.
\ee
From the requirement (\ref{stat-req}) it follows that $\delta \df
W(\phi)$ must vanish for all solutions $\phi$ that are stationary at
${\cal B}_j$, and for all linearized solutions $\delta \phi$.
If one additionally assumes that $\df W$ depends analytically on the
fields, it follows that $\df W=0$ at future null infinity $\scri^+$ in
vacuum general relativity \cite{WZ}.  We give a similar argument in
Sec.\ \ref{sec:wz} below to show that the ambiguity $\df W$ vanishes
at finite null surfaces, if we assume that the maximum number of derivatives
appearing in $\df W$ is one less than the number of derivatives
appearing in the Lagrangian.

Third, one can redefine the Lagrangian by an exact form,
\(\df L \to \df L + d \df K\), without changing the equations of motion
of the theory.  The corresponding transformations of the presymplectic potential \(\df\theta\),
presymplectic current $\df \omega$, Noether charge $d-2$-form $\df Q_\xi$ and
the integrands $\delta \df Q_\xi - i_\xi \df \theta$ and $d \df {\mc Q}_\xi^{\rm loc}$ of the symmetry generator
charge (\ref{Hamiltonian Definition}) and localized charge (\ref{localcharge1}) are given by
\begin{subequations}
\label{amb1}
\begin{eqnarray}
\df \theta(\phi,\delta \phi) &\to& \df \theta(\phi,\delta \phi) + \delta
\df K(\phi), \\
\df \omega(\phi,\delta_1 \phi,\delta_2 \phi) &\to& \df
\omega(\phi,\delta_1 \phi,\delta_2 \phi), \\
\df Q_\xi(\phi) &\to&  \df Q_\xi(\phi) + i_\xi \df K(\phi), \\
\delta \df Q_\xi - i_\xi \df \theta &\to&  \delta \df Q_\xi - i_\xi \df \theta, \\
{\df {\mc Q}}_\xi^{\rm loc}(\phi)
&\to&
{\df {\mc Q}}_\xi^{\rm loc}(\phi).
\end{eqnarray}
\end{subequations}
While this transformation does affect the Noether charge, it does not affect the symmetry generator charge
$\mc Q_\xi$ and localized charge $\mc Q_\xi^{\rm loc}$ that are of the most interest for this paper.


\section{Review of the local geometry of null hypersurfaces}
\label{sec:null-geometry}

\subsection{Foundations}

In this section we review the local geometry of null hypersurfaces
\cite{Ashtekar:2001jb,Gourgoulhon:2005ng}, in order to fix our
notations and conventions.
For the remainder of the paper we specialize to $3+1$ spacetime dimensions.
Suppose we are given a spacetime $(M, g_{ab})$, and a null hypersurface ${\cal N}$ in $M$ whose topology is
${\cal Z} \times \mathbb{R}$ for some base space ${\cal Z}$.  We denote
by $\ell_a$ a choice of future directed, null normal to the surface
${\cal N}$.  This normal is not unique but
can be rescaled according to
\be
\ell_a \goesto e^\sigma \ell_a,
\label{rescale}
\ee
where $\sigma$ is any smooth function on ${\cal N}$.
We define the non-affinity $\kappa$, a function on ${\cal N}$, by
\be
\ell^a \nabla_a \ell^b \hateq \kappa l ^b.
\label{kappadef}
\ee
As a reminder here we are using
$\hateq$ to mean equality when restricted to ${\cal N}$.
The non-affinity transforms under the rescaling
(\ref{rescale}) as
\be
\kappa \goesto e^\sigma(\kappa + \lie_{\ell} \sigma).
\label{rescale1}
\ee
We will adopt the terminology that any quantity $f$ which transforms under the transformation
(\ref{rescale})  as
\be
f \goesto e^{-n \sigma} f
\label{scalingweight}
\ee
has {\it scaling weight} $n$.

We can identify the tangent
space $T_p({\cal N})$ to ${\cal N}$ at a point $p$ with the subspace
of the tangent space $T_p(M)$ consisting of vectors $v^a$ with $v^a
\ell_a=0$.
Since $\ell^a \equiv g^{ab} \ell_b$ lies in this subspace we can identify it with a
vector field $\ell^i$ on ${\cal N}$, the integral curves of which are the
null generators of the null surface.
(Recall that we use lowercase Roman indices
$i,j, \ldots$ to
denote tensors intrinsic to ${\cal N}$.)
Next, the pullback map takes covectors $w_a$ on $M$ evaluated on ${\cal N}$ to
covectors $w_i$ on ${\cal N}$.
We denote this pullback map
by
\be
w_a \goesto \pullback_i^a w_a,
\ee
thereby defining the quantity $\pullback_i^a$.  The pullback of the
null
normal covector $\ell_a$ vanishes identically by
definition, since all vectors on ${\cal N}$ are orthogonal to $\ell_a$:
\be
\pullback_i^a \ell_a =0.
\label{lAvanish}
\ee

A question that often arises in computations is when can a contraction
$w_a v^a$ of spacetime tensors be replaced by a corresponding
contraction $w_i v^i$ of tensors intrinsic to ${\cal N}$.  First,
given $w_a$ and $v^a$,
while $w_i$ can be defined using the pullback, the quantity $v^i$ is
not necessarily well defined; it is defined only when
$ \ell_a v^a=0$.  When this condition is satisfied, the contractions coincide:
\be
\ell_a v^a=0 \ \ \ \  \implies \ \ \ \  w_a v^a = w_i v^i.
\label{downshift}
\ee

A similar issue arises in going from three dimensions down to two dimensions.
We denote by $W_p$ the two dimensional subspace of the dual space
$T_p({\cal N})^*$ consisting of covectors $w_i$ that satisfy $w_i \ell^i
=0$. We will denote by abstract indices $A$, $B$ etc. tensors built on
$W_p$.  When can a contraction
$w_i v^i$ of tensors on ${\cal N}$ be replaced by a corresponding
contraction $w_A v^A$ of tensors in $W_p$ and $W_p^*$?  The answer in
this case is the opposite of that for going from four to three
dimensions.  First, given $w_i$ and $v^i$, the quantity $v^A$ is
always well defined by considering $v^i$ as a linear map on $T_p({\cal
  N})^*$ and restricting its action to $W_p$ (we shall call this
operation a pullback). On the other hand,
it is necessary that $w_i \ell^i =0$ in order that $w_A$ be defined.
When this condition is satisfied, the contractions coincide:
\be
w_i \ell^i =0 \ \ \ \  \implies \ \ \ \ w_i v^i = w_A v^A.
\label{downshift1}
\ee

\subsection{Geometric fields defined on a null hypersurface}

We denote by $q_{ij}$ the induced metric on ${\cal N}$
\be
q_{ij} = \pullback_i^a \pullback_j^b g_{ab},
\label{inducedq}
\ee
which has signature $(0,+,+)$.  Taking the pullback of the relation
$\ell_a = g_{ab} \ell^b$ and using Eq.\ (\ref{downshift}) gives
\be
q_{ij} \ell^j = 0,
\label{eigen}
\ee
{\it i.e.}, $\ell^i$ is a eigenvector of the induced metric with
eigenvalue zero.  It follows that we can regard $q_{ij}$ as a tensor
in $W_p \otimes W_p$, which we write as $q_{AB}$.  This has a unique
inverse in $W_p^* \otimes W_p^*$ which we write as $q^{AB}$.
We will use $q_{AB}$ and $q^{AB}$ to freely raise and lower capital Roman indices.

The second fundamental form of the surface ${\cal N}$ is given by
\be
K_{ij} = \pullback_i^a \pullback_j^b \nabla_a \ell_b.
\label{eq:sff}
\ee
Since $\ell_a$ is normal to a hypersurface we have $\ell_{[a} \nabla_{b} \ell_{c]}
  \hateq 0$ or $\nabla_{[a} \ell_{b]} \hateq \ell_{[a} w_{b]}$ for some
  $w_b$, and taking the pullback and using (\ref{lAvanish}) gives
\be
K_{[ij]} =0.
\label{sffp}
\ee
Similarly, lowering the index in Eq.\ (\ref{kappadef}), taking the
pullback and using Eqs.\ (\ref{lAvanish}) and (\ref{downshift}) gives
\be
\ell^i K_{ij} =0.
\label{sffp1}
\ee
It follows that $K_{ij}$ lies in $W_p \otimes W_p$ and so can be
written as $K_{AB}$.  We can uniquely
decompose the second fundamental form as
\be
K_{AB} = \frac{1}{2} \theta q_{AB} + \sigma_{AB},
\label{sffd}
\ee
where $\theta$ is the expansion and the shear $\sigma_{AB}$ is traceless, $q^{AB} \sigma_{AB}=0$.
This equation can also be written as $K_{ij} = \theta q_{ij}/2 + \sigma_{ij}$.

The second fundamental form is related to the Lie derivative of the
induced metric.  Taking the pullback of the identity
$
\lie_\ell g_{ab} = 2 \nabla_{(a} \ell_{b)}
$
and using the fact that the pullback commutes with the Lie derivative gives
\be
K_{ij} = \frac{1}{2} \lie_\ell q_{ij}.
\label{identity22}
\ee

Consider next the object
\be
\Pi^a_i \nabla_a \ell^b.
\label{weingarten}
\ee
This tensor is orthogonal to the normal on the $b$ index, since
$\ell_b \Pi^a_i \nabla_a \ell^b = \Pi^a_i \nabla_a (\ell_b \ell^b)/2 = 0$,
since $\ell_b \ell^b=0$ on ${\cal N}$ and the derivative is along the surface.
Therefore this quantity is an
intrinsic tensor which we write as
\be
{\cal K}_i^{\ j},
\ee
called the Weingarten map \cite{Gourgoulhon:2005ng}.  From Eqs.\
(\ref{kappadef}) and (\ref{downshift}) it follows that
\be
{\cal K}_i^{\ j} \ell^i = \kappa \ell^j.
\label{sffp2}
\ee
Similarly taking the pullback of the relation $\nabla_a \ell^b g_{bc}
= \nabla_a \ell_c$ and using (\ref{downshift}) and (\ref{inducedq}) gives that
\be
{\cal K}_i^{\ j} q_{jk} = K_{ik}.
\label{sffp3}
\ee
It follows from Eqs.\ (\ref{sffp}), (\ref{sffp1}), (\ref{sffp2}) and
(\ref{sffp3}) that the Weingarten map ${\cal K}_i^{\ j}$ has six independent nonzero
components in general in four spacetime dimensions, three of which are determined by the second
fundamental form $K_{ij}$, and one of which is determined by the
non-affinity $\kappa$, leaving two additional independent components [see
Appendix \ref{app:translate} for more details, especially Eqs.\ (\ref{wme}) and (\ref{rofe})].

Next, a choice of volume form $\varepsilon_{abcd}$ on spacetime determines a volume form
$\varepsilon_{ijk}$ on ${\cal N}$ as follows.
We consider three-forms
${\bar \varepsilon}_{abc}$ on ${\cal N}$ which satisfy
\be
4 {\bar \varepsilon}_{[abc} \ell_{d]} \hateq \varepsilon_{abcd},
\label{barepsilon}
\ee
and then take the pullback of these three-forms:
\be
\varepsilon_{ijk} = \Pi^a_i \Pi^b_j \Pi^c_k \ {\bar
  \varepsilon}_{abc}.
\label{ve3}
\ee
Although ${\bar \varepsilon}_{abc}$ is not unique, its pullback
$\varepsilon_{ijk}$ is.  We define the antisymmetric tensor $\varepsilon^{ijk}$ by
\be
\varepsilon^{ijk} \varepsilon_{ijk} = 3!,
\ee
and the two-form $\varepsilon_{ij}$ by
\be
\varepsilon_{ij} = - \varepsilon_{ijk} \ell^k.
\label{ve2}
\ee


Under the scaling transformation (\ref{rescale}) the various quantities
defined in this subsection transform as
\begin{subequations}
\label{rescale00}
\begin{eqnarray}
q_{ij} &\to& q_{ij}, \\
K_{ij} &\to& e^\sigma K_{ij}, \\
{\cal K}_i^{\ j} &\to& e^\sigma \left( {\cal K}_i^{\ j} + D_i \sigma
  \, \ell^j \right),
\label{calKtrans}\\
\theta &\to& e^{\sigma}
\theta, \label{thetascale}\\
\varepsilon_{ijk} &\to& e^{-\sigma}
\varepsilon_{ijk}, \label{vescale}\\
\varepsilon^{ijk} &\to& e^{\sigma}
\varepsilon^{ijk}, \\
 \varepsilon_{ij} &\to&  \varepsilon_{ij},
\end{eqnarray}
\end{subequations}
where $D_i$ is any derivative operator on ${\cal N}$.

\subsection{Divergence operator}

Although there is no preferred derivative operator on ${\cal N}$, one
can define a divergence operation
$
v^i \to {\hat D}_i v^i
$
on vector fields via
\be
{\hat D}_i v^i = \frac{1}{2} \varepsilon^{ijk} D_k ( \varepsilon_{ijm} v^m),
\label{divergenceop}
\ee
where $D_i$ is again any derivative operator on ${\cal N}$.  The right hand
side is independent of the choice of $D_i$ since it enters as an
exterior derivative.  

We can relate this divergence operator to the four dimensional
divergence operator as follows.
A vector field $v^i$ on ${\cal N}$ corresponds to a unique vector field $v^a$
on ${\cal N}$ with $v^a \ell_a \hateq 0$.  Now choose an extension of
$v^a$ to a neighborhood of ${\cal N}$ in $M$.
The linearized diffeomorphism associated with $v^a$ maps ${\cal N}$
into itself, and therefore preserves the
normal $\ell_a$ up to a rescaling.  Therefore there exists a function
$\varpi$ on ${\cal N}$ which depends on $v^a$ such that
\be
\lie_v \ell_a \hateq \varpi \ell_a.
\label{varpidef}
\ee
The relation between the two divergence operators is\footnote{
This relation can be derived by specializing to a coordinate system
$(r, y^1, y^2, y^3) = (r, y^\Gamma)$ for which the hypersurface ${\cal
  N}$ is given by $r=0$ and with $\ell_a \hateq (d r)_a$.  Writing the
volume form as $\df \varepsilon = e^\Upsilon dr \wedge dy^1 \wedge
dy^2 \wedge dy^3$ for some function $\Upsilon$, the left hand side of
Eq.\ (\ref{dividentity}) can be written as
$$
e^{-\Upsilon} \partial_r (e^\Upsilon v^r) +
e^{-\Upsilon} \partial_\Gamma (e^\Upsilon v^\Gamma) =
 \partial_r v^r +
e^{-\Upsilon} \partial_\Gamma (e^\Upsilon v^\Gamma).
$$
The first term on the right hand side here is $\varpi$, while the
second term is the intrinsic divergence ${\hat D}_i v^i$, by Eqs.\
(\ref{barepsilon}), (\ref{ve3}) and (\ref{divergenceop}).}
\be
\nabla_a v^a \hateq {\hat D}_i v^i + \varpi.
\label{dividentity}
\ee

The divergence of the normal is
\be
{\hat D}_i \ell^i =  \theta.
\label{divell0}
\ee
This follows from the relation (\ref{dividentity}), the definition
(\ref{varpidef}) of $\varpi$, and the trace of Eq.\ (\ref{ffff}).

\subsection{Stationary regions of null hypersurfaces}
\label{sec:stationary}

As discussed in Sec.\ \ref{sec:WZ-charge} above, we shall call a
region of a null surface {\it stationary} if there is a choice of
normal
covector $\tau_a$ in that region which satisfies Killings equation on the surface and to first order in deviations off the surface,
\begin{subequations}
\label{killing}
\begin{eqnarray}
\label{killing0}
\lie_\tau g_{ab} &\hateq& 0, \\
\nabla_c \lie_\tau g_{ab} &\hateq& 0.
\label{killing1}
\end{eqnarray}
\end{subequations}
We will denote the corresponding value of $\kappa$ by $\kappa_\tau$,
the surface gravity.  Taking the pullback of Eq.\ (\ref{killing0})
and using the fact that the pullback commutes with the Lie derivative gives
\be
\lie_\tau q_{ij} =0,
\label{stat00}
\ee
and it follows from Eq.\ (\ref{identity22}) that
\be
K_{ij} =0,
\label{stationary1}
\ee
i.e. that the surface is shear free and expansion free.

It follows from the condition (\ref{stationary1})
together with Eqs.\ (\ref{r1ff})
that the rotation one-form defined by
\be
\omega_i = - {\cal K}_i^{\ j} n_j,
\label{r1f0}
\ee
where $n_i$ is any covector with $n_i \ell^i =-1$, is independent of
the choice of $n_i$.  This is true only for null surfaces that satisfy
(\ref{stationary1}).  Under the transformation (\ref{rescale})
$\omega_i$ transforms as $\omega_i \to \omega_i +
D_i \sigma$, from Eqs.\ (\ref{calKtrans}) and (\ref{r1ff}).

We define
\be
\omega_{\tau\, i} = \omega_i {|}_{{\vec \ell} = {\vec \tau}}
\label{r1fstat}
\ee
to be the rotation one-form $\omega_i$
specialized to the choice of representative $\ell^i=\tau^i$ \footnote{See
Ashtekar \cite{Ashtekar:2001jb} for an alternative method of
defining $\omega_{\tau\,i}$.}.  Now Eq.\ (\ref{killing}) together with
Eq.\ (C.3.6)
of Wald \cite{Wald-book} imply that
$\lie_\tau \nabla_a \tau^b \hateq 0$, and taking a pullback yields $\lie_\tau
{\cal K}_i^{\ j}=0$.  Combining this with Eqs.\ (\ref{twistprop}) and (\ref{r1ff}) now shows
that the nonaffinity and rotation one-form are Lie transported along
the null surface:
\be
\lie_\tau \kappa_\tau =0, \ \ \ \ \ \lie_\tau \omega_{\tau\,i} =0.
\label{stationary2}
\ee
More generally, the Bardeen-Carter-Hawking derivation
\cite{Bardeen:1973gs} of the zeroth
law of black hole thermodynamics,
\be
D_i \kappa_\tau=0,
\label{zeroth}
\ee
applies in this context, assuming the Einstein equations and the
dominant energy condition.  In the remainder of the paper
we will be working in the context of vacuum general relativity,
for which (\ref{zeroth}) will be satisfied in stationary regions.

\subsection{Orthonormal basis formalism}

Finally, it is sometimes useful for computational purposes to choose an auxiliary
null vector field $n^a$ on ${\cal N}$ which together with $\ell^a$ forms
part of an orthonormal basis.  Some aspects of the formalism
described above simplify when described in the language of an
orthonormal basis, although that language
does carry the baggage of an arbitrary choice.  While the
main results of this paper will not require a choice of auxiliary
null vector, we will translate our results into the language of the
orthonormal basis formalism since it is widely used.  Details of the relation
between the covariant and orthonormal basis formalisms for null
surfaces are given in Appendix \ref{app:translate}.

\section{Universal intrinsic structure of a null hypersurface}
\label{sec:universal}

In this section we will describe an intrinsic geometric structure
on null hypersurfaces ${\cal N}$ that is determined by the spacetime geometry.
It is universal in the sense that for a given ${\cal N}$ any two such structures are diffeomorphic.
We will define the structure in Sec.\ \ref{sec:uis}, and in Sec.\ \ref{sec:uiss} we
will describe the symmetry group of diffeomorphisms from ${\cal N}$ to ${\cal N}$ that
preserve the structure.  The corresponding Lie algebra is described in
Sec.\ \ref{sec:uiss2}; we will show in Sec.\ \ref{sec:cpsnb}
that this symmetry algebra coincides
with that obtained from a particular definition of covariant phase space for general relativity with a null boundary
in the Wald-Zoupas approach.
Section \ref{sec:subalg0} discusses preferred subalgebras associated
with stationary regions of the null hypersurface.
Finally in Sec.\ \ref{sec:uiss1} we discuss how the group and algebra
are modified in the case where the null hypersurface has a boundary
$\partial {\cal N}$ in $M$.

\subsection{Definition of intrinsic structure}
\label{sec:uis}

Consider a manifold ${\cal N}$ which is equipped with a smooth, nowhere vanishing vector field
$\ell^i$ and a smooth function $\kappa$.
Letting ${\cal Z}$ denote the manifold of integral curves, we assume that
${\cal N}$ is diffeomorphic to the product ${\cal Z} \times \mathbb{R}$.
We define an equivalence relation on such pairs $(\ell^i,\kappa)$
by saying that two pairs
are equivalent if they are related by a
rescaling of the form [cf.\ Eqs.\ (\ref{rescale}) and (\ref{rescale1}) above]
\begin{subequations}
\label{rescale-equiv}
\begin{eqnarray}
\ell^i &\goesto& e^\sigma \ell^i, \\
\kappa &\goesto& e^\sigma(\kappa + \lie_{\ell} \sigma)
\end{eqnarray}
\end{subequations}
where $\sigma$ is a smooth function on ${\cal N}$.
We denote by
\be
\frak{u} = [\ell^i,\kappa]
\label{eq:gs}
\ee
the equivalence class associated with $(\ell^i,\kappa)$.
A choice of equivalence class is the desired intrinsic geometric structure on ${\cal N}$.

Suppose now we are given a spacetime $(M,g_{ab})$ with null boundary ${\cal N}$.  The spacetime geometry
then determines a structure $[\ell^i,\kappa]$ in the manner described
in Sec.\ \ref{sec:null-geometry} above: the vector $\ell^i$ is
obtained by raising the index on a choice of normal covector, and $\kappa$
is the non-affinity of that vector.  The resulting equivalence class
$[\ell^i,\kappa]$ is independent of the choice of normalization of the
covector, by the equivalence relation (\ref{rescale-equiv}).

The intrinsic structure determines a class of foliations of ${\cal N}$ as follows.
Choose a cross section ${\cal S}$ of ${\cal N}$, a surface which each integral curve
intersects exactly once, which will be diffeomorphic to the base space ${\cal Z}$.
Out of the equivalence class $[\ell^i,\kappa]$, pick a member
$(\ell_0^i,0)$ for which the non-affinity vanishes, by starting with a
general member $(\ell^i,\kappa)$ and solving the differential equation
$\kappa + \lie_\ell \sigma=0$ for the scaling function $\sigma$.
Now Lie drag the cross section ${\cal S}$ along integral curves of $\ell_0^i$.
The resulting foliation\footnote{The class of foliations generated in this way has considerable freedom.
One can pick the initial cross section ${\cal S}$ arbitrarily, and in addition
one can pick a second arbitrary cross section ${\cal S}'$ disjoint from ${\cal S}$ and arrange for it
to belong to the foliation, by exploiting the rescaling freedom
$\ell_0^i \to e^\sigma \ell_0^i$ with $\lie_{\ell_0} \sigma=0$.
However, once ${\cal S}$ and ${\cal S}'$ are specified, the foliation
is uniquely determined.}
 will be level sets of a coordinate $u$ which
is determined by the properties that $u=0$ on ${\cal S}$ and $\ell_0^i D_i u =1$.
In addition, if $\theta^A$ is any coordinate system on ${\cal S}$, one
can extend the definition of these coordinates to ${\cal N}$ by
demanding that they be constant along the integral curves, thereby
generating a coordinate system $(u,\theta^A)$ on ${\cal N}$ for which
${\vec \ell}_0 = \partial_u$.

We will say that an intrinsic structure is {\it complete} if all of
the generators of ${\cal N}$ can be extended to arbitrary values of
affine parameter in both directions (where $u$ is an affine parameter
if ${\vec \ell} = \partial_u$ with $\kappa=0$).
For example, the future light cone of a point ${\cal P}$ in Minkowski spacetime
(with ${\cal P}$ itself removed)
is not complete when the intrinsic structure induced by the flat
Minkowski metric is used, since all of the generators start at ${\cal
  P}$.  By contrast, the event horizon in maximally extended
Schwarzschild is complete (see Appendix \ref{app:background}).
We will study both types of intrinsic structure later in this paper.

Given two different complete intrinsic structures $\frak{u} = [\ell^i,\kappa]$ and
$\frak{u}' = [\ell^{\prime\,i},\kappa']$ on ${\cal N}$, there exists a
diffeomorphism $\varphi : {\cal N} \to {\cal N}$ which maps $\frak{u}$
onto $\frak{u}'$. In this sense the complete intrinsic structure is {\it universal},
in the same way that an intrinsic structure of a different kind on future null infinity
is universal in the BMS construction \cite{Ashtekar:2014zsa}.
The existence of the
diffeomorphism $\varphi$ can be shown as follows.
Choose a cross section ${\cal S}$ of ${\cal N}$, and using $\frak{u}$
construct a coordinate $u$ on ${\cal N}$ in the manner discussed
above.  Define the diffeomorphism
\be
\Phi = (u, \pi ) : {\cal N} \to \mathbb{R} \times {\cal
  Z},
\ee
where $\pi: {\cal N} \to {\cal Z}$ is the natural projection obtained
by taking each point to the corresponding integral curve.
Starting from the intrinsic
structure $\frak{u}'$ one can similarly define a diffeomorphism $\Phi'$,
and then $\varphi = \Phi^{-1} \circ \Phi'$ maps $\frak{u}$ onto $\frak{u}'$.

\subsection{Symmetry group of a complete intrinsic structure}
\label{sec:uiss}

We now turn to a discussion of the symmetry group $G_{\frak{u}}$ of
diffeomorphisms $\varphi : {\cal N} \to {\cal N}$ which preserve
a universal structure $\frak{u}$.
Remarkably, the structure of this group is very similar to that of
the BMS group at null infinity, but with two important differences.
First, the Lorentz group at null infinity is replaced by the group
${\rm Diff}({\cal Z})$ of diffeomorphisms of the base space ${\cal Z}$, typically the
two-sphere $S^2$. This replacement is not
surprising, since the conformal freedom that is used at null
infinity to map the induced metric onto a metric of constant curvature
is not present for general null surfaces.
Second, the abelian subgroup of supertranslations at null
infinity is replaced by a nonabelian subgroup, which contains
angle-dependent displacements of affine parameters and
rescalings of affine parameters.


From the definition (\ref{rescale-equiv}) of the
equivalence class, it follows that a diffeomorphism $\varphi : {\cal
  N} \to {\cal N}$ is a symmetry in
$G_{\frak{u}}$ if, for a given representative $(\ell^i,\kappa)$ in
$\frak{u}$, the pullback $\varphi_*$ acts as a scaling transformation for
some smooth scaling function $\beta = \beta(\varphi)$ on ${\cal N}$
[cf.\ Eqs.\ (\ref{rescale-equiv}) above]:
\begin{subequations}
\label{symmetrydef}
\begin{eqnarray}
\varphi_* \ell^i &=& e^\beta \ell^i, \\
\varphi_* \kappa &=& e^\beta ( \kappa + \lie_\ell \beta ).
\end{eqnarray}
\end{subequations}
If we choose a different representative $(\ell^{\prime\,i},\kappa')$
with $\ell^{\prime\,i} = e^{\sigma} \ell^i$, then we find from
(\ref{symmetrydef}) that
\begin{subequations}
\label{symmetrydef1}
\begin{eqnarray}
\varphi_* \ell^{\prime\,i} &=& e^{\beta'} \ell^{\prime\,i}, \\
\varphi_* \kappa' &=& e^{\beta'} ( \kappa' + \lie_{\ell'} \beta' ),
\end{eqnarray}
\end{subequations}
where
\be
\beta' = \beta + \varphi_* \sigma - \sigma.
\label{betap}
\ee
Hence $\varphi$ will be a symmetry if (\ref{symmetrydef}) is satisfied for
any choice of representative.

Specialize now to a choice of coordinate system $(u,\theta^A)$ and
representative of the kind discussed in Sec.\ \ref{sec:uis} above, where $\kappa =0$ and
${\vec \ell} = \partial_u$.  Then the general solution for a
diffeomorphism that satisfies (\ref{symmetrydef}) is $(u,\theta^A) \to
({\bar u}, {\bar \theta}^A)$, where
\begin{subequations}
\label{groupt}
\begin{eqnarray}
\label{groupta}
{\bar u}(u,\theta^A) &=& \alpha(\theta^A) + e^{-\beta(\theta^A)} u \\
{\bar \theta}^A(u,\theta^B) &=& {\bar \theta}^A(\theta^B).
\end{eqnarray}
\end{subequations}
This group of transformations contains a number different subgroups:
\begin{itemize}
\item The subgroup with $\alpha=0$, $\beta=0$, which consists of
  arbitrary diffeomorphisms on the base space ${\cal Z}$, ${\rm
    Diff}({\cal Z})$.  In many applications this will be ${\rm
    Diff}(\twosphere)$, the diffeomorphisms of the two-sphere.  These
  transformations have also been called {\it superrotations}
  \cite{Hawking:2016sgy}.

\item The subgroup with ${\bar \theta}^A = \theta^A$, parameterized by
  $\alpha(\theta^A)$ and $\beta(\theta^A)$.  These transformations
  consist of reparameterizations of the generators of the null
  surface\footnote{This supertranslation subgroup of symmetries played
an important role in Wall's proof of the generalized second law
\cite{Wall:2011hj}.}.  We will call these transformations {\it
    supertranslations}, following common use
  \cite{Hawking:2016msc,Hawking:2016sgy,Koga:2001vq,Hotta:2000gx,Cai:2016idg,Lust:2017gez,Hou:2017pes,Eling:2016xlx,Donnay:2016ejv,Blau:2015nee},
  and because of the analogy with the supertranslations of the BMS group.

\item The subgroup of the supertranslation group with $\beta=0$, ${\bar \theta}^A = \theta^A$, which
is parameterized by $\alpha(\theta^A)$.
We will call these transformations
{\it affine supertranslations} since they consist of angle-dependent
displacements in affine parameter (as opposed to angle-dependent
displacements in Killing parameter or Killing supertranslations
\cite{Hawking:2016sgy,Koga:2001vq,Hotta:2000gx,Cai:2016idg,Lust:2017gez,Hou:2017pes,Eling:2016xlx,Donnay:2016ejv}, to be
discussed in Sec.\ \ref{sec:subalg0} below.)

\item The subgroup of the supertranslation group with $\alpha=0$, ${\bar \theta}^A = \theta^A$,
  which is parameterized by $\beta(\theta^A)$.  These transformations consist of
  constant {\it rescalings} of affine parameter on each generator.
 (Note however that if $\tau = \ln(u)/\kappa$ is a Killing parameter,
 the transformations consist of angle-dependent displacements in
 $\tau$; see Sec.\ \ref{sec:subalg0}.)


\end{itemize}
The first subgroup preserves the foliation associated with the
coordinate system $(u,\theta^A)$, while the last three
preserve the integral curves.
The affine supertranslation and supertranslation subgroups do
not depend on the choice of coordinate system or
representative, and can be invariantly defined.  The rescaling and ${\rm Diff}({\cal Z})$ subgroups,
by contrast, do depend on these choices.  Their status is analogous to
that of Lorentz subgroups of the BMS group: there are many such
subgroups, but no natural or unique choice.

The symmetry algebra associated with the group of transformations is given by
the linearization of Eq.\ (\ref{groupt}), which yields the vector field
\be
{\vec \chi} = \left[ \alpha(\theta^A) - \beta(\theta^A) u
\right] \partial_u + X^A(\theta^B) \partial_A,
\label{groupt1}
\ee
where $X^A$ is arbitrary.  The algebra of these generators under Lie brackets is
\be
\left[ (\alpha_1 - \beta_1 u) \partial_u + X_1^A \partial_A, (\alpha_2
  - \beta_2 u) \partial_u + X_2^A \partial_A \right] =
 (\alpha_3 - \beta_3 u) \partial_u + X_3^A \partial_A
\ee
with
\begin{subequations}
\label{algexplicit}
\begin{eqnarray}
\alpha_3 &=& - \alpha_1 \beta_2 + X_1^A \partial_A \alpha_2 + \alpha_2
\beta_1 - X_2^A \partial_A \alpha_1, \\
\beta_3 &=& -X_1^A \partial_A \beta_2 + X_2^A \partial_A \beta_1,\\
X_3^A &=& X_1^B \partial_B X_2^A - X_2^B \partial_B X_1^A.
\end{eqnarray}
\end{subequations}
While these explicit coordinate expressions are convenient, it can be difficult
to discern which aspects of the structures are specific to the choice of coordinate system.
We now turn to an analysis of the symmetry algebra which is covariant and does
not depend on a choice of coordinates.

\subsection{Symmetry algebra of a complete intrinsic structure}
\label{sec:uiss2}

The Lie algebra ${\mf g}_\frak{u}$ of infinitesimal symmetries in
$G_\frak{u}$ consists of vector fields $\chi^i$ on ${\cal N}$ which
obey
the linearized versions of Eqs.\ (\ref{symmetrydef}):
\begin{subequations}
\label{symmetrydeflinear}
\begin{eqnarray}
\label{betadef}
\lie_\chi \ell^i &=& \beta \ell^i, \\
\lie_\chi \kappa &=& \beta  \kappa + \lie_\ell \beta.
\label{sdl1}
\end{eqnarray}
\end{subequations}
As before, if these equations are satisfied for one representative
$(\ell^i,\kappa)$ of the equivalence class, they will be satisfied for
all representatives.  The function $\beta$ depends on both the
symmetry $\chi^i$ and on the representative $\ell^i$, $\beta =
\beta(\chi^i,\ell^i)$, and the dependence on the normalization is
given by the linearized version of Eq.\ (\ref{betap}):
\be
\beta(\chi^i,e^\sigma \ell^i) = \beta(\chi^i,\ell^i) + \lie_\chi \sigma.
\label{betap1}
\ee
The general solution of Eqs.\ (\ref{symmetrydeflinear}) for $\chi^i$,
with a choice of representative and coordinate system $(u,\theta^A)$
for which $\kappa=0$ and ${\vec \ell} = \partial_u$, is given by Eq.\ (\ref{groupt1}) above.

The algebra ${\mf g}_\frak{u}$ inherits the Lie bracket structure of
the space of vector fields on ${\cal N}$.  From the definition of the
symmetry group $G_\frak{u}$ as a subgroup of ${\rm Diff}({\cal N})$, it
follows that ${\mf g}_\frak{u}$ is closed under this Lie bracket.
This closure was also shown in Eq.\ (\ref{algexplicit}) above, and
can also be checked directly in the covariant context: if ${\vec \chi}_1$ and
${\vec \chi}_2$ are two vector fields which satisfy Eqs.\
(\ref{symmetrydeflinear}),
then ${\vec \chi}_3 = \left[ {\vec \chi}_1, {\vec \chi}_2 \right]$
also satisfies Eqs.\ (\ref{symmetrydeflinear}) with
\be
\beta({\vec \chi}_3) = \lie_{\chi_1} \beta_2 - \lie_{\chi_2} \beta_1,
\label{alg}
\ee
where $\beta_1 = \beta({\vec \chi}_1)$ and $\beta_2 = \beta({\vec
  \chi}_2)$.

We now argue that the symmetry algebra has the structure
\be
\mf g_\frak{u} \cong  \diff({\cal Z}) \ltimes (\mf
b \ltimes \mf s_0 ),
\label{algebrastructure}
\ee
where $\ltimes$ denotes semidirect sum and the various algebras are as follows:
\begin{itemize}
\item $\diff({\cal Z})$ is the algebra of linearized diffeomorphisms
  of the base space ${\cal Z}$, {\it i.e.}, vector fields on ${\cal
    Z}$.
\item $\mf s_0$ is the abelian algebra of linearized affine supertranslations,
  consisting of vector fields of the form
\be
\chi^i = f \ell^i
\label{les}
\ee
where the function $f$ on ${\cal N}$ satisfies
\be
\lie_\ell f + \kappa f=0.
\ee

\item $\mf b$ is an abelian algebra of linearized rescalings such that
$\mf b \ltimes \mf s_0 \cong \mf s$, where $\mf s$ is the algebra of
linearized supertranslations.  This is the algebra consisting of vector fields of
the form (\ref{les})
where the function $f$ satisfies
\be
\lie_\ell (\lie_\ell f + \kappa f)=0.
\label{estdef}
\ee

\end{itemize}

We now turn to the derivation of the structure (\ref{algebrastructure}).
We define the subspace
\be
\mf s = \left\{ \chi^i \in \mf g_\frak{u} | \chi^i = f \ell^i {\rm \ \ for
 \ some\ } f\right\}.
\label{mfsdef}
\ee
By comparison with Eqs.\ (\ref{groupt}) and (\ref{groupt1}), we see that this subspace consists of
the linearized supertranslations.  Inserting the definition
(\ref{mfsdef}) into Eqs.\ (\ref{symmetrydeflinear}) yields that the function $f$
satisfies the condition (\ref{estdef}) with
\be
\beta(f {\vec \ell}) = - \lie_\ell f.
\label{betaff}
\ee
The condition (\ref{estdef}) is invariant
under the scaling transformations
\be
{\vec \ell} \to e^\sigma {\vec \ell}, \ \ \ \ \ f \to e^{-\sigma} f,
\label{rescale2}
\ee
so the subspace $\mf s$ is
parameterized by functions of scaling weight $1$ on ${\cal N}$ [cf.\
Eq.\ (\ref{scalingweight})].
The subspace $\mf s$ is closed
under the Lie bracket and so is a subalgebra; we have
\be
[ f_1 {\vec \ell}, f_2 {\vec \ell} ] = (f_1 \lie_\ell f_2 - f_2
\lie_\ell f_2) {\vec \ell}.
\label{lieons}
\ee
Since the right hand side is nonvanishing in general, the subalgebra
is nonabelian.  Finally, for any $f {\vec \ell} \in \mf s$ and any
${\vec \chi} \in \mf g_{\frak{u}}$, we have from Eqs.\
(\ref{symmetrydeflinear}) that
\be
[ f {\vec \ell}, {\vec \chi} ]  = - \left[ \lie_\chi f + \beta({\vec
    \chi}) f \right] {\vec \ell}.
\ee
Hence \([\mf s, \mf g_{\frak{u}} ] \subseteq \mf s\), so
\(\mf s\) is a Lie ideal of $\mf g_{\frak{u}}$.

Next, we define the subalgebra $\mf s_0$ of $\mf s$ by
\be
\mf s_0 = \left\{ f \ell^i | \lie_\ell f + \kappa f =0 \right\}.
\label{mfs0def}
\ee
By comparison with Eqs.\ (\ref{groupt}) and (\ref{groupt1}), we see that this subalgebra consists of
the linearized affine supertranslations, and it follows from
Eq.\ (\ref{lieons}) that it is abelian.  The definition
(\ref{mfs0def}) is invariant under the rescalings (\ref{rescale2}).
If we choose a representative $(\ell^i,\kappa)$ of the equivalence
class with $\kappa=0$, it follows that $f$ is constant along generators
and so can be regarded as a function on ${\cal Z}$.  There is a
residual rescaling freedom of the form (\ref{rescale2}) with
$\lie_\ell \sigma=0$ that preserves $\kappa=0$.  Hence, the algebra
$\mf s_0$ can be identified with functions on the base space ${\cal
  Z}$ of scaling weight $1$, from Eq.\ (\ref{scalingweight}), just like supertranslations on
$\scri$\footnote{Unlike the case with the BMS algebra, there is no
  preferred translation subalgebra of the affine supertranslation
algebra $\mf s_0$.
Even if \({\cal Z} \) is topologically \(\twosphere\), there is no universal
metric on \({\cal Z}\), so it is not possible to single out a
\(4\)-dimensional subalgebra of  translations by the first four
spherical harmonics. Also, there is no scaling-invariant notion of
constant functions on \({\cal Z}\), so there is not even a natural way to single out
``time-translations''.}.

Next, if $f_1 {\vec \ell}$ and $f_2 {\vec \ell}$ are elements of $\mf
s$, it follows from Eq.\ (\ref{estdef}) that
\be
(\lie_\ell + \kappa) \left( f_1 \lie_\ell f_2 - f_2 \lie_\ell f_2
\right) =0.
\ee
Combining this with Eq.\ (\ref{lieons}) shows that
\be
[\mf s, \mf s] \subseteq \mf s_0,
\label{babelian}
\ee
so \(\mf s_0\) is a Lie ideal of \(\mf s\).
We define the quotient algebra \(\mf b \cong \mf s / \mf s_0\).  This
consists of equivalence classes of elements of $\mf s$, where
$f_1 {\vec \ell} \sim f_2 {\vec \ell}$ if $\lie_\ell f_1 + \kappa f_1 = \lie_\ell
f_2 + \kappa f_2$.  Elements of $\mf b$ can be parameterized in terms of
functions\footnote{Essentially the functions $\lie_\ell f + \kappa f$
  projected to ${\cal Z}$, where $ f {\vec \ell} \in \mf s$.}
on ${\cal Z}$ of scaling weight zero,
and they correspond to linearized rescalings, cf.\ Eq.\
(\ref{groupt1}) above.
It follows from Eq.\ (\ref{babelian}) that $\mf b$ is abelian,
and so we obtain
\be
        \mf s \cong \mf b \ltimes \mf s_0
\label{algebrastructure1}
\ee
where both \(\mf b\) and \(\mf s_0\) are abelian.

We next argue that the quotient algebra $\mf g_{\frak{u}} / \mf s$ is isomorphic
to the algebra of linearized diffeomorphisms on the base space ${\cal
  Z}$,
\be
\mf g_{\frak{u}} / \mf s \cong \diff({\cal Z}),
\label{algebrastructure2}
\ee
which when combined with Eq.\ (\ref{algebrastructure1}) gives the
algebra structure\footnote{The subalgebra $\mf s_0$ is also a Lie
  ideal of $\mf g_\frak{u}$, but $\mf g / \mf s_0 \not\cong
  \diff({\cal Z}) \ltimes \mf b$.}
 (\ref{algebrastructure}).
The algebra $\mf g_{\frak{u}} / \mf s$ consists of equivalence classes $[\chi^i]$
of vector fields $\chi^i$ in $g_{\frak{u}}$, where two vector fields
are equivalent if they differ by an element $f \ell^i$ in $\mf s$.
Pick a cross section ${\cal S}$ of ${\cal N}$, and denote by $n_i$ the
unique normal covector to ${\cal S}$ whose normalization is fixed by
$n_i \ell^i=-1$.  Given an equivalence class $[\chi^i]$, one can find
a member $\chi^i$ with $\chi^i n_i =0$ on ${\cal S}$, by using the
freedom to add terms of the form $f \ell^i$ and using the fact that
solutions to Eq.\ (\ref{estdef}) can be freely specified on an initial cross
section ${\cal S}$.  This member $\chi^i$ can then be regarded as a
vector field $\chi^A$ on ${\cal S}$, and by using the natural identification of
${\cal S}$ and ${\cal Z}$, as a vector field on ${\cal Z}$.
We have thus defined a mapping from $\mf g_{\frak{u}} / \mf s$ to $\diff({\cal Z})$.
One can check that this mapping is onto, and it
follows from Eqs.\ (\ref{symmetrydeflinear}) that the identification of
$\mf g_{\frak{u}} / \mf s$ and $\diff({\cal Z})$ is independent
of the choice of cross section ${\cal S}$.
Thus we have derived the decomposition (\ref{algebrastructure}) of the algebra $\mf g_{\mf u}$.

For the computations of charges in Sec.\ \ref{sec:wzcharges} below, it
will be useful to use an explicit decomposition of symmetry generators
${\vec \chi}$ into different pieces.  However, because of the semi-direct structure \(\mf g
\cong \diff({\cal Z}) \ltimes \mf s\), there is no natural way to
decompose a generator \(\chi^i\) into a \(\mf s\)-part and a
\(\diff({\cal Z})\)-part.  Such a decomposition requires an arbitrary choice of
origin in $\mf s$.  We make such a choice by choosing a smooth
covector $n_i$ on ${\cal N}$, normalized so that
\be
n_i \ell^i =-1.
\label{normalized1}
\ee
The generator $\chi^i$ can then be uniquely decomposed as
\be
\chi^i = f \ell^i + X^i,
\label{unde}
\ee
where
\be
X^i n_i = 0.
\label{ortho1}
\ee
Here the first term $f \ell^i$ parameterizes the
supertranslations, and the second term $X^i$ parameterizes
the diffeomorphisms on the base space.

In order for both terms on the right hand side of Eq.\ (\ref{unde})
to belong to $\mf g_{\frak{u}}$, from Eq.\ (\ref{estdef}) it is necessary
that
\be
 \lie_\ell ( \lie_\ell + \kappa) (\chi^i n_i) =0.
\label{require}
\ee
Using Eq.\ (\ref{symmetrydeflinear}), this will be automatically satisfied if $n_i$
obeys the equation
\be
\lie_\ell (\lie_\ell + \kappa) n_i + D_i \kappa =0,
\label{nieqn}
\ee
where $D_i$ is any derivative operator on ${\cal N}$.
This equation is invariant under the rescalings (\ref{rescale-equiv}),
since $n_i$ transforms as $n_i \to
e^{-\sigma} n_i$ from Eq.\ (\ref{normalized1}).
If we choose $n_i$ to be the normal covector to a
foliation of surfaces in the natural class of foliations discussed
in Sec.\ \ref{sec:uis} above, normalized according to (\ref{normalized1}),
then the condition (\ref{nieqn}) is satisfied.

\subsection{Preferred subalgebra for stationary regions of a null
  hypersurface: Killing supertranslations}
\label{sec:subalg0}

Stationary regions of the hypersurface ${\cal N}$ that intersect all
the generators determine
a preferred subalgebra $\mf t$ of the supertranslation algebra $\mf s$.  This algebra is the set
of vector fields $\chi^i$ in $\mf s$ for which
\be
\lie_\tau \chi^i =0
\label{mftdef}
\ee
in the stationary region,
where $\tau^a$ is the Killing vector field which is normal to ${\cal N}$.  Since $(\tau^i,\kappa_\tau)$
is a representative of the equivalence class, we have from Eq.\ (\ref{estdef}) that
all elements $\chi^i$ of $\mf s$ satisfy
\be
\lie_\tau (\lie_\tau + \kappa_\tau) \chi^i=0.
\ee
Hence it follows from Eqs.\ (\ref{stationary2}) and (\ref{mftdef}) that
all solutions of Eq.\ (\ref{mftdef}) in the stationary region can be
extended to vector fields on all of ${\cal N}$ which lie in $\mf s$.

To get some insight into the nature of this subalgebra\footnote{The
 pullback $\tau^i$ of the  Killing field is itself a member of the
 subspace $\mf t$, giving a preferred one-dimensional subspace of ``translations''.}, specialize to a
representative $(\ell^i,\kappa)$ and a coordinate system
$(u,\theta^A)$ where ${\vec \ell} = \partial_u$ and $\kappa=0$,
where the general solution for $\chi^i$ is given by Eq.\ (\ref{groupt1}).
Then the Killing field $\tau^i$ will be of the form
$
\tau^i = \kappa_\tau (u - u_0) \ell^i,
$
by Eqs.\ (\ref{rescale1}), (\ref{stationary2}) and (\ref{zeroth}),
where $\kappa_\tau$ is a constant and $u_0$ is a function of $\theta^A$ but
independent of $u$.  The subalgebra $\mf t$ is then given by the
condition
\be
\alpha - \beta u_0 =0,
\label{mftcondt}
\ee
and consists of vector fields of the form
$
\chi^i = - (\beta/\kappa_\tau) \, \tau^i.
$
The corresponding transformation (\ref{groupt}) can be expressed as
\be
{\bar \tau} = \tau - \frac{\beta}{\kappa_\tau},
\ee
where we have defined a Killing parameter $\tau$ by ${\vec \tau} = d/d\tau$.
We will call these angle-dependent displacements of Killing parameter
{\it Killing supertranslations}.  They have been studied in Refs.\
\cite{Hawking:2016sgy,Koga:2001vq,Hotta:2000gx,Cai:2016idg,Lust:2017gez,Hou:2017pes,Eling:2016xlx,Donnay:2016ejv}
(although they are often called just supertranslations).
The intersection of the Killing supertranslation subalgebra $\mf t$
with the affine supertranslation
subalgebra $\mf s_0$ will generically have dimension $0$.

We note that the Killing supertranslation subalgebra $\mf t$ can be defined under the
slightly weaker hypothesis that the region of ${\cal N}$ is {\it
  weakly isolated} in the sense of Ashtekar \cite{Ashtekar:2001jb},
which implies that it is shear and expansion free, satisfies Eqs. (\ref{stationary2}),
 and possesses a preferred choice of normal up to constant rescalings.

\subsection{Symmetry groups of null hypersurfaces with boundaries}
\label{sec:uiss1}

Our analysis so far has been restricted by the assumptions that
the null hypersurface ${\cal N}$ has topology ${\cal Z} \times \mathbb{R}$,
and that the intrinsic structure is complete, that is, that the
generators of the null surface extend to infinite affine parameters in
both directions.  We now discuss how the symmetry group is modified
when these assumptions are relaxed.
Specifically, we will consider incomplete intrinsic structures.  These
generally occur when the null hypersurface ${\cal N}$ has a
nontrivial topological boundary $\partial {\cal N}$ in
$M$\footnote{The hypersurface ${\cal N}$ can have a nontrivial
  boundary only when ${\cal N}$ is a proper subset of
  the boundary $\partial M$ of $M$, as it will be in typical
  applications, since $\partial \partial M = \{ \}$.}.  Rather than give a general
analysis of the different possibilities, we will discuss two specific
examples.

The first example is the future light cone of a point ${\cal P}$ in a
spacetime which is spherically symmetric about ${\cal P}$.
This could be the future event horizon of a black hole in a
spherically symmetric gravitational collapse spacetime.  Or, it could be the
future light cone of a point in Minkowski spacetime.
The null hypersurface still has topology ${\cal Z} \times \mathbb{R}
\simeq \twosphere \times \mathbb{R}$ (if the point ${\cal P}$ is
excluded), but has the nontrivial boundary $\partial {\cal N}  = \{
{\cal P} \}$.  The induced intrinsic structure is incomplete if the
metric is smooth in a neighborhood of ${\cal P}$,
as all the generators start at ${\cal P}$.

The second example is the future event horizon in the maximally
extended Schwarzschild spacetime, on one of the two branches.
In this case the boundary of ${\cal N}$ is the bifurcation twosphere,
and the induced intrinsic structure is again incomplete,
as all the generators start on the bifurcation twosphere.

In these cases, the definition of the symmetry group is modified to
include the requirement that it preserve the boundary:
\be
G_{\mf u} = \left\{ \varphi: {\cal N} \to {\cal N} \ | \ \varphi_* \mf
  u = \mf u, \ \ \varphi(\partial
  {\cal N}) = \partial {\cal N} \right\}.
\ee
In the first case of a single point, $\partial {\cal N}  = \left\{
  {\cal P} \right\}$, the corresponding Lie algebra consists of the vector
fields $\chi^i$ which satisfy Eqs.\ (\ref{symmetrydeflinear}) and in addition the
condition
\be
\left. \chi^i \right|_{\partial {\cal N}} =0.
\label{collapse0}
\ee
This removes the affine supertranslations and
but not the rescalings or $\diff(\twosphere)$ diffeomorphisms.
If one chooses an affine coordinate system
$(u,\theta^A)$ of the type described in Sec.\ \ref{sec:uis}, specialized so
that $u=0$ on $\partial {\cal N}$, then the transformation group (\ref{groupt}) is
modified by the condition.
\be
\alpha=0.
\label{collapse}
\ee
In the second case of the bifurcation twosphere, the condition
(\ref{collapse0}) is replaced by the requirement
that the vector field be
tangent to $\partial {\cal N}$ on $\partial {\cal N}$,
\be
\left. \chi^i n_i  \right|_{\partial {\cal N}} =0,
\label{condt3}
\ee
where $n_i$ is the normal to $\partial {\cal N}$.
The modification to the algebra is the same as in the first case, given by the condition
(\ref{collapse}).

We note that in this context the Killing supertranslation subalgebra $\mf t$
associated with stationary regions of the null surface will
generically have dimension $0$, by Eqs.\ (\ref{mftcondt}) and (\ref{collapse}).
This is discussed further in Sec.\ \ref{sec:globalcons} below
(footnote \ref{caveat}).


\section{General relativity with a null boundary: covariant phase space}\label{sec:cov-phase-null}
\label{sec:cpsnb}

As discussed in Sec.\ \ref{sec:cov-phase}, the starting point of the Wald-Zoupas
framework is the definition of a field configuration space $\ms F$
of kinematically allowed field configurations, and the corresponding
covariant phase space ${\bar{ \ms F}} \subset \ms F$ obtained by restricting attention
to on-shell field configurations.  In this section we give a
particular version of these
definitions for general relativity in the presence of a null
boundary in $3+1$ dimensions.  The definition is given in
Sec.\ \ref{sec:cps}, and in Sec.\ \ref{sec:symgroup} we show that the symmetry group and
algebra associated with this field configuration space coincide with
those of the universal intrinsic structure of the null surface
discussed in Sec.\ \ref{sec:universal}.

\subsection{Definition of field configuration space}
\label{sec:cps}

Consider a manifold $M$ with boundary, for which a manifold ${\cal N}$
is a portion of the boundary.  We would like to consider the space
$\ms F_0$ consisting of smooth metrics $g_{ab}$ on $M$ for which the
boundary ${\cal N}$ is null and for which the induced boundary
structure on ${\cal N}$ is complete.  This space $\ms F_0$ is not the field
configuration space $\ms F$ we seek, since it contains a considerable amount
of diffeomorphism redundancy.  We will obtain our definition of $\ms F$ by
fixing some of this freedom.

The kinds of fixing of diffeomorphism freedom that we will allow
will be restricted by three general considerations:
\begin{itemize}
\item They must be global on the field configuration space, not restricted
  to on-shell configurations.
\item They must be local to the boundary in the sense that the diffeomorphisms
  needed to enforce the gauge condition can be computed
  from degrees of freedom on the boundary.
\item Field configurations (metrics in this case) and their derivatives evaluated on the boundary induce on the boundary
 certain geometric structures, which can be divided into universal and
 non-universal structures.  The universal structures are the same for
 all field configurations (up to boundary diffeomorphisms), while the
 non-universal ones depend on the field configuration.  We restrict
 attention to fixings of the diffeomorphism freedom that involve only the universal
 structures.
\end{itemize}
The diffeomorphism (and conformal freedom) fixings used at future null infinity by Wald and Zoupas
\cite{WZ} are also of this type.

As a side note, as discussed in Sec.\ \ref{sec:presymform} above, gauge in this context is not synonymous with
diffeomorphism freedom, since there are some diffeomorphisms that act on the boundary which
do not correspond to degeneracies of the symplectic form on phase space (a more fundamental notion of gauge).
Some of the diffeomorphism freedom we fix in going from $\ms F_0$ to $\ms F$ is not gauge in this sense.
For this reason, it would be desirable to consider a larger field configuration space that includes all metric variations
that are not degeneracy directions of the symplectic form.  In
Appendix \ref{app:alternative} we explore a modification of our
definition of the field configuration space
which yields a modified and larger algebra of symmetries and a modified set of charges.  The main
drawback of this modification is that it is no longer possible to obtain uniqueness of the prescription
for defining localized charges by demanding that fluxes vanishes for stationary solutions, as discussed
in Sec.\ \ref{sec:WZ-charge} above. It is possible that a unique prescription may be obtained from some other criterion.

Our definition of the field configuration space $\ms F$ proceeds as follows.
We start by defining a particular geometric structure on ${\cal N}$ which
we will call a {\it boundary structure}.  We consider triples $(\ell^a,
\kappa, {\hat \ell}_a)$ of fields on ${\cal N}$, where $\ell^a$ is a smooth,
nowhere vanishing vector field, $\kappa$ is a smooth function, ${\hat
  \ell}_a$ is a choice of normal covector\footnote{This normal
  covector was denoted $\ell_a$ earlier in the paper.  We introduce
  the separate notation ${\hat \ell}_a$ because the context here
  of the definition of a
  boundary structure does not involve a metric, and to clarify that
  there are two independent tensor fields in the definition.}
  to ${\cal N}$, and
\be
\ell^a {\hat \ell}_a \hateq 0.
\label{ortho}
\ee
Recall that we are using $\hateq$ to mean equality when restricted to
${\cal N}$.  We define two such triples $(\ell^a,\kappa,{\hat \ell}_a)$ and
$(\ell^{\prime\,a},\kappa',{\hat \ell}^\prime_a)$ to be equivalent if they
are related by the rescaling
\begin{subequations}
\label{rescale-equiv1}
\begin{eqnarray}
\ell^{\prime\,a} &\hateq& e^\sigma \ell^a, \\
\label{kappaequiv}
\kappa' &\hateq& e^\sigma(\kappa + \lie_{\ell} \sigma), \\
{\hat \ell}^{\prime}_a &\hateq& e^\sigma {\hat \ell}_a,
\end{eqnarray}
\end{subequations}
where $\sigma$ is a smooth function on ${\cal N}$.
We denote by
\be
\frak{p} = [\ell^a,\kappa,{\hat \ell}_a]
\label{frakpdef}
\ee
the equivalence class associated with $(\ell^a,\kappa,{\hat \ell}_a)$.
A choice of equivalence class is the desired boundary
structure on ${\cal N}$.

It is clear that a choice of boundary structure
$\frak{p} = [\ell^a,\kappa,{\hat \ell}_a]$ determines a unique
universal intrinsic structure $\frak{u}$:  choose a
representative $(\ell^a,\kappa,{\hat \ell}_a)$, discard ${\hat
  \ell}_a$, and note that from Eq.\ (\ref{ortho}) that $\ell^a$ can be
regarded as an intrinsic vector field $\ell^i$.
Then from $(\ell^i,\kappa)$ form the equivalence class
$\frak{u} = [\ell^i,\kappa]$ under the equivalence relation
(\ref{rescale-equiv}).  From Eqs.\ (\ref{rescale-equiv}) and
(\ref{rescale-equiv1})
the result is independent of the representative $(\ell^a,\kappa,{\hat
  \ell}_a)$ initially chosen.  We will denote this induced intrinsic structure by $\mf u(\mf p)$.
Our boundary structures contain more information than the intrinsic
structures, which will be necessary for the definition of the field
configuration space.
We will say that a boundary structure $\frak{p}$ is complete if the
corresponding intrinsic structure $\mf u$ is complete.

In addition, a metric $g_{ab}$ on $M$ for which the boundary ${\cal N}$ is null
determines a unique boundary structure $\mf p$, just as for
intrinsic structures discussed in Sec.\ \ref{sec:uis}.  Pick a normal
covector ${\hat \ell}_a$ to ${\cal N}$, raise the index to obtain
$\ell^a = g^{ab} {\hat \ell}_b$, and compute the non-affinity $\kappa$
using the metric via Eq.\ (\ref{kappadef}).  Then from the triple $(\ell^a,\kappa, {\hat
  \ell}_a)$ form the equivalence class $\mf p =  [\ell^a,\kappa, {\hat
  \ell}_a]$.  The result is independent of the choice of initial
normal covector, by the equivalence relation (\ref{rescale-equiv1}).

Given a boundary structure $\mf p$, we now define the field
configuration space $\ms F_{\mf p}$ to be the set of smooth metrics
$g_{ab}$ on $M$ which satisfy on ${\cal N}$ the relations
\begin{subequations}
\label{calFdef}
\begin{eqnarray}
\label{calFdef1}
\ell^a \ &{\hateq}& \ g^{ab} {\hat \ell}_b, \\
\label{calFdef2}
\ell^a \nabla_a \ell^b \ &{\hateq}& \ \kappa \ell^b.
\end{eqnarray}
\end{subequations}
From Eqs.\ (\ref{ortho}) and (\ref{calFdef1}) it follows that the
boundary ${\cal N}$ is null with respect to $g_{ab}$, so that $\ms
F_{\mf p} \subset \ms F_0$.  Also if the conditions (\ref{calFdef})
are satisfied by one representative $(\ell^a,\kappa, {\hat \ell}_a)$,
they will be satisfied by all representatives, from Eqs.\ (\ref{rescale1}) and
(\ref{rescale-equiv1}).  Hence $\ms F_{\mf p}$ is well defined and
depends only on $\mf p$.  (An equivalent definition of $\ms F_{\mf p}$
is the set of smooth metrics on $M$
for which ${\cal N}$ is null and
whose associated boundary structures agree with $\mf
p$).  We define the corresponding covariant phase space $\bar{\ms
F}_{\mf p}$ to be the set of metrics in $\ms F_{\mf p}$ which satisfy the
equations of motion.

Note that the order of definitions being used in this construction is the opposite of
that which is normally used.  Normally, one first picks the spacetime
metric, then defines the covariant version of the null normal by
raising the index as in Eq.\ (\ref{calFdef1}), and defines
the non-affinity function $\kappa$ via
Eq.\ (\ref{calFdef2}).
Here, instead, we first choose the quantities $\ell^a$, ${\hat
\ell}_a$, and $\kappa$, and then specialize the spacetime metric
$g_{ab}$ to enforce Eqs.\ (\ref{calFdef}).

It may appear that the conditions (\ref{calFdef}) we are imposing on the metric are
overly restrictive.  In fact, they do not restrict the physical degrees
of freedom in the sense that $\ms F_{\mf p}$ is obtained from $\ms F_0$ by a
fixing of the diffeomorphism freedom.  More precisely, given a complete boundary structure $\mf p$,
and given any metric $g_{ab}$ on $M$ for which ${\cal N}$ is
null and for which the boundary structure induced by $g_{ab}$ is complete,
one can find a diffeomorphism $\psi : M \to M$ which takes ${\cal N}$ into ${\cal N}$
for which $\psi_* g_{ab}$ satisfies the conditions
(\ref{calFdef}).  This is proved in Appendix \ref{app:proofs}.

We next show that the mapping $\mf p \to \ms F_{\mf p}$ is injective,
so that if $\ms F_{\mf p} = \ms F_{{\mf p}'}$ then
${\mf p} = {\mf p}'$.  This property will be used in Sec.\ \ref{sec:symgroup} below.
Let $(\ell^a,\kappa,{\hat \ell}_a)$ be a representative of $\mf p$,
and $(\ell^{\prime\,a},\kappa',{\hat \ell}'_a)$ be a representative of $\mf p'$.
Since ${\hat \ell}_a$ and ${\hat \ell}'_a$ are both normals to ${\cal
  N}$, they are related by a rescaling, and hence by adjusting our choice
of representative we can without loss of generality take
${\hat \ell}_a = {\hat \ell}'_a$.  Now pick a metric $g_{ab}$ which
belongs to both $\ms F_{\mf p}$ and $\ms F_{{\mf p}'}$.
Applying Eq.\ (\ref{calFdef1}) to both $\mf p$ and $\mf p'$ we find that $\ell^a = \ell^{\prime\,a}$,
and it follows from Eq.\ (\ref{calFdef2}) that $\kappa = \kappa'$.
Hence we have $\mf p = \mf p'$.

\subsection{Symmetry algebra of the field configuration space}
\label{sec:symgroup}

We now show that for a complete boundary structure $\mf p$, the symmetry
algebra associated with the field configuration space
$\ms F_{\mf p}$ coincides with
the algebra $\mf g_{\mf u}$ of the universal intrinsic structure $\mf u$ of the null surface
discussed in Sec.\ \ref{sec:universal}, where $\mf u = \mf u(\mf p)$ is the intrinsic
structure obtained from $\mf p$ discussed in Sec.\ \ref{sec:cps}.

We start by defining the group of diffeomorphisms on $M$ whose pullbacks preserve
the boundary and the field configuration space:
\be
H_{\mf p} = \left\{ \psi : M \to M \ |  \ \psi({\cal N}) = {\cal
    N}, \ \psi_* {\ms F}_{\mf p} = {\ms F}_{\mf p} \ \right\}.
\label{Hpdef}
\ee
These diffeomorphisms induce diffeomorphisms of the boundary: for any $\psi$ in $H_{\mf p}$ we define
\be
\varphi = \psi |_{\cal N},
\label{induced}
\ee
and since $\psi$ preserves the boundary, $\varphi$ is a diffeomorphism
from ${\cal N}$ to ${\cal N}$.  Next, since $\psi$ preserves the
boundary, the pullback of the normal must be a rescaling of the
normal, so we have
\be
\psi_* {\hat \ell}_a = e^{\gamma} {\hat \ell}_a,
\label{gammadef}
\ee
where $\gamma = \gamma(\psi,{\hat \ell}_a)$ is a smooth function on ${\cal N}$
which depends on the diffeomorphism and on the normalization of the
normal covector ${\hat \ell}_a$.  From Eq.\ (\ref{gammadef}) we find
for the dependence on the normalization
[{\it cf.}\ Eq.\ (\ref{betap}) above]
\be
\gamma(\psi,e^\sigma {\hat \ell}_a) = \gamma(\psi,{\hat \ell}_a) +
\psi_* \sigma - \sigma,
\label{gammascaling}
\ee
for any smooth function $\sigma$ on ${\cal N}$.

Next from the definition (\ref{Hpdef}) we have
\be
{\ms F}_{\mf p} = \psi_* {\ms F}_{\mf p} = {\ms F}_{\psi_* {\mf p}},
\ee
where the action of the pullback $\psi_*$ on the boundary structure
$\mf p$ is defined by its action on a representative
$(\ell^a,\kappa,{\hat \ell}_a)$.  Now using the injectivity property
of the mapping $\mf p \to {\ms F}_{\mf p}$ proved in Sec.\
\ref{sec:cps}, we obtain
\be
{\psi_* {\mf p}} = \mf p.
\ee
From the definition (\ref{rescale-equiv1}) of the
equivalence class, it follows that
for a given representative $(\ell^a,\kappa,{\hat \ell}_a)$ in
$\mf p$, the pullback $\psi_*$ acts as a scaling transformation for
some smooth scaling function $\beta = \beta(\psi)$ on ${\cal N}$ [cf.\
Eq.\ (\ref{symmetrydef}) above]:
\begin{subequations}
\label{symmetrydef11}
\begin{eqnarray}
\psi_* \ell^a &\hateq& e^\beta \ell^a, \\
\psi_* \kappa &\hateq& e^\beta ( \kappa + \lie_\ell \beta ), \\
\psi_* {\hat \ell}_a &\hateq& e^\beta {\hat \ell}_a.
\label{ree}
\end{eqnarray}
\end{subequations}
In the first two of these equations we can replace $\ell^a$ with
$\ell^i$, by Eq.\ (\ref{ortho}), and we can replace $\psi_*$ with
$\varphi_*$.  These two equations then coincide with the defining
equations (\ref{symmetrydef}) for the group $G_{\mf u}$ of boundary symmetries
$\varphi : {\cal N} \to {\cal N}$ that preserve the intrinsic
structure $\mf u$ associated with $\mf p$.  Combining Eqs.\ (\ref{gammadef})
and (\ref{ree}) yields that
\be
\gamma(\psi) = \beta(\varphi).
\label{condition}
\ee
Hence we have shown that
\be
H_{\mf p} = \left\{ \psi : M \to M \ |  \ \psi({\cal N}) = {\cal
    N}, \ \varphi \in G_{\mf u}, \ \beta(\varphi) = \gamma(\psi) \ \right\},
\label{Hpdef1}
\ee
where $\varphi$ is the diffeomorphism (\ref{induced}) induced on the
boundary.  A bulk diffeomorphism $\psi$ is a symmetry if it preserves
the boundary, if the induced boundary diffeomorphism is a symmetry of
the intrinsic structure on the boundary, and if the scaling
function $\gamma(\psi)$ defined by Eq.\ (\ref{gammadef}) satisfies Eq.\ (\ref{condition}).

We next specialize these results to infinitesimal diffeomorphisms.
Linearized diffeomorphisms on $M$ are parameterized in terms of vector
fields $\xi^a$ on $M$, and the boundary is preserved if these vector
fields are tangent to the boundary,
\be
\xi^a {\hat \ell}_a \hateq 0.
\label{tangent}
\ee
We define
\be
{\vec \chi} = {\vec \xi}|_{\cal N},
\label{restriction}
\ee
and it follows from the condition  (\ref{tangent}) that we can regard
${\vec \chi}$ as an intrinsic vector field $\chi^i$ on ${\cal N}$, as
in Sec.\ \ref{sec:uiss2} above.  The definition (\ref{gammadef}) of
the scaling function $\gamma$ becomes
\be
\lie_\xi {\hat \ell}_a \hateq \gamma(\xi^a,{\hat \ell}_a) {\hat \ell}_a,
\label{gammadef1}
\ee
while the dependence (\ref{gammascaling}) on the normalization of the
normal becomes
\be
\gamma(\xi^a,e^\sigma {\hat \ell}_a) = \gamma(\xi^a,{\hat \ell}_a) +
\lie_\xi \sigma.
\label{gammascaling1}
\ee
The linearized version of the constraint (\ref{condition}) is
\be
\gamma(\xi^a) = \beta(\chi^i).
\label{condition1}
\ee
Defining $\mf h_{\mf p}$ to be the Lie algebra corresponding to the
group $H_{\mf p}$, we find by linearizing the result (\ref{Hpdef1}) that
\be
{\mf h}_{\mf p} = \left\{ \xi^a \ {\rm on}\  M |  \ \xi^a {\hat \ell}_a
  \hateq 0, \ \chi^i \in {\mf g}_{\mf u}, \ \beta(\chi^i) = \gamma(\xi^a) \ \right\},
\label{Hpdef2}
\ee
where $\chi^i$ is given by the restriction (\ref{restriction}) to the
boundary and $\beta(\chi^i)$ is defined by Eq.\ (\ref{betadef}).

Finally, to obtain the physical symmetry algebra, we need to factor
out the trivial diffeomorphisms for which the symmetry generator
charge variation (\ref{Hamiltonian Definition}) vanishes,
using the equivalence relation $\sim$ defined in Eq.\ (\ref{eq:boundary-symm-equiv}).
In Sec.\ \ref{sec:wzcharges} below we compute the charge
variation (\ref{Hamiltonian Definition})
explicitly, and in Appendix \ref{app:trivial} we show that it
vanishes for all metric perturbations if and only
if $\chi^i$ and $\gamma(\xi^a)$ both vanish.  Hence the quotient set
${\mf h}_{\mf p} / \sim$ is parameterized by $\chi^i$ and $\gamma$,
but from Eq.\ (\ref{condition1}) $\gamma$ is determined by $\chi^i$.
We conclude from Eq.\ (\ref{Hpdef1}) that
\be
{\mf h}_{\mf p} / \sim \ \ \cong \ \ {\mf g}_{\mf u},
\label{res1}
\ee
as claimed.

To summarize, infinitesimal symmetries are in one-to-one
correspondence with symmetries $\chi^i \in \mf g_{\mf u}$ of the intrinsic structure
$\mf u$.  However, all representatives $\xi^a$ whose restriction to the
boundary is $\chi^i$ must also obey the constraint (\ref{condition1}).

\subsection{Boundary conditions on the variation of the metric}
\label{sec:bcs}

In our application of the Wald-Zoupas formalism we will need to
consider variations of the metric of the form
\be
g_{ab} \to g_{ab} + \delta g_{ab} = g_{ab} + h_{ab}.
\label{vary}
\ee
We assume that both the original and varied metric lie in the
configuration space $\ms F_{\mf p}$, so that they both satisfy
conditions (\ref{calFdef}).
In this subsection, we will derive some resulting boundary
conditions on the metric variation $h_{ab}$
that will be useful in later
sections of the paper.  Specifically these conditions are
\begin{subequations}
\label{zcondts}
\begin{eqnarray}
\label{z1}
h_{ab} \ell^b \  &{\hateq}& \ 0, \\
\label{z2}
\nabla_c (h_{ab} \ell^a \ell^b ) \ &{\hateq}&\  0.
\end{eqnarray}
\end{subequations}
Note that the condition (\ref{z2}) is independent of the definition of $\ell^a$
off the surface, because of Eq.\ (\ref{z1}).  As a consistency check of our computations,
we show in Appendix \ref{sec:consistency} that the conditions (\ref{zcondts}) are automatically satisfied
for a metric perturbation of the form
\be
h_{ab} = \lie_\xi g_{ab}
\label{metricsym}
\ee
generated by a representative $\xi^a$ of a symmetry in the algebra discussed in the previous section.

We now turn to the derivation of Eqs.\ (\ref{zcondts}). Equation (\ref{z1})
follows from taking the variation of Eq.\ (\ref{calFdef1}) and noting
that $\ell^a$ and ${\hat \ell}_a$ are fixed under the variation.
By varying the definition (\ref{calFdef2}) of non-affinity and noting that
$\kappa$ is fixed under the variation we find
\be
\nabla_a h_{bc} \ell^a \ell^c - \nabla_b h_{ac} \ell^a \ell^c /2 =0.
\ee
We can rewrite the first term as $\ell^a \nabla_a (\ell^c h_{bc}) -
(\ell^a \nabla_a \ell^c) h_{bc}$.  The first term here vanishes by
Eq.\ (\ref{z1}) since the derivative is along the surface ${\cal N}$,
while the second vanishes by Eqs.\ (\ref{calFdef2}) and (\ref{z1}).
Thus we obtain $\nabla_b h_{ac} \ell^a \ell^c=0$, which is equivalent
to Eq.\ (\ref{z2}) by Eq.\ (\ref{z1}).

It follows from Eq.\ (\ref{z1}) that we can regard $h^{ab}$ restricted
to ${\cal N}$ as an intrinsic tensor $h^{ij}$ on ${\cal N}$.
We can also construct the down index versions
\begin{subequations}
\begin{eqnarray}
h_{i}^{\ j} & = & q_{ik} h^{kj} \\
h_{ij} &=& q_{ik} q_{jl} h^{kl} = \Pi^a_i \Pi^b_j h_{ab}.
\end{eqnarray}
\end{subequations}
These quantities satisfy
\be
h_{ij} \ell^i \hateq h_i^{\ \,j} \ell^i \hateq 0,
\label{iddd}
\ee
from Eqs.\ (\ref{downshift}) and (\ref{z1}).
In four spacetime dimensions, $h^{ij}$ contains six independent
components, $h_i^{\ j}$ five, and $h_{ij}$ three.
We will express charge variations in Sec.\ \ref{sec:wzcharges} below in terms of $h_i^{\ \,j}$.

A useful quantity involving the metric perturbation that will appear
in the charge variations can be defined as follows.  Defining $h_a =
h_{a}^{\ \,b} {\hat \ell}_b$, we have from Eqs.\ (\ref{calFdef1}) and (\ref{z1})
that $h_a$ vanishes on ${\cal N}$. Hence there exists a one-form $\Gamma_a$
on ${\cal N}$
so that
\be
\nabla_{[a} h_{b]} \hateq {\hat \ell}_{[a} \Gamma_{b]}.
\label{Gammadef}
\ee
The quantity $\Gamma_a$ depends linearly on $h_{a}^{\ \,b}$ and its
first derivatives, including in directions off the surface ${\cal N}$,
but is independent of the background metric and
connection.  It does depend on how one extends the definition of
${\hat \ell}_a$ off the surface ${\cal N}$.  However if we impose on
this extension the condition
\be
\nabla_{[a} {\hat \ell}_{b]} \hateq 0,
\label{impose}
\ee
then $\Gamma_a$ is uniquely determined up to the transformation
$\Gamma_a \to \Gamma_a + \zeta {\hat \ell}_a$ for some $\zeta$.
It follows that the pullback
\be
\Gamma_i = \Pi^a_i \Gamma_a
\ee
is unique.
The quantity $\Gamma_i$ 
is invariant under a rescaling of the normal $\ell^a \to e^\sigma \ell^a$.
It also satisfies
\be
\Gamma_i \ell^i =0,
\label{Gammaconstraint}
\ee
from Eq.\ (\ref{z2}).

\section{Global and localized charges for a null boundary component}
\label{sec:wzcharges}

From the perspective of the covariant phase space, we have seen in the
last two sections that general relativity
in the presence of null boundaries has quite a rich structure as
encapsulated in
the infinite dimensional symmetry algebra $\mathfrak{g}_{\mf u}$. With these
symmetries at hand, in this section we move on to the
calculation of the corresponding charges and fluxes.
We compute the Noether charge $Q_\xi$ in Sec.\ \ref{sec:noether}, its
variation $\delta Q_\xi$ in Sec.\ \ref{sec:noether1}, the boundary
symmetry generator ${\cal Q}_\xi$ and its variation in
Sec.\ \ref{sec:ham},
and the localized charge or Wald-Zoupas charge ${\cal Q}^{\rm
  loc}_\xi$ and its flux in Sec.\
\ref{sec:wz}.

Appendix \ref{app:alternative} computes the corresponding charges for a modified definition of field
configuration space.  As mentioned in the previous section the drawback of the modification is that
one looses uniqueness in the prescription for defining localized charges.

\subsection{Noether charge}
\label{sec:noether}

For general relativity in vacuum, the Lagrangian, presymplectic
potential \(3\)-form and Noether charge \(2\)-form are given by \cite{WZ}
\begin{subequations}\begin{align}
        L_{abcd} & = \tfrac{1}{16\pi} \varepsilon_{abcd}~ R, \\
        \theta_{abc} & = \tfrac{1}{16\pi} \varepsilon_{abc}{}^d~  ( g^{ef} \nabla_{d}h_{ef} - \nabla^{e}h_{de} ), \label{Presymplectic form} \\
        Q_{\xi\,\, ab} & = -\tfrac{1}{16\pi}\varepsilon _{abcd}~ \nabla^c \xi^d. \label{Noether form}
\end{align}\end{subequations}
The ambiguity (\ref{amb2}) in the presymplectic potential can be resolved in the case
 of general relativity by demanding that the total number of derivatives of
 the metric $g_{ab}$ or metric perturbation $h_{ab}$ in the 2-form
 $Y_{ab}$ be two less than the number of derivatives appearing in the
 Lagrangian.  One can readily convince oneself that there is no 2-form
 $Y_{ab}$ that depends on $g_{ab}$, $\varepsilon_{abcd}$ and $h_{ab}$
 that depends linearly on $h_{ab}$ and has no derivatives.

We now evaluate the pullback of the Noether charge $2$-form to ${\cal
  N}$.  From Eqs.\ (\ref{barepsilon}) and (\ref{lAvanish})
we find
\be
Q_{\xi\,\, ij} = - \frac{1}{16 \pi} \Pi^a_i \Pi^b_j (
 {\bar \varepsilon}_{abc} {\hat \ell}_d
- {\bar \varepsilon}_{dab} {\hat \ell}_c
+ {\bar \varepsilon}_{cda} {\hat \ell}_b
- {\bar \varepsilon}_{bcd} {\hat \ell}_a) \nabla^c \xi^d = -
\frac{1}{16 \pi} \Pi^a_i \Pi^b_j  {\bar \varepsilon}_{abc} q^c
\ee
where $q^c = 2 {\hat \ell}_d \nabla^{[c} \xi^{d]}$.  Since ${\hat
  \ell}_a q^a=0$ it follows from Eqs.\ (\ref{downshift}) and
(\ref{ve3}) that we can rewrite this expression in terms of tensors
intrinsic to ${\cal N}$.
\be
Q_{\xi\,\, ij} = - \frac{1}{16 \pi} \varepsilon_{ijk} q^k.
\label{qij}
\ee
We can rewrite $q^c$ as
\be
q^c = g^{cd} \lie_\xi {\hat \ell}_d + \lie_\xi \ell^c - 2 \xi^b
\nabla_b \ell^c,
\ee
where we have used the validity of Eq.\ (\ref{calFdef1}) on ${\cal N}$
and the fact that $\xi^b \nabla_b$ differentiates along the surface,
by Eq.\ (\ref{tangent}).  Next, using the definitions (\ref{betadef}) and
(\ref{gammadef}) of $\beta$ and $\gamma$ and the condition (\ref{condition1}) we obtain
\be
q^c = (\beta + \gamma) \ell^c - 2 \xi^b \nabla_b \ell^c = 2 \beta
\ell^c - 2 \xi^b \nabla_b \ell^c.
\label{qcc}
\ee
From Eqs.\ (\ref{downshift}) and (\ref{tangent}) the contracted $b$ index in the
second term can be replaced by an intrinsic index $k$, and we can then use the
definition (\ref{weingarten}) of the Weingarten map.  Inserting the result into
(\ref{qij}) and using (\ref{restriction}) gives
\be
Q_{\xi\,\, ij} = \frac{1}{8 \pi} \varepsilon_{ijk} \left[ \chi^l {\cal K}_l^{\
    k} - \beta(\chi^i) \ell^k \right].
\label{noether1}
\ee
This expression is invariant under the scaling transformation (\ref{rescale-equiv}),
from the transformation properties (\ref{calKtrans}), (\ref{betap1}) and (\ref{vescale}).

Suppose now that we are given a cross section ${\cal S}$ of ${\cal
  N}$.  The Noether charge associated with that cross section is given
by integrating the two form (\ref{noether1}).  Letting $n_i$ denote
the unique normal covector to ${\cal S}$ in ${\cal N}$ with the
normalization (\ref{normalized1}), we obtain from the definition (\ref{ve2})
\be
Q_\xi({\cal S}) = \int_{\cal S} \df Q_\xi = \frac{1}{8 \pi} \int_{\cal S} \varepsilon_{ij} \left[
  \chi^l n_k {\cal K}_l^{\ k} + \beta(\chi^i) \right].
\label{noether2}
\ee

\subsection{Variation of Noether charge}
\label{sec:noether1}

We next turn to computing the variation of the Noether charge under
a variation of the metric of the form (\ref{vary}).  Of all the
quantities which appear in the expression (\ref{noether1}) for the pullback of
the Noether charge two-form, only the volume form $\varepsilon_{ijk}$
and the Weingarten map ${\cal K}_l^{\ k}$ vary as the metric is varied.
Using $\delta \varepsilon_{abcd} = h {\varepsilon}_{abcd}/2$ with $h =
g^{ab} h_{ab} = q^{AB} \delta q_{AB}$ we obtain
\be
16 \pi \delta Q_{\xi\,\, ij} = \varepsilon_{ijk} \left[ h \chi^l {\cal K}_l^{\
    k} - h \beta(\chi^i) \ell^k + 2 \chi^l \delta {\cal K}_l^{\ k} \right].
\label{vnoether}
\ee

To compute the variation of the Weingarten map we define $K_a^{\ b} =
\nabla_a \ell^b$, which when we pullback the $a$ index is orthogonal
to ${\hat \ell}_b$ on the $b$ index and reduces to ${\cal K}_l^{\ k}$.
Taking a variation we find
\be
2 \Pi^a_i \delta K_a^{\ b} = \Pi^a_i \left[ - \nabla^b h_{ac} + \nabla_a h^b_{\ c} +
  \nabla_c h_a^{\ b} \right] \ell^c.
\label{kkk}
\ee
We now use the definition (\ref{Gammadef}) of $\Gamma_a$ to rewrite
the first two terms in (\ref{kkk}), and rewrite the last term in terms
of a Lie derivative.  This yields
\be
\label{kkk00}
2 \Pi^a_i \delta K_a^{\ b} = \Pi^a_i \left[
{\hat \ell}_a \Gamma^b  - \Gamma_a {\hat \ell}^b - h^{bc}
\nabla_a {\hat \ell}_c + h_a^{\ \,c} \nabla^b {\hat \ell}_c
+ \lie_\ell h_a^{\ b} + h^c_{\ a} \nabla_c \ell^b -
h_c^{\ b} \nabla_a \ell^c \right].
\ee
The first term here vanishes by Eq.\ (\ref{lAvanish}).
We can replace the ${\hat \ell}^a$ with $\ell^a$ in the second and
third terms, using the condition (\ref{calFdef1}) and the fact that the
derivative is along the surface.  We rewrite the fourth term using the
condition (\ref{impose}) as $h_a^{\ \,c} \nabla_c {\hat \ell}^b =
h_a^{\ \,c} \nabla_c \ell^b$, where we have used the fact that the
derivative is along the surface by Eq.\ (\ref{z1}).  We thus obtain
\be
2 \Pi^a_i \delta K_a^{\ b} = \Pi^a_i \left[
  - \Gamma_a \ell^b
+ \lie_\ell h_a^{\ b} + 2 h^c_{\ a} \nabla_c \ell^b -
2 h^{b}_{\ \,c} \nabla_a \ell^c \right].
\ee
Now the individual terms on the right hand side are all
orthogonal\footnote{For the second term this is because
${\hat \ell}_b \lie_\ell h_a^{\ \,b} = \lie_\ell (h_a^{\ \,b} {\hat
  \ell}_b) - h_a^{\ \,b} \lie_\ell {\hat \ell}_b = - h_a^{\ \,b}
\ell^c \lie_\ell g_{bc} = -h_a^{\ \,b} \nabla_b (\ell_c \ell^c)/2 -
\kappa h_a^{\ \,b} \ell_b=0$.}
to ${\hat \ell}_b$, by Eqs.\ (\ref{ortho}) and (\ref{z1}).
Hence they all give rise to tensors intrinsic to ${\cal N}$.
Also the contractions on the $c$ index in the last two terms can be
replaced by intrinsic contractions, by Eqs.\ (\ref{downshift}) and (\ref{z1}).
Using the definition (\ref{weingarten}) finally gives
\be
\delta {\cal K}_i^{\ \,j} = - \frac{1}{2} \Gamma_i \ell^j +
\frac{1}{2} \lie_\ell h_i^{\ \,j} + h^{k}_{\ \,i} {\cal K}_k^{\ \,j} -
h^{j}_{\ \,k} {\cal K}_{i}^{\ \,k}.
\label{deltaWein}
\ee
From Eq.\ (\ref{vnoether}) we obtain for the variation of the pullback of the
Noether charge two-form
\be
 \delta Q_{\xi\,\, ij} = \frac{1}{16 \pi} \varepsilon_{ijk} \left[ h \chi^l {\cal K}_l^{\
    k} - h \beta(\chi^i) \ell^k -  \chi^l \Gamma_l \ell^k +
\chi^l \lie_\ell h_l^{\ \,k}
+ 2 \chi^l h^{m}_{\ \ l} {\cal K}_m^{\ \,k} -
2 \chi^l h^{k}_{\ \,m} {\cal K}_{l}^{\ \,m} \right].
\label{vnoether1}
\ee

\subsection{Global charges that generate boundary symmetries}
\label{sec:ham}

As described in Sec.\ \ref{sec:globalcharges} above, the charge ${\cal Q}_\xi$ that
generates a boundary symmetry $\xi^a$ has a variation $\delta {\cal
  Q}_\xi$ which is an integral of the form (\ref{Hamiltonian presymplectic}) over a
Cauchy surface $\Sigma$, and can be expressed as the surface integral
(\ref{Hamiltonian Definition}) over the boundary $\partial \Sigma$.
We now assume that a cross section ${\cal S}$ of the null surface
${\cal N}$ is a component of the boundary $\partial \Sigma$.
This gives
\be
\delta {\cal Q}_\xi = \int_{\cal S} \delta {\cal Q}_{\xi\,\, ab} + \ldots
\label{hehe}
\ee
where the ellipses represent integrals over the remaining components of
$\partial \Sigma$ (for example spatial infinity).
Here the integrand is the two-form
\be
\delta {\cal Q}_{\xi\,\, ab} = \delta Q_{\xi\,\, ab} - \xi^c \theta_{cab}
\label{H2}
\ee
and $\theta_{abc}$ is the presymplectic potential three-form
(\ref{Presymplectic form}).  Pulling back this expression to ${\cal
  N}$ using Eqs.\ (\ref{barepsilon}) and (\ref{ve3}) gives
\be
\theta_{ijk} = \frac{1}{16 \pi} \varepsilon_{ijk} \ell^f \left(
  \nabla_f h - \nabla_e h_f^{\ \,e} \right).
\label{theta11}
\ee
The second term can be rewritten using the boundary conditions
(\ref{z1}) and (\ref{z2}) as $-l^f \nabla_e h_{f}^{\ \,e} =  h_f^{\
  \,e} \nabla_e \ell^f$, which then allows using the definition
(\ref{weingarten}).  This yields
\be
\theta_{ijk} = \frac{1}{16 \pi} \varepsilon_{ijk} \left[ \lie_\ell h +
  h_i^{\ \,j} {\cal K}_j^{\ \,i} \right].
\label{theta22}
\ee
In this expression the trace $h$ of the metric perturbation can be
written as $q^{AB} h_{AB}$, from the condition (\ref{z1}), while the
second term in the brackets can be written as\footnote{An alternative
  form is $h_{ij} K_{kl} q^{ik} q^{jl}$ where $q^{ij}$ is any tensor
  that satisfies $q^{ij} q_{ik} q_{jl} = q_{kl}$, from Eq.\ (\ref{downshift1}).}
\be
h_{AB} {\cal K}^{AB} = h_{AB} \left( \frac{1}{2} \theta q^{AB} +
  \sigma^{AB} \right),
\label{simple}
\ee
from Eqs.\ (\ref{downshift1}), (\ref{sffp1}), (\ref{sffd}),
(\ref{sffp3}) and (\ref{iddd}).  Finally using Eqs.\
(\ref{vnoether1}) and (\ref{theta22}) yields for
the pullback of the perturbed symmetry generator two-form (\ref{H2})
\begin{eqnarray}
 \delta {\cal Q}_{\xi\,\, ij} &=& \frac{1}{16 \pi} \varepsilon_{ijk} \bigg[ h \chi^l {\cal K}_l^{\
    k} - h \beta(\chi^i) \ell^k -  \chi^l \Gamma_l \ell^k +
\chi^l \lie_\ell h_l^{\ \,k}
+ 2 \chi^l h^{m}_{\ \ l} {\cal K}_m^{\ \,k} \nonumber \\
&&
- 2 \chi^l h^{k}_{\ \,m} {\cal K}_{l}^{\ \,m}
-\chi^k \lie_\ell h - \chi^k h_i^{\ \,j} {\cal K}_j^{\ \,i}
\bigg].
\label{H2a}
\end{eqnarray}

In general this expression is not a total variation, and so cannot be
integrated up to
compute a finite charge corresponding to the first term in the
symmetry generator (\ref{hehe}).  To see this, we compute the pullback
(\ref{Hamiltonian Condition}) to
${\cal N}$ of the presymplectic current, contracted with $\chi^i$.
As shown in Sec.\ \ref{sec:cov-phase} above, when this quantity is
nonzero the variation (\ref{H2a}) is not a total variation.
Taking a variation of the expression (\ref{theta22}) for the pullback
of the presymplectic potential, and using the formula (\ref{deltaWein}) for the
variation of the Weingarten map, we obtain
\begin{eqnarray}
\label{omegapb}
\chi^i \omega_{ijk}(h^{lm},h^{\prime \, pq}) &=& \frac{1}{16 \pi} \chi^i
\varepsilon_{ijk} \bigg[ \frac{1}{2} h \lie_\ell h'
+ \frac{1}{2} \lie_\ell h_m^{\ \ l} h_l^{\prime \ m}
+ \frac{1}{2}  h h_l^{\prime\ m}  {\cal K}_m^{\ \,l}
+ 2 h_m^{\ \, p} h_l^{\prime \ m} {\cal K}_p^{\ \,l}
\bigg] \nonumber\\ && - (h \leftrightarrow h').
\end{eqnarray}
The integral of this quantity over a cross section ${\cal S}$ is
nonvanishing in general.  It does vanish in the case
when $\chi^i$ is tangent to ${\cal S}$, that is, when it is a generator
of the diffeomorphism symmetries.
It also vanishes if we demand that both the background and perturbed
configurations are shear and expansion free on ${\cal S}$, and in
particular if they are stationary on ${\cal S}$.

We now specialize to the case where the null surface ${\cal N}$ is the
future event horizon ${\cal H}^+$ of a black hole.  The boundary of
the horizon consists of the asymptotic boundary ${\cal H}_+^+$ at
future timelike infinity, together with a bifurcation twosphere ${\cal
  H}^+_-$ in the case of an eternal black hole.  In Appendix
\ref{sec:bh} we show that the obstruction (\ref{omegapb}) vanishes on
the bifurcation twosphere ${\cal H}^+_-$.  We also show that the
obstruction vanishes on the future boundary ${\cal H}_+^+$, assuming
certain fall off conditions on the shear along the horizon towards
future timelike infinity, and we argue that these fall off conditions
are physically reasonable.  Hence for horizons, Eq.\ (\ref{hehe}) can be
used directly to compute the contribution from the horizon to global charges.

We also show in Appendix \ref{sec:bh} that the correction term
$i_\xi \df\Theta$ in the definition (\ref{localcharge1}) of the
localized charge vanishes on the boundaries ${\cal H}^+_\pm$, under
the same assumptions as above.  Hence from Eq.\ (\ref{locglobal})
the contribution from the horizon to the global charge can be written
as
\be
\mc Q_\xi^{\rm loc}({\cal H}^+_+) - \mc Q_\xi^{\rm loc}({\cal H}^+_-).
\label{softhair}
\ee
Here $\mc Q_\xi^{\rm loc}$ is the localized charge which is computed
explicitly in the next subsection, cf.\ Eq.\ (\ref{GenHam}).

Global conservation laws involving these global charges are discussed
further in Sec.\ \ref{sec:globalcons} below.

\subsection{Localized (Wald-Zoupas) charges and fluxes}
\label{sec:wz}

We now turn to a computation of localized charges $\mc Q_\xi^{\rm
  loc}({\cal S})$ for cross sections ${\cal S}$ of a null surface
${\cal N}$.
As explained in Sec.\ \ref{sec:cov-phase} above, the integrand in the
expression (\ref{localcharge1}) for this charge is given by adding to
the pullback of the right hand side
of Eq.\ (\ref{H2}) a term $\chi^k \Theta_{ijk}$, where the
presymplectic potential
$\Theta_{ijk}$ is of the form [cf.\ Eq.\ (\ref{Generalized Potential Prescription})]
\be
\Theta_{ijk} = \theta_{ijk} - \delta \alpha_{ijk}
\label{iid}
\ee
that is necessary for the right hand side to be a total variation.
In addition, $\Theta_{ijk}$
is required to have the property that it vanish on backgrounds for
which
the null surface is shear free and expansion free.
Then the charge is given by the expression (\ref{localcharge}) where the
integrand is
\be
 {\mc Q}_{\xi\ ij}^{\rm loc} = Q_{\xi\,\, ij} - \chi^k \alpha_{ijk},
\label{anss}
\ee
up to an overall constant of integration on phase space.
We verify that this constant of integration vanishes by showing that
the right hand side of Eq.\ (\ref{anss}) vanishes on our reference
solution, and by assuming that the left hand side vanishes
on this solution.  This computation is carried out in Appendix
\ref{app:background}.

We choose the \(3\)-form $\df \alpha$ on \(\mc N\) given by
\be\label{eq:alpha-gr}
\alpha_{ijk} = \frac{1}{8\pi} \theta \varepsilon_{ijk},
\ee
where $\theta$ is the expansion (\ref{divell0}).
Computing its variation yields
\be
        \delta \alpha_{ijk} =  \frac{1}{16\pi} \varepsilon_{ijk} ( h
        \theta + \lie_\ell h).
\ee
Combining this with Eqs.\ (\ref{theta22}), (\ref{iid}) and
(\ref{simple}) gives for the
presymplectic potential on ${\cal N}$
\be
        \Theta_{ijk} =  \frac{1}{16\pi} \varepsilon_{ijk} \left[
          h_i^{\ \,j} {\cal K}_j^{\ \,i} - h \theta \right]
          = \frac{1}{16 \pi}
         \varepsilon_{ijk} h^{AB} \left( \sigma_{AB} - \frac{1}{2} q_{AB}
         \theta \right).
\label{Theta11}
\ee
This choice of presymplectic potential on a null surface was
independently previously suggested in Eq. (8.2.20) of a thesis by Morales
\cite{Morales:2008}.  For backgrounds for which the null surface is shear free and expansion
free,
it follows from Eq.\ (\ref{Theta11})
that $\df \Theta$
vanishes, as required.  The two-form (\ref{anss}) is now obtained by combining Eqs.\
(\ref{noether1}) and (\ref{eq:alpha-gr}), which gives
\be
 {\mc Q}_{\xi\ ij}^{\rm loc} = \frac{1}{8 \pi} \varepsilon_{ijk} \left[ \chi^l {\cal K}_l^{\
    k} - \theta \chi^k - \beta(\chi^i) \ell^k \right].
\label{GenHam}
\ee
It follows from the transformation properties (\ref{rescale00}) and
(\ref{betap1}) that this two-form
is invariant under the rescaling (\ref{rescale}).

We next argue that the expression (\ref{GenHam}) we have derived
for the localized charge is unique.
As discussed in Sec.\ \ref{sec:ambiguities} above, the presymplectic
potential $\df
\Theta$ will be unique if there does not exist a 3-form $\df W(\phi)$
on the boundary ${\cal N}$ that is locally and covariantly
constructed out of the fields and of the universal structure,
with the property that its variation $\delta \df W$ vanishes
identically on solutions for which the null boundary is shear free and
expansion free.
We assume that $\df W$ depends analytically on the fields, and that
the maximum number of derivatives in the expression for $\df W$ is
one less than the number of derivatives appearing in the Lagrangian,
or one in this case.

The various geometrical quantities on which $\df W$ can depend are
reviewed in Sec.\ \ref{sec:null-geometry} above.  The restriction on
the number of derivatives in $\df W$ eliminates other quantities, not
reviewed in Sec.\ \ref{sec:null-geometry}, that can be used to
construct candidate expressions for $\df W$, such as $\varepsilon_{ijk}
\lie_\ell R$ where $R$ is the Ricci scalar.  Using the finite number
of quantities in Sec.\ \ref{sec:null-geometry} one can show by inspection
that there are no expressions with the right properties.  For example
the expressions $\kappa \varepsilon_{ijk}$ and $\varepsilon_{[ij}
D_{k]}\theta$ not invariant under the transformation (\ref{rescale00}),
while the expression $(\sigma_{AB} \sigma^{AB}/\theta)
\varepsilon_{ijk}$ is invariant but does not depend analytically on
the fields.  We conclude that $\df W=0$ and so $\df \Theta$ and $\mc
Q_\xi^{\rm loc}({\cal S})$ are uniquely determined by our assumptions
and by the Lagrangian $\df L$.


Finally, the on-shell flux $d \df {\mc Q}_\xi^{\rm loc}$ associated with the
localized charge
is given by the symplectic potential $\df \Theta$ evaluated at $h_{ab}
= \lie_\xi g_{ab}$, from Eq.\ (\ref{eq:WZ-flux}).  From the expression
(\ref{Theta11}) for $\df \Theta$, combined with Eqs.\ (\ref{downshift})
and (\ref{weingarten}) to transform from three dimensional notation to
four dimensional notation, we obtain
\be
(d \df {\mc Q}_\xi^{\rm loc})_{ijk}
= \frac{1}{8 \pi} \varepsilon_{ijk} \, \nabla_a \xi_b
\left( \nabla^{(a} \ell^{b)} - g^{ab} \theta \right).
\label{flux1}
\ee
Alternatively, the flux can be obtained by taking an exterior
derivative of the two-form (\ref{GenHam})
\be
(d \df {\mc Q}_\xi^{\rm loc})_{ijk}
= \frac{1}{8 \pi} \varepsilon_{ijk} {\hat D}_p \left[
  \chi^m {\cal K}_m^{\ \,p} - \theta \chi^p - \beta \ell^p \right],
\label{flux2}
\ee
where ${\hat D}_p$ is the divergence operator (\ref{divergenceop}).
It follows from the transformation properties (\ref{rescale00}) and
(\ref{betap1}) that this flux
is invariant under the rescaling (\ref{rescale}).
We show in Appendix \ref{app:consistent1} that the two expressions (\ref{flux1}) and (\ref{flux2}) for the
flux coincide.  This serves as a consistency check of the formalism.

\subsection{Charges and fluxes for specific symmetry generators}
\label{sec:specific}
We now specialize as before to a cross section ${\cal S}$ with normal $n_i$.
For the special case of a supertranslation with $\chi^k = f
\ell^k$, integrating the 2-form (\ref{GenHam}) over
${\cal S}$ and using Eqs.\ (\ref{sffp2}), (\ref{betaff}), (\ref{ve2})
and (\ref{normalized1}) gives the charge
\be
\mc Q_f^{\rm loc}({\cal S})
 = \frac{1}{8 \pi} \int_{\cal S} \varepsilon_{ij} \left[ \theta f -
  \lie_\ell f - \kappa f \right].
\label{GenHam3}
\ee
For stationary null surfaces, these charges vanish identically for
affine supertranslations, for which $\lie_\ell f + \kappa f=0$.
The corresponding flux through a region $\Delta {\cal N}$ of ${\cal
  N}$ is given by integrating the expression (\ref{flux2}):
\be
\Delta \mc Q_f^{\rm loc}
= \frac{1}{8 \pi} \int_{\Delta {\cal N}} \varepsilon_{ijk}
{\hat D}_p \left[ ( f \kappa - \theta f + \lie_\ell f) \ell^p \right].
\ee
This can be simplified using the formula (\ref{divell0}), the symmetry
condition (\ref{estdef}) and Raychaudhuri's equation in vacuum to give
\be
\Delta \mc Q_f^{\rm loc}
 = \frac{1}{8 \pi} \int_{\Delta {\cal N}} \varepsilon_{ijk}
f  ( \theta \kappa - \theta^2  - \lie_\ell \theta)
= \frac{1}{8 \pi} \int_{\Delta {\cal N}} \varepsilon_{ijk}
f  \left( \sigma_{AB} \sigma^{AB} - \frac{1}{2} \theta^2 \right).
\ee

We next consider $\diff({\cal Z})$ generators of the form $\chi^i =
X^i$ where $X^i n_i=0$, making use of the decomposition (\ref{unde}).
Here $n_i$ is the normal to the cross section ${\cal S}$, and we also
demand that it obey the differential equation (\ref{nieqn}) on ${\cal N}$,
in order that $X^i$ be an element of the symmetry algebra $\mf g_{\mf
  u}$.
For such generators the pullback to ${\cal S}$ of $\varepsilon_{ijk} X^k$
vanishes, and so from Eqs.\ (\ref{anss}) and (\ref{eq:alpha-gr}) the localized
charge and Noether charge coincide.  From Eq.\ (\ref{GenHam})
the localized charge is
\be
\mc Q_X^{\rm loc}({\cal S})
= \frac{1}{8 \pi} \int_{\cal S} \varepsilon_{ij} \left[ X^l
  {\cal K}_l^{\ \,k} n_k + \beta \right],
\label{GenHamX}
\ee
and the corresponding flux from Eq.\ (\ref{flux2}) is
\be
\Delta \mc Q_X^{\rm loc}= \frac{1}{8 \pi} \int_{\Delta {\cal N}} \varepsilon_{ijk} {\hat D}_p \left[
  X^m {\cal K}_m^{\ \,p} - \theta X^p  -\beta \ell^p \right].
\ee


\subsection{Stationary regions of the null surface}
\label{sec:stat}

We now specialize to stationary regions of the null surface to obtain
explicit forms for the various charges.  In stationary regions the general charge (\ref{GenHam})
reduces to
\be
\mc Q_\xi^{\rm loc}({\cal S}) = - \frac{1}{8 \pi} \int_{\cal S}
\varepsilon_{ij} (\chi^l \omega_l - \beta),
\label{statform}
\ee
by Eqs.\ (\ref{stationary1}), (\ref{r1ff}) and
(\ref{normalized1}).  The integrand here has a vanishing Lie derivative with respect to the
Killing field $\tau^i$, so the result is independent of ${\cal S}$ as one
would expect.  To see this, take the Lie derivative of the integrand
with respect to $\ell^i$, and simplify using Eqs.\ (\ref{stat00}),
(\ref{symmetrydeflinear}) and (\ref{twistprop}) to obtain $-
\varepsilon_{ij} [ \chi^k \lie_\ell \omega_k - \lie_\chi \kappa ]/(8
\pi)$.
Now specializing without loss of generality to the choice of representative $\ell^i = \tau^i$ and
using Eqs.\ (\ref{stationary2}) and (\ref{zeroth}) shows that the
expression vanishes.

We next specialize to the choice of normal $\ell^i = \tau^i$, and
to a coordinate
system $(\tau,\theta^A)$ for which the Killing field is ${\vec \tau}
= \partial/\partial \tau$, and we write the rotation one-form (\ref{r1fstat})
and
symmetry generator as
\be
\df \omega_{\tau} = \omega_{\tau\,A}(\theta^B) d\theta^A + \kappa_\tau d\tau
\ee
and
\be
{\vec \chi} = \left[{\hat \alpha}(\theta^A) e^{-\kappa_\tau \tau}
+ {\hat \beta}(\theta^A)
 \right] \frac{\partial}{\partial \tau} +
X^A(\theta^B) \frac{\partial}{\partial \theta^A}.
\label{param1}
\ee
Here ${\hat \alpha}$ parameterizes the affine supertranslations,
${\hat \beta}$ the Killing supertranslations\footnote{Note that these
  parameters ${\hat \alpha}$ and ${\hat \beta}$ do not coincide with
  the parameters $\alpha$ and $\beta$ of Eq.\ (\ref{groupt1}).}, and $X^A$ the ${\rm
  diff}(S^2)$ transformations or superrotations.  From Eq.\ (\ref{betadef})
we obtain $\beta = \kappa_\tau {\hat \alpha} \exp[-\kappa_\tau \tau]$
and substituting into Eq.\ (\ref{statform}) gives
\be
\mc Q_\xi^{\rm loc}({\cal S}) = - \frac{1}{8 \pi} \int_{\cal S}
\varepsilon_{ij} (X^A \omega_{\tau \, A} + {\hat \beta} \kappa_\tau).
\label{statform1}
\ee

\section{Global conservation laws involving black holes}
\label{sec:globalcons}

Although our analysis has considered arbitrary null surfaces
in the preceding sections, our main interest lies with black
holes. Accordingly, in the next few sections, we take $\mathcal{N}$ to be the future event
horizon ${\cal H}^+$ of a black hole. This can either be a black hole
formed in gravitational collapse, or an eternal black hole, as in
Fig.\ \ref{fig:spacetimes} above.  We note that our analysis is
limited to smooth horizons, and that generic horizons are not smooth
because of generators that join the horizon.  We leave the
analysis of charges and symmetries associated with nonsmooth horizons
for future work.

In this section we consider global conservation laws involving black
hole horizons.
As discussed in the introduction, we distinguish between {\it localized}
conservation laws that involve only one component ${\cal B}_j$ of the
spacetime boundary, and {\it global} conservation laws that involve
entire Cauchy surfaces.  The foundation for both types of laws is the
fact that the expression (\ref{Hamiltonian presymplectic}) for the
variation of the global charge $\mc Q_\xi$ is invariant under local
deformations of the hypersurface $\Sigma$ when on shell, from Eq.\
(\ref{cc}).  In situations where the charge variation is a total
variation and the condition (\ref{Hamiltonian Condition}) is satisfied
(which happens for internal symmetries), one does not need to
distinguish between these two types of conservation laws.  More
generally, the localized conservation laws require the application of
the Wald-Zoupas procedure, while the global laws do not, as argued in
Sec.\ \ref{sec:globalcharges} above.

As discussed in the introduction, in the past few years an infinite
set of new global conservation laws in gauge theories have been discovered,
associated with ``large'' gauge transformations which are not trivial at
infinity \cite{He:2014cra,Strominger:2017zoo}.  Similar conservation
laws have been argued for in the gravitational case
\cite{Strominger:2013jfa,Strominger:2017zoo}, although completely
rigorous derivations have yet to be given.
One of the key motivations for studying
horizon symmetries and charges is the realization that the associated
global conservation laws place constraints on black hole evaporation, and
that the (electric parity superrotation) charges constitute ``soft hair'' that may play a role in how
information is released as a black hole evaporates \cite{Strominger:2013jfa, Kapec:2015ena,
  Strominger:2014pwa, Hawking:2016sgy, Hawking:2016msc,
  Strominger:2017aeh} (see also \cite{Strominger:2017zoo} for a
complete review).  In this section we review the status of these
conservation laws and the implications of our results for their formulation.

Consider for example a spacetime with no horizons for which the only
components of the boundary are $\scri^+$, $\scri^-$ and the points at
infinity $i^-$, $i^0$ and $i^+$, and specialize to vacuum general
relativity.  Since the charge variation (\ref{Hamiltonian
  presymplectic}) is invariant under local deformations of the Cauchy
surface $\Sigma$, one can deform $\Sigma$ into the
distant past and also into the distant future.
Then with appropriate sign
conventions one obtains a conservation law of the form
\be
\delta \mc Q_\xi(\scri^-) + \delta \mc Q_\xi(i^-) + \delta \mc
Q_\xi(i^0) = \delta \mc Q_\xi(\scri^+) + \delta \mc Q_\xi(i^+),
\label{conslaweg}
\ee
where each term is an integral of the form (\ref{Hamiltonian presymplectic}) over the
corresponding hypersurface or an appropriate limit of such integrals
converging to one of the points at infinity, assuming such limits exist.
If we specialize to
spacetimes for which the Bondi mass vanishes at $\scri^+_+$, the
future limit of $\scri^+$, and at $\scri^-_-$, the past limit of
$\scri^-$, then the terms at future and past timelike infinity should
vanish \cite{Strominger:2013jfa} giving
\be
\delta \mc Q_\xi(\scri^-)  + \delta \mc
Q_\xi(i^0) = \delta \mc Q_\xi(\scri^+).
\label{conslaw11}
\ee
The contribution from spatial infinity in this equation need not
vanish in general\footnote{For
example, suppose that in Eq.\ (\ref{Hamiltonian presymplectic}) $\phi$ is
the Minkowski metric, $\delta \phi$ is the linearized Schwarzschild
solution, and $\xi^a$ is a vector
field which asymptotes to one timelike Killing vector field
$\tau^a_-$ of the Minkowski background at $\scri^-$ and to another $\tau^a_+$ at $\scri^+$.  Then
$\delta \mc Q_\xi(i^0)$ is proportional to $P_a (\tau^a_+ - \tau^a_-)$,
where $P_a$ is the ADM 4-momentum.}.

To derive a global conservation law one needs to show that the
various limiting integrals exist, and that the contribution $\delta
\mc Q_\xi(i^0)$ vanishes.  Then integrating in phase space would yield a
relation of the form
\be
\mc Q_\xi(\scri^-)  =  \mc Q_\xi(\scri^+),
\label{conslaw12}
\ee
which constrains gravitational scattering \cite{Strominger:2013jfa}.
We expect that imposing suitable boundary conditions at $i^0$ in the definition of
$\ms F$ should eliminate the term $\delta \mc Q_\xi(i^0)$ (this is
closely related to the matching conditions proposed in Ref.\ \cite{Strominger:2017zoo}).
In addition these boundary conditions should reduce
the global symmetry algebra to a diagonal subalgebra of ${\rm BMS}{}^-
\oplus {\rm BMS}{}^+$, with an appropriate identification of
${\rm BMS}{}^-$ and ${\rm BMS}{}^+$, as argued by Strominger
\cite{Strominger:2013jfa}.  See Refs.\
\cite{PhysRevLett.43.181,Campiglia:2015lxa,Campiglia:2017mua,Troessaert:2017jcm,Prabhu2018}
for more detailed analyses of spatial infinity and of the validity of
conservation laws of the form (\ref{conslaw12}).

Consider now the generalization of this discussion to include horizons
\cite{Hawking:2016msc,Hawking:2016sgy}.  For the black hole
formed from gravitational collapse shown in Fig.\
\ref{fig:spacetimes}, and for a representative $\xi^a$ of the global
symmetry algebra, following the argument that led to Eq.\
(\ref{conslaweg}) we obtain
\be
\delta \mc Q_\xi(\scri^-) + \delta \mc Q_\xi(i^-) + \delta \mc
Q_\xi(i^0) = \delta \mc Q_\xi(\scri^+) + \delta \mc Q_\xi(i^+) +
\delta \mc Q_\xi({\cal H}^+).
\label{conslaweg1}
\ee
Here each term is an integral of the form (\ref{Hamiltonian presymplectic}) over the
corresponding hypersurface or an appropriate limit of such integrals
converging to one of the points at infinity. The term $\delta \mc Q_\xi(i^-)$
can be eliminated as described above.
{\it A priori}, the symmetry
generator $\xi^a$ appearing in this equation can have independent
limits at the horizon ${\cal H}^+$ and at null infinity $\scri^+ /
\scri^-$.
However, just as for $i^0$, we expect that imposing appropriate boundary
conditions at future timelike infinity $i^+$ should eliminate the term $\delta \mc Q_\xi(i^+)$,
and impose the appropriate relation between the limits of $\xi^a$ at
${\cal H}^+$ and at $\scri^+/\scri^-$.  Note that this viewpoint
differs from that of Ref.\ \cite{Hawking:2016sgy},
which used the specific prescription of maintaining global Bondi
coordinates to link generators at $\scri^-$ to
those at ${\cal H}^+$.  However the specific identification for
supertranslations obtained there seems inevitable for the case of
spherical symmetry\footnote{\label{caveat}With the following minor adjustment:
assuming that a supertranslation on $\scri^-$ corresponds to some specific element of the
symmetry algebra on ${\cal H}^+$, as found in Ref.\ \cite{Hawking:2016sgy},
the restriction (\ref{collapse}) implies that the algebra element is a linear combination of a
Killing supertranslation and an affine supertranslation,
instead of a pure Killing supertranslation.  In the notation of Eq.\
(\ref{param1}) near ${\cal H}^+_+$, this linear combination will be of the form
$$
{\vec \chi} = {\hat \beta} ( 1 + u_0 e^{-\kappa_\tau \tau}
) \partial_\tau,
$$
where $u_0 = u_0(\theta^A)$ is determined by the conditions
(\ref{collapse0}) or (\ref{condt3}) at early times.  The affine
supertranslation correction term does not contribute to localized charges in
stationary regions or to global charges.}.  For more general
generators and situations, the appropriate
identification of the generators is an interesting question for future
study, and will need to be resolved in order to obtain the general
form of the global conservation law.

Finally, assuming such an identification has been derived, our
explicit expressions for localized charges can be used to
obtain an explicit and nonperturbative form of the resulting global conservation law.
Integrating Eq.\ (\ref{conslaweg1}) in phase space
and making use of Eq.\ (\ref{locglobal}), this form is
\be
\mc Q_\xi^{\rm loc}(\scri^+_-) - \mc Q_\xi^{\rm loc}(\scri^+_+) - \mc
Q_\xi^{\rm loc}({\cal H}^+_+) =
\mc Q_\xi^{\rm loc}(\scri^-_+) - \mc Q_\xi^{\rm loc}(\scri^-_-).
\label{conslaw14a}
\ee
Note that the third term on the left hand side is the operator that
creates soft graviton hair on a black hole horizon in the quantum
theory, when $\xi^a$ is a Killing supertranslation associated with the asymptotic Killing field near future timelike
infinity, as explained by Hawking, Perry and Strominger
\cite{Hawking:2016msc,Hawking:2016sgy}.  Our result (\ref{GenHam})
for this operator improves on existing treatments
\cite{Hawking:2016msc,Hawking:2016sgy}
in that it is nonperturbative and not a variation\footnote{The
  explicit form of this operator is given by Eq.\ (\ref{statform1})
  above, since the horizon is asymptotically stationary,
  assuming the fall-off conditions on the shear of Appendix \ref{sec:bh}.}.

\section{Algebra of symmetry generator charges and central charges}
\label{sec:central}

As is well known, the algebra of the global symmetry
generator charges
$\mc Q_\xi$ under Dirac brackets need not coincide with the
symmetry algebra $\mf g$ of the vector fields $\xi^a$ under Lie
brackets,
and can instead
be a central extension of that algebra \cite{brown1986,LW,Barnich:2001jy,Barnich:2007bf}.
This phenomena already arises in classical mechanics \cite{Arnold}.
For example, there is a nontrivial central extension
for 2+1
dimensional gravity with a negative cosmological constant with a
certain choice of AdS boundary conditions, as shown by Brown and
Henneaux \cite{brown1986}.  There is no central extension for BMS generators in 3+1
dimensional general relativity \cite{Barnich:2011ct}, and we show in
this section that
the same is true for the symmetry algebra of charges at event horizons in general
relativity, assuming certain fall off conditions on the shear near
future timelike infinity.  Thus, there is no central extension of the algebra for the
symmetry algebra of global charges derived in this paper.

\subsection{Algebra of symmetry generator charges in general contexts}
\label{centralextension1}

We first review in this subsection the theory of central extensions
\cite{brown1986,LW,Barnich:2001jy,Barnich:2007bf} in general contexts, and in the
following subsection we will apply it to black hole event horizons.

The first step in the computation of the algebra of global charges
$\mc Q_\xi$ is the computation of the Dirac bracket.  For the specific
case of vacuum general relativity at a future event horizon, a careful
derivation of the Dirac bracket including the effects of zero modes
has been given by Hawking, Perry and Strominger \cite{Hawking:2016sgy}.
Here, for the discussion in a general context, we will assume that a
Dirac bracket can be found for which the global charges implement the
symmetries in the sense\footnote{Our sign convention for Eq.\
(\ref{represent}) is the opposite of that of Ref.\
\cite{Hawking:2016sgy} and agrees with that of Ref.\ \cite{2002clme.book.....G}.}
\be
\left\{ F[\phi], \mc Q_\xi \right\} = \delta_\xi F[\phi],
\label{represent}
\ee
where the variation $\delta_\xi$ is defined by
\be
\delta_\xi F[\phi] = F[\phi + \lie_\xi \phi] - F[\phi].
\label{deltaxidef}
\ee
Here $F$ is any function on the covariant phase space ${\bar{ \ms
    F}}$ (i.e. functional of field configurations $\phi$), and the right
hand side is understood to be linearized in $\xi^a$.  Combining this
with the definition (\ref{Hamiltonian presymplectic}) yields
\be
\left\{ \mc Q_\xi, \mc Q_{\tilde \xi} \right\} = - \Omega_\Sigma(\phi,
\lie_\xi \phi, \lie_{\tilde \xi} \phi).
\label{xitxi}
\ee

An alternative formal derivation of Eq.\ (\ref{xitxi}) is as follows.
We write the presymplectic form (\ref{Omegadef})
as $\Omega_{{\mathscr A}{\mathscr B}}$, where the indices $\mathscr A,
\mathscr B, \ldots $ represent tensors on $\bar{\ms F}$.
The definition (\ref{Hamiltonian presymplectic}) of global charges can
be written in this notation as
\be
\nabla_{\mathscr A} \mc Q_\xi = \Omega_{{\mathscr A}{\mathscr B}}
v_\xi^{{\mathscr B}},
\label{chargedef}
\ee
where $v_\xi^{{\mathscr B}}$ is the vector field on the covariant phase
space that assigns to each solution $\phi$ the linearized solution
$\lie_\xi \phi$.  We assume the existence of a Dirac bracket on functions
$F, G: \bar{\ms F} \to {\bf R}$ of the form
\be
\left\{ F,G \right\} = \Omega^{{\mathscr A}{\mathscr B}}
\nabla_{\mathscr A} F \nabla_{\mathscr B} G
\label{bracketdef}
\ee
where $\Omega^{{\mathscr A}{\mathscr B}}$
satisfies
\be
\Omega_{{\mathscr A}{\mathscr B}} \Omega^{{\mathscr B}{\mathscr C}}
\Omega_{{\mathscr C}{\mathscr D}}  = \Omega_{{\mathscr A}{\mathscr
    D}}.
\label{Poissondef}
\ee
Now inserting the charge definition (\ref{chargedef}) into the bracket
(\ref{bracketdef}) and using Eq.\ (\ref{Poissondef}) gives $\left\{
  \mc Q_\xi, \mc Q_{\tilde \xi} \right\} = - \Omega_{{\mathscr
    A}{\mathscr B}} v_\xi^{\mathscr A} v_{\tilde \xi}^{\mathscr B}$,
which is equivalent to Eq.\ (\ref{xitxi}).

Next, the relation (\ref{xitxi}) can be rewritten
using the formulae (\ref{Omegadef1}) and (\ref{deltaxidef}) as
\be
\left\{ \mc Q_\xi, \mc Q_{\tilde \xi} \right\} = - \int_{\partial
  \Sigma} \left[ \delta_\xi \df Q_{\tilde \xi}(\phi) -
  i_{\tilde \xi} \df \theta(\phi,\lie_\xi \phi) \right].
\label{cm1}
\ee
We now specialize to situations where the presymplectic potential $\df
\Theta$ exists, and where the correction term $i_\xi \df
\Theta$ in the definition (\ref{localcharge1}) of the localized charge vanishes on
$\partial \Sigma$ for all $\xi^a$.  As discussed in Sec.\ \ref{sec:WZ-charge} above, we expect this to be generically valid when
$\Sigma$ is a Cauchy surface.  Taking a variation of Eq.\
(\ref{locglobal}) and combining with Eqs.\ (\ref{Hamiltonian
  Definition}) and (\ref{cm1}) gives
\be
\left\{ \mc Q_\xi, \mc Q_{\tilde \xi} \right\} = - \int_{\partial
  \Sigma}  \delta_\xi \df {\mc Q}^{\rm loc}_{\tilde \xi}(\phi).
\label{cm2}
\ee

Now let $\psi_\varepsilon : M \to M$ be the one parameter family of diffeomorphisms
that move points along integral curves of $\xi^a$.  Since these diffeomorphisms
preserve the boundaries and the universal structures on the boundaries, and since by construction
$\df {\mc Q}_{\tilde \xi}^{\rm loc}$ is local and covariant in the
sense of footnote \ref{localcovariant}, it
follows from the argument in that footnote that\footnote{Note that it
  is important for this argument that $\df {\mc Q}_{\tilde \xi}^{\rm
    loc}$ does not depend on arbitrary choices such as a choice of
  representative of an equivalence class in the universal structure.}
\be
\psi_{\varepsilon*} \df {\mc Q}_{\tilde \xi}^{\rm loc}(\phi) =
 \df {\mc Q}_{\psi_{\varepsilon*} \tilde \xi}^{\rm loc}(\psi_{\varepsilon*}
 \phi).
\ee
Now differentiating with respect to $\varepsilon$ and setting
$\varepsilon=0$ gives the identity
\be
\lie_\xi \df {\mc Q}_{\tilde \xi}^{\rm loc}(\phi) = \delta_\xi \df {\mc
  Q}^{\rm loc}_{\tilde \xi}(\phi) +  \df {\mc
  Q}^{\rm loc}_{\lie_\xi \tilde \xi}(\phi).
\ee
Inserting this into Eq.\ (\ref{cm2}) finally gives
\be
\left\{ \mc Q_\xi, \mc Q_{\tilde \xi} \right\} = \mc Q_{[\xi,{\tilde
    \xi}]} + K_{\xi, {\tilde \xi}},
\label{kkk2}
\ee
where $[ \xi, {\tilde \xi}]^a = \lie_\xi {\tilde \xi}^a$ is the Lie bracket
and
\be
K_{\xi,{\tilde \xi}} =
-  \int_{\partial \Sigma}
  \lie_\xi \df {\mc Q}_{\tilde \xi}^{\rm loc}.
\label{centralcharge1}
\ee
Equation (\ref{kkk2}) shows that when the quantity $K_{\xi,{\tilde \xi}}$ is non vanishing, the algebra
of charges will differ from the algebra of vector fields.

A priori the quantity $K_{\xi,{\tilde \xi}}$ could depend on the
background solution $\phi$.  However, a theorem due to
Brown and Henneaux \cite{1986JMP....27..489B} shows that there is no
such dependence, and so the algebra of charges consists at most of a
central extension of the algebra of vector fields.  A formal version of the argument is as
follows \cite{LW,Seraj:2016cym}:
\be
\nabla_{\mathscr A} \mc Q_{[\xi,{\tilde \xi}]} =
\Omega_{{\mathscr A}{\mathscr B}} v_{[\xi,{\tilde \xi}]}^{\mathscr B}
= -\Omega_{{\mathscr A}{\mathscr B}} \lie_{v_\xi} v_{\tilde
  \xi}^{\mathscr B} = -\lie_{v_\xi} \left(\Omega_{{\mathscr A}{\mathscr B}} v_{\tilde
  \xi}^{\mathscr B} \right) = -\lie_{v_\xi} \nabla_{\mathscr A} \mc
Q_{\tilde \xi}.
\ee
Here we have used the charge definition (\ref{chargedef}), then the
fact that the mapping $\xi^a \to -v_\xi^{\mathscr A}$ is a Lie algebra
homomorphism\footnote{This follows from the fact that $v_\xi$ maps any
  functional $F[\phi]$ to $F[\phi + \lie_\xi \phi] - F[\phi]$ to
  linear order, so $\left[ v_\xi, v_{\tilde \xi} \right] F[\phi] =
  F[\phi + (\lie_{\tilde \xi} \lie_\xi - \lie_\xi \lie_{\tilde \xi})
  \phi] - F[\phi]$.}, then the fact that $\Omega_{{\mathscr
    A}{\mathscr B}}$ is
a closed two form on $\bar {\ms F}$, and finally the definition
(\ref{chargedef}) again.  Continuing we obtain
\be
\nabla_{\mathscr A} \mc Q_{[\xi,{\tilde \xi}]} =
-\nabla_{\mathscr A} \left( v_\xi^{\mathscr B} \nabla_{\mathscr B} \mc
  Q_{\tilde \xi} \right) = -\nabla_{\mathscr A} \left( v_\xi^{\mathscr
    B} \Omega_{{\mathscr B}{\mathscr C}} v_{\tilde \xi}^{\mathscr C}
\right) =  \nabla_{\mathscr A} \left\{ \mc Q_\xi , \mc Q_{\tilde \xi} \right\},
\ee
where we have used Eqs.\ (\ref{chargedef}), (\ref{bracketdef}) and (\ref{Poissondef}).
It follows from Eq.\ (\ref{kkk2}) that $\nabla_{\mathscr A}
K_{\xi,{\tilde \xi}} = 0$, as claimed.

\subsection{Symmetry algebra of global charges at event horizons}
\label{centralextension2}

We now show that the
contribution\footnote{As discussed in Secs.\ \ref{sec:globalcharges} and \ref{sec:algebra} above, a
  symmetry $\xi^a$ can have different limiting forms on different
  boundaries ${\cal B}_j$, and more than one can contribute to the
  central charge (\ref{centralcharge1}), depending on the Cauchy
  surface $\Sigma$.}
to the central charges (\ref{centralcharge1}) from
a future event horizon vanishes, assuming certain fall off conditions on the shear near
future timelike infinity.  This generalizes a result of Guo, Hwang and Wu who
show that the central charges vanish on the horizon of a stationary,
axisymmetric black hole for a large class of generators \cite{Guo:2002ed}.

Consider a connected component ${\cal S}$ of $\partial \Sigma$ which
lies in the event horizon $\mc H^+$.
Using Cartan's formula together with Eq.\ (\ref{flux2}) we find that
the contribution from ${\cal S}$ to the
central charges (\ref{centralcharge1}) can be written as
\be
-\int_{\cal S} i_\xi d \df
{\mc Q}_{\tilde \xi}^{\rm loc} = - \frac{1}{8 \pi} \int_{\cal S} \chi^i \varepsilon_{ijk} {\hat D}_p \left[ {\tilde \chi}^m
  {\cal K}_m^{\ p} - \theta {\tilde \chi}^p - {\tilde \beta} \ell^p
\right].
\label{centralcharge3}
\ee
Now ${\cal S}$ cannot lie in the interior of ${\cal H}^+$, otherwise
$\Sigma$ would not be a Cauchy surface.  We consider two different
cases:
\begin{itemize}
\item The cross section ${\cal S}$ coincides with a component of the
  boundary ${\cal H}^+$, for example the bifurcation
  two-sphere ${\cal H}^+_-$ in an eternal black hole spacetime.  Since $\chi^i$ must
  be tangent to ${\cal S}$ in this case, as argued in Sec.\ \ref{sec:uiss1} above, it
  follows that the quantity (\ref{centralcharge3}) vanishes.

\item The cross section ${\cal S}$ represents the future asymptotic boundary ${\cal
H}^+_+$ of ${\cal H}^+$.  Now as discussed in Appendix
\ref{sec:bh}, event horizons are asymptotically stationary.
Assuming exact stationarity and using the
condition (\ref{stationary1}), the quantity (\ref{centralcharge3}) reduces to
\be
- \frac{1}{8 \pi} \int_{\cal S} \chi^i \varepsilon_{ijk} {\hat D}_p \left[ {\tilde \chi}^m
  \omega_m \ell^p - {\tilde \beta} \ell^p \right] = - \frac{1}{8 \pi} \int_{\cal S} \chi^i \varepsilon_{ijk} \lie_\ell \left[ {\tilde \chi}^m
  \omega_m  - {\tilde \beta} \right].
\label{centralcharge4}
\ee
Here $\omega_m$ is the rotation one-form (\ref{r1f0}) and we have used
Eqs.\ (\ref{divell0}), (\ref{sffd}) and (\ref{stationary1}).  The Lie derivative in the
integrand on the right hand side of Eq.\ (\ref{centralcharge4}) vanishes, as shown after Eq.\ (\ref{statform}), and so the result vanishes.
In this analysis we have set the shear $\sigma_{ij}$ and expansion
$\theta$ to zero.  Assuming instead the falloff conditions
$\sigma_{ij},\theta \sim v^{-p}$ with $p>1$ of Appendix \ref{sec:bh},
where $v$ is affine parameter,
one can show by an analysis similar to that of Appendix \ref{sec:bh}
that the contribution of the shear and expansion to the expression
(\ref{centralcharge3}) vanishes in the limit $v \to \infty$.
Hence the contribution from ${\cal H}^+_+$ to the central charge (\ref{centralcharge1}) vanishes.

\end{itemize}

\subsection{Symmetry algebras of localized charges}
\label{centralextension3}

One can also consider the algebra of
localized charges $\mc Q_\xi^{\rm loc}({\cal S})$.
The Poisson bracket of two such charges
$\mc Q_\xi^{\rm loc}({\cal S})$ and $\mc Q_{{\tilde \xi}}^{\rm
  loc}({\tilde {\cal S}})$
will in general depend on the
two-surfaces ${\cal S}$ and ${\tilde {\cal S}}$, but if one specializes to a stationary region
of the null surface ${\cal N}$ the bracket becomes independent of the
two-surfaces.  It will be of the form\footnote{Barnich and Troessaert have
  shown that
  the anomalous term $K^{\rm loc}_{\xi, {\tilde \xi}}(\phi)$ vanishes
  for the case of BMS generators at null infinity in
  3+1 general relativity \cite{Barnich:2011mi}.}
\be
\left\{ \mc Q_\xi^{\rm loc}(\phi), \mc Q_{\tilde \xi}^{\rm loc}(\phi) \right\} = \mc Q_{[\xi,{\tilde
    \xi}]}^{\rm loc}(\phi) + K^{\rm loc}_{\xi, {\tilde \xi}}(\phi),
\label{kkk16}
\ee
where the anomalous term $K^{\rm loc}_{\xi, {\tilde \xi}}(\phi)$ will in general
depend on the background solution $\phi$ \cite{Barnich:2011mi},
in contrast to the situation
(\ref{kkk2}) for the global charges.
While we do not consider the algebra (\ref{kkk16}) in this paper, we
note that a recent paper by Haco, Hawking, Perry and Strominger has
computed
such an algebra for Kerr black holes, and used it to derive the
Bekenstein-Hawking entropy \cite{Andy2018talk,Haco:2018ske}.
The symmetry algebra $\mf g$ of vector fields used
there is not the same as the algebra (\ref{symmetrydeflinear}) used in this paper, and may be
related to the extended algebra we discuss in Appendix \ref{app:alternative}.

\section{Discussion, applications and future directions}
\label{sec:discussion}

In this final section we recap our main results, discuss some
implications and applications, and discuss some open questions and
future directions.

\subsection{Recap}

In this paper, we have applied the covariant phase space formalism to
general relativity with a null boundary. By an appropriate gauge-fixing
at the boundary we defined a field configuration space, and derived
the conditions for linearized diffeomorphisms to preserve this configuration space.
Factoring out by the degeneracies left us with the infinite
dimensional symmetry algebra $\mathfrak{g} = \diff(\twosphere) \ltimes
\mathfrak{s}$, where $\mathfrak{s}$ is the set of supertranslations at
$\mathcal{N}$ i.e. vector fields $\chi^i = f \ell^i$ satisfying
$\lie_{\ell}(\lie_{\ell}f + \kappa f) = 0$.
Supertranslations were therefore found to be
symmetries of general relativity at general null boundaries. We then
calculated the general form of the global conserved charges, and the
localized charges and fluxes associated to $\mathfrak{g}$ by way of the Wald-Zoupas
prescription. In particular, we found explicit expressions for the
supertranslation localized charges and fluxes.  These expressions are unique when we
impose the condition that the potential $\df \Theta$ for the
presymplectic current on the null surface vanish when the surface is
shear free and expansion free.

\subsection{Black holes: localized conservation laws and horizon memory}
\label{sec:applic:local}

We next discuss the implications and
interpretation in the event horizon context
of the localized conservation laws that we have derived.

As discussed in Sec.\ \ref{sec:WZ-charge} above, given any two cross sections ${\cal
  S}$ and ${\cal S}'$ of the event horizon, we have for each symmetry
generator a localized conservation law of the form
\be
\int_{\Delta {\cal N}} d \df {\mc Q}_\xi^{\rm loc} =
\int_{{\cal S}'} \df {\mc Q}_\xi^{\rm loc} - \int_{{\cal S}'} \df {\mc
  Q}_\xi^{\rm loc},
\label{lcl}
\ee
where $\Delta {\cal N}$ is the region of ${\cal N}$ between ${\cal S}$
and ${\cal S}'$ and explicit expressions for the charge and flux are
given in Eqs.\ (\ref{GenHam}) and (\ref{flux2}).  Now since the event
horizon has a boundary (either an initial event ${\cal P}$ or a
bifurcation two-sphere), some of the symmetry generators $\chi^i$
of the algebra discussed in Sec.\ \ref{sec:universal} do not preserve
the boundary.  As discussed in Sec.\ \ref{sec:uiss1}, those generators must be
excluded from the global algebra $\mf g$ that is relevant for global
conservations laws.  Nevertheless, the conservation law (\ref{lcl}) is
valid for all generators.  This is because the derivation of the law
(\ref{lcl}) is local, and is not invalidated if the vector field
violates the required boundary conditions at $\partial {\cal N}$ if
$\partial {\cal N}$ is disjoint from $\Delta {\cal N}$.

In order to get some insight into the physical interpretation of the
charges in (\ref{lcl}), we specialize to stationary regions.
The three different types of generators are:
\begin{itemize}

\item {\it Affine supertranslations:} The associated charges vanish
  identically in stationary regions, as noted in Sec.\
  \ref{sec:specific} above.

\item {\it Superrotations or ${\rm diff}(S^2)$ generators:}  The
  corresponding
charges in stationary regions are given by the first term in Eq.\
(\ref{statform1}) above.  The curl (magnetic parity) piece of $X^A$ yields the horizon
angular momentum multipoles of Ashtekar \cite{Ashtekar:2004gp}, while
the gradient (electric parity) piece gives additional charges.

\item {\it Killing supertranslations:} The charge in this case is
  given by the second term in Eq.\ (\ref{statform1}).

\end{itemize}

These charges all vanish for a Schwarzschild black hole, except for the
$l=m=0$ component of the Killing supertranslation charge in
(\ref{statform1})\footnote{Here we define the splitting of a general
  generator into supertranslation and superrotation pieces
  by identifying the coordinates in
(\ref{param1}) with ingoing Eddington-Finkelstein coordinates.}.
However, as explained in Ref.\ \cite{Hawking:2016sgy},
one can turn on an infinite number of non trivial charges by acting
on the metric with symmetry transformations.  If we write the charges
as $\mc Q_\xi^{\rm loc}({\cal S},g_{ab})$, including the dependence on
the metric $g_{ab}$, then it follows from covariance and the fact that
the charges are independent of ${\cal S}$ in stationary regions that
\be
\mc Q_\xi^{\rm loc}({\cal S},g_{ab} + \lie_{\tilde \xi} g_{ab}) =
\mc Q_\xi^{\rm loc}({\cal S},g_{ab}) - \mc Q_{\lie_{\tilde \xi} \xi}^{\rm loc}({\cal S},g_{ab})
\ee
to linear order in ${\tilde \xi}^a$.  Hence one can compute the charges
on a transformed background in terms of the charges on the original background
by making use of the algebra (\ref{algexplicit}) of
symmetry generators.  It follows from this algebra that acting on the Schwarzschild metric with a
superrotation turns on an infinite number of Killing supertranslation
charges, and similarly acting with a Killing supertranslation turns on
an infinite number of superrotation charges.

We next turn to a consideration of {\it stationary to stationary
  transitions}, which helps to clarify the nature of the charges and
conservation laws just as at future null infinity.
Suppose that there are two different stationary regions of the horizon
separated by a region which is non-stationary\footnote{Actually it is
not possible to have the first region be exactly stationary, since by
Raychaudhuri's equation in vacuum the expansion $\theta$ must
monotonically decrease to zero in affine parameterization; it can only
be approximately stationary.}.  Then the stationary regions are associated
with two different Killing supertranslation algebras $\mf t_1$ and
$\mf t_2$. This is analogous to
the status of Poincar\'e subalgebras of the BMS algebra at null
infinity.
%
%
Just as there, one can find a finite supertranslation
$\sigma$ for which
\be
\mf t_2 = \sigma \mf t_1 \sigma^{-1},
\ee
so that the two subalgebras are related by a supertranslation.
Specifically, in the notation of Eq.\ (\ref{mftcondt}), if the two
subalgebras are given by $\alpha - \beta u_1=0$ and $\alpha - \beta
u_2 =0$, where $u_1$ and $u_2$ are functions just of $\theta^A$, then
one can take $\sigma$ to be the affine supertranslation $u \to u +
(u_2-u_1)$.  This supertranslation is presumably is related to an
analog of gravitational wave memory on the horizon \cite{Hawking:2016msc}.
The details of how such memory can be defined and measured is an
interesting topic for future study.
The transition is also associated with net changes in (electric parity)
superrotation charges, as at $\scri^+$.

Finally, our formalism does not furnish an analog of the Bondi mass on black
hole horizons, that is, a prescription for computing the mass of the
black hole at an arbitrary cross section ${\cal S}$ of the horizon.
This is so for two reasons.  First, it would be necessary to specify a
preferred symmetry generator (or preferred four-dimensional subgroup
of translations for a 4-momentum) from the algebra in order to obtain such
a definition.  While there is a preferred generator for each
stationary region (the Killing vector), in general horizons are
non-stationary, and there is no preferred generator or preferred
four-dimensional subgroup of translations.
Second, even when given a generator associated with a stationary region, the
corresponding charge is proportional to the area of the black hole (as
used in derivations of the first law), not the mass.
In this sense horizons are not similar to future null infinity.

\subsection{The limit to future null infinity}
\label{sec:limittoscri}

The symmetry algebra for a general null surface that we have derived
is larger than the BMS algebra which applies to the asymptotic
boundary of future null infinity $\scri^+$.  An interesting question
is how the symmetries and charges of the two algebras are related,
for a family of null surfaces that limit to $\scri^+$ in an
asymptotically flat spacetime.  One might expect that the localized
charges $\mc Q_\xi^{\rm loc}$ have
finite limits for a subalgebra of the symmetry algebra isomorphic to
the BMS algebra, and that the limits of those charges coincide with
the BMS charges.  In fact, this does not occur, and none of the
localized charges $\mc Q_\xi^{\rm loc}$ have finite limits.
This occurs because of our choice of reference solution, in effect a
different choice of reference solution is necessary in order for
finite limiting charges to be obtained at $\scri^+$.  Details of this
comparison will be discussed elsewhere \cite{CentralExt}.

\subsection{Generalizations}

While our results are specific to $d=4$ spacetime dimensions, they
generalize straightforwardly to all spacetime
dimensions $d \ge 4$, with appropriate changes in numerical
coefficients.  Our analysis does not depend on details of Greens
functions or on asymptotic fall off conditions which can be dimension
dependent.  This is in contrast to the situation at future null
infinity, where the generalization of the symmetry group, charges and memory
to higher dimensions is much more involved
\cite{Kapec:2015vwa,Hollands:2016oma,Pate:2017fgt,Hollands:2003ie}.
Thus supertranslation and
base-space diffeomorphism (superrotation) symmetries are universal symmetries of all null
surfaces in vacuum general relativity.

It would also be useful to generalize our analysis to allow for the
presence of matter.  We expect that the symmetry algebra and
expressions for charges will not be modified, but that the flux
expressions will acquire corrections involving the stress-energy tensor,
as in the BMS context.

Generalizations to other theories of gravity will be more involved.
In particular, the symmetry algebra obtained from the Wald-Zoupas
procedure can depend on the Lagrangian through the explicit expression
for the charge variation in Eq.\ (\ref{eq:boundary-symm-equiv}), and
may no longer coincide with the specific intrinsic symmetry algebra of
Sec.\ \ref{sec:universal} (although it may still posess an intrinsic
characterization).

Our symmetry algebra is analogous to the BMS symmetry algebra at future null infinity.
In that context it has been suggested that the BMS algebra can be usefully extended to include
additional symmetries, which do correspond to soft theorems and to new types of gravitational wave memory
\cite{Barnich:2009se,Strominger:2017zoo,Campiglia:2015yka,Compere:2018ylh}.
However these generators are not obtained from the Wald-Zoupas
construction and
their status as symmetries on phase space is still unclear.  Perhaps the algebra
computed here of symmetries on finite null surfaces could be similarly extended.

Finally, as discussed in Sec.\ \ref{sec:globalcons}, a key open
question in the black hole context is the restriction on the global algebra of symmetry
generators imposed by boundary conditions near future
timelike infinity, that should determine the identification of
symmetry generators on the horizon and at future null infinity.  This
identification is necessary in order to formulate the general form of
the global conservation law associated with the global charges on the horizon.

\section*{Acknowledgments}

We thank Robert M. Wald for helpful discussions, and an anonymous
referee for some useful comments.
V.C. is supported in part by the Berkeley Center for Theoretical
Physics, by the National Science Foundation (award numbers 1214644,
1316783, and 1521446), by FQXI grant RFP3-1323, and by the US
Department of Energy under Contract DE-AC02-05CH11231.  \'E.\'E.F. and K.P. are
supported in part by NSF grants PHY-1404105 and PHY-1707800 to Cornell
University.

\appendix

\section{Orthonormal basis formalism for null surfaces}
\label{app:translate}

In this appendix we translate the definitions and formalism described
in Sec.\ \ref{sec:null-geometry}, and some of the results of Sec.\
\ref{sec:wzcharges}, into the language of
components on an orthonormal basis.  This specialization is often
useful in computations, although it does depend on arbitrary choices.
We first describe the specializations that
occur when one chooses an auxiliary null vector, and then the
specializations associated with a complete orthonormal basis.

\subsection{Review of structures associated with a choice of auxiliary null vector}

We choose an auxiliary null vector field $n^a$ on ${\cal N}$ which satisfies
\begin{subequations}
\label{eq:n-defn}
\begin{eqnarray}
\label{nnull}
n_a n^a &=& 0, \\
n_a \ell^a &=& -1.
\label{northog}
\end{eqnarray}
\end{subequations}
The pullback of the covector field $n_a$ yields a covector on ${\cal N}$
\be
n_i = \pullback_i^a n_a
\label{pb1}
\ee
which from Eqs.\ (\ref{downshift}) and (\ref{northog}) satisfies $n_i
\ell^i = -1$ \footnote{Given a covector $n_i$ on ${\cal N}$ with
  $n_i \ell^i=-1$, $n_a$
  is uniquely determined by the conditions (\ref{nnull}) and (\ref{pb1}).}.  We define the projection tensor
\begin{align}
\pi^{a}{}{}_b = \delta^a {}{}_b + n^a \ell_b. \label{Projector}
\end{align}
At a given point $p$ the mapping $v^a \goesto \pi^{a}_{\ \,b} v^b$ maps vectors into the space
of vectors orthogonal to $\ell_a$, {\it i.e.}, into
the tangent space $T_p({\cal N})$.
We write this mapping from $T_p(M)$ to $T_p({\cal N})$ as
\be
v^a \goesto \pushback^i_a v^a.
\ee
The quantities $\pushback^i_a$ and $\pullback_i^a$ satisfy
\be
\delta^i_j = \pushback^i_a \, \pullback^a_j, \ \ \ \ \ \ \pi^a_{\ \,b} = \pullback^a_i \,
\pushback^i_b.
\ee

We can now define spacetime tensors that correspond to the induced
metric
\be
q_{ab} = \pushback^i_a \pushback^j_b q_{ij} = g_{ab} + 2 \ell_{(a} n_{b)}, \label{Induced Metric}
\ee
and shear tensor
\be
\sigma_{ab} = \pushback^i_a \pushback^j_b \sigma_{ij}.
\ee
These quantities depend on the choice of auxiliary null vector $n_a$.
We can also define a derivative operator $D_i$ on ${\cal N}$ by,
for a given vector field $v^i$ on ${\cal N}$,
\be
D_i v^j = \pullback_i^a \pushback^j_b \nabla_a v^b.
\ee
Here, on the right hand side, $v^a$ is any choice of vector field on
$M$ for which $v^a = \pullback^a_i v^i$ when evaluated on ${\cal N}$.
It can be checked that this prescription yields a well defined
derivative operator, which depends on the choice of $n_a$.

We define the rotation one-form $\omega_i$ by
\be
\omega_i = - n_j {\cal K}_i^{\ j}.
\label{Definition twist 1-form}
\ee
From Eq.\ (\ref{sffp2}) this satisfies
\be
\omega_i \ell^i = \kappa.
\label{twistprop}
\ee
As noted in Sec.\ \ref{sec:stationary} above the rotation one-form depends on $n_i$
except when $K_{ij}=0$.

\subsection{Geometric fields on an orthonormal basis}
\label{onb}

We choose on ${\cal N}$ a set of basis vectors
\be
{\vec e}_{\hat \alpha} = \left( {\vec e}_{\hat 0}, {\vec e}_{\hat 1}, {\vec e}_{\hat A} \right) = \left( {\vec \ell}, {\vec n}, {\vec e}_{\hat A} \right),
\ee
where ${\hat A} = 2, 3$, and where ${\vec \ell}^2 = {\vec n}^2 = {\vec \ell} \cdot {\vec e}_{\hat A} = {\vec n} \cdot {\vec e}_{\hat A} =0$, ${\vec \ell} \cdot {\vec n}=-1$, ${\vec e}_{\hat A} \cdot {\vec e}_{\hat B} = \delta_{{\hat A}{\hat B}}$ on ${\cal N}$.  We extend the definition of these vectors off ${\cal N}$ but do not require them to be orthonormal off ${\cal N}$.

We can decompose the covariant derivative of the normal on this basis as
\be
\nabla_a \ell_b = \gamma \ell_a \ell_b + \eta \ell_a n_b + \tau_{\hat A} \ell_a e^{\hat A}_b
+ \epsilon n_a \ell_b + \zeta n_a n_b + \kappa_{\hat A} n_a e^{\hat A}_b + \alpha_{\hat A} e^{\hat A}_a \ell_b + \iota_{\hat A} e^{\hat A}_a n_b + \left( \frac{1}{2} \theta \delta_{{\hat A}{\hat B}} + \sigma_{{\hat A}{\hat B}} \right) e^{\hat A}_a e^{\hat B}_b.
\label{ffff}
\ee
where $\sigma_{{\hat A}{\hat B}}$ is traceless.  Imposing the orthonormality of the basis on the hypersurface gives
$\zeta = \iota_{\hat A} =0$, while imposing (\ref{kappadef}) gives $\epsilon = - \kappa$, $\kappa_{\hat A}=0$.
The induced metric, second fundamental form,
Weingarten map and rotation one-form in terms of these quantities are
\begin{subequations}
\label{r1ff}
\begin{eqnarray}
q_{ij} &=& \delta_{{\hat A}{\hat B}} e^{\hat A}_i e^{\hat B}_j, \\
K_{ij} &=& \left( \frac{1}{2} \theta \delta_{{\hat A}{\hat B}} + \sigma_{{\hat A}{\hat B}} \right) e^{\hat A}_i e^{{\hat B}}_j, \\
{\cal K}_i^{\ j} &=& - \kappa n_i \ell^j + \alpha_{\hat A} e^{\hat A}_i \ell^j +
\left( \frac{1}{2} \theta \delta_{{\hat A}{\hat B}} + \sigma_{{\hat A}{\hat B}} \right) e^{\hat A}_i e^{{\hat B}\,j}, \label{wme}\\
\omega_i &=& -\kappa n_i + \alpha_{\hat A} e^{\hat A}_i.\label{rofe}
\label{r1f}
\end{eqnarray}
\end{subequations}

\subsection{Expressions for charges}

A simple expression for the Noether charge in terms of the orthonormal
basis can be found by combining Eqs.\ (\ref{noether2}),
(\ref{weingarten}), (\ref{betadef}), (\ref{downshift}) and (\ref{z1proof}):
\be
Q_\xi({\cal S}) = \frac{1}{8 \pi} \int_{\cal S} \varepsilon_{ij}  \left[ n_b
\ell^a \nabla_a \xi^b \right].
\label{noether2onb}
\ee
Here the null vector $n_a$ has been chosen so that its pullback $n_i$
to ${\cal N}$ is normal to the cross section ${\cal S}$.
A similar calculation starting from the localized charge
(\ref{GenHam})
gives
\be
\mc Q_\xi^{\rm loc}({\cal S}) = \frac{1}{8 \pi} \int_{\cal S} \varepsilon_{ij}  \left[ n_b
\ell^a \nabla_a \xi^b - \theta \xi^a n_a\right].
\label{GenHamonb}
\ee
For $\diff({\cal Z})$ generators we have $\xi^a n_a\hateq 0$, and this charge
can be rewritten as
\be
\mc Q_X^{\rm loc}({\cal S})
= -\frac{1}{8 \pi} \int_{\cal S} \varepsilon_{ij}
\left[ \ell^a \xi^b \nabla_a n_b  \right].
\label{GenHamonb1}
\ee

\section{Gauge fixing in the definition of field configuration space}
\label{app:proofs}

In this appendix we show that the field configuration space $\ms
F_{\mf p}$ that we defined is obtained from the larger space $\ms F_0$
by a gauge
fixing.  Specifically, given a manifold $M$ with boundary ${\cal N}$, a
complete boundary structure $\mf p$ on ${\cal N}$, and a metric $g_{ab}$ on $M$ for which ${\cal N}$ is
null and for which the boundary structure induced by $g_{ab}$ is
complete, we show that
one can find a diffeomorphism $\psi : M \to M$ which takes ${\cal N}$ into ${\cal N}$
for which $\psi_* g_{ab}$ lies in $\ms F_{\mf p}$.

Let $\mf u$ be the intrinsic structure induced by $\mf p$, and $\mf
u'$ be the intrinsic structure induced by the metric $g_{ab}$.  By
hypothesis, both $\mf u$ and $\mf u'$ are complete.  Hence by the
argument given in Sec.\ \ref{sec:uis} there exists a diffeomorphism $\varphi:
{\cal N} \to {\cal N}$ which takes $\mf u$ to $\mf u'$.
Now choose a diffeomorphism $\psi : M \to M$ whose restriction to
${\cal N}$ is $\varphi$.  By acting with $\psi$ on the metric we can
without loss of generality assume that $\mf u = \mf u'$.

Now let $\mf p'$ be the boundary structure induced by $g_{ab}$, and
choose representatives $(\ell^a, \kappa, {\hat \ell}_a)$  and
$(\ell^{\prime\,a}, \kappa', {\hat
  \ell}'_a)$ of $\mf p$ and $\mf p'$.  Since $\mf u = \mf u'$ we can,
by adjusting the choice of representative if necessary, take $\ell^a =
\ell^{\prime\,a}$ and $\kappa = \kappa'$.  The two normal covectors
must be related by some rescaling of the form ${\hat \ell}_a =
e^\lambda {\hat \ell}'_a$ for some smooth function $\lambda$ on ${\cal
  N}$.  We thus have
\be
g^{ab} {\hat \ell}_a \hateq e^\lambda \ell^b,
\label{qq0}
\ee
and we want to show that there exists a diffeomorphism $\psi$ that
preserves $\mf u'$ so that
\be
(\psi_* g^{ab}) {\hat \ell}_a \hateq \ell^b.
\label{qq}
\ee
By applying $\psi^{-1}_*$ to both sides of Eq.\ (\ref{qq}),
specializing the diffeomorphism so that the induced diffeomorphism
$\varphi$ on ${\cal N}$ is the identity, and using (\ref{qq0}),
we find that a sufficient condition for (\ref{qq}) is that
\be
\psi^{-1}_* {\hat \ell}_a \hateq e^{-\lambda} {\hat \ell}_a.
\label{pb}
\ee

To find a diffeomorphism $\psi$ satisfying (\ref{pb}), we need only
specify its action to linear order in
deviation off the surface ${\cal N}$.  We can parameterize points near
${\cal N}$ to linear order by specifying a point ${\cal P}$ on ${\cal
  N}$ and a vector $v^a$ at ${\cal P}$.  We define $\psi$ to be the
mapping that takes
\be
\psi:  ({\cal P}, v^a) \to ({\cal P}, v^a + \zeta^a({\cal P}) {\hat
  \ell}_b v^b),
\label{psidef}
\ee
where $\zeta^a$ is some vector field defined on ${\cal N}$.
This mapping is well defined despite the fact that representing points
near ${\cal N}$ as pairs $({\cal P},v^a)$ is not unique, since
components of $v^a$ along the surface are annihilated by the term
proportional to $\zeta^a$.  Now computing the pullback of the mapping (\ref{psidef})
we find that the condition (\ref{pb}) will be satisfied if we choose the vector
field $\zeta^a$ to satisfy
\be
1 + \zeta^a {\hat \ell}_a = e^\lambda.
\ee

\section{Characterization of trivial diffeomorphisms at a null boundary}
\label{app:trivial}

In this appendix we show that the charge variation
(\ref{Hamiltonian Definition}) vanishes for all cross sections ${\cal
  S}$ of a null boundary, and for all solutions and variations of solutions,
if and only if the symmetry $\xi^a$ satisfies $\chi^i = 0$ and
$\gamma(\xi^a) =0$, where $\chi^i$ is defined by Eq.\ (\ref{restriction}) and
$\gamma$ by Eq.\ (\ref{gammadef1}).

The charge variation is given by Eq.\ (\ref{H2a}), but with $\beta$ replaced by
$(\beta + \gamma)/2$ from Eq.\ (\ref{qcc}):
\begin{eqnarray}
 \delta {\cal Q}_{\xi} &=& \frac{1}{16 \pi} \int_{\cal S} \varepsilon_{ijk} \bigg[ h \chi^l {\cal K}_l^{\
    k} - h \beta(\chi^i) \ell^k/2 - h \gamma(\xi^a) \ell^k/2 -  \chi^l \Gamma_l \ell^k +
\chi^l \lie_\ell h_l^{\ \,k}
+ 2 \chi^l h^{m}_{\ \ l} {\cal K}_m^{\ \,k} \nonumber \\
&&
- 2 \chi^l h^{k}_{\ \,m} {\cal K}_{l}^{\ \,m}
-\chi^k \lie_\ell h - \chi^k h_i^{\ \,j} {\cal K}_j^{\ \,i}
\bigg].
\label{H2aa}
\end{eqnarray}
This expression vanishes if $\chi^i$ and $\gamma$ vanish, from Eq.\
(\ref{betadef}).  Conversely, we want to show that the vanishing of
the expression (\ref{H2aa}) for all solutions and variations of
solutions forces $\chi^i = \gamma=0$.

Fix a cross section ${\cal S}$.  We make use of the explicit form
of the general solution to the vacuum Einstein equations on a null
surface given by Hayward \cite{1993CQGra..10..773H}.  It follows from
this solution that, on shell, we can freely specify $h_i^{\ j}$ on
${\cal S}$ subject to the constraint (\ref{iddd}), $\lie_\ell h_i^{\
  j}$ subject to the constraint
\be
\ell^i \lie_\ell h_i^{\ j} =0,
\label{constraint14}
\ee
and the quantity $\Gamma_i$ defined by Eq.\ (\ref{Gammadef}) subject
to the constraint (\ref{Gammaconstraint}).  We now choose $h_i^{\
  j}=0$ and $\lie_\ell h_i^{\ j}=0$.  In this case the charge
variation (\ref{H2aa}) reduces to
\begin{eqnarray}
 \delta {\cal Q}_{\xi} &=& \frac{1}{16 \pi} \int_{\cal S} \varepsilon_{ij}
\, \chi^l \Gamma_l.
\label{H2aa1}
\end{eqnarray}
Since $\Gamma_l$ can be chosen arbitrarily on ${\cal S}$ subject to
Eq.\ (\ref{Gammaconstraint}), this forces $\chi^i = f \ell^i$ on
${\cal S}$ for some function $f$.  Returning now to Eq.\ (\ref{H2aa}),
choosing $h_i^{\ j}=0$, and making use of the constraint (\ref{constraint14})
gives the charge variation
\begin{eqnarray}
 \delta {\cal Q}_{\xi} &=& \frac{1}{16 \pi} \int_{\cal S} \varepsilon_{ij}
\, f \lie_\ell h.
\label{H2aa2}
\end{eqnarray}
Since $\lie_\ell h$ can be chosen arbitrarily on ${\cal S}$, this
forces $f$ to vanish on ${\cal S}$.  Since ${\cal S}$ was
chosen arbitrarily, $f$ (and therefore $\chi^i$) must vanish on all of ${\cal N}$, and so
$\beta=0$ from Eq.\ (\ref{betaff}).
Now reverting to a general $h_i^{\ j}$ and $\lie_\ell h_i^{\ j}$ in
Eq.\ (\ref{H2aa}), we obtain the charge variation
\begin{eqnarray}
 \delta {\cal Q}_{\xi} &=& \frac{1}{32 \pi} \int_{\cal S} \varepsilon_{ij}
\, h \gamma.
\label{H2aa2}
\end{eqnarray}
Since $h$ can be chosen arbitrarily on ${\cal S}$, this forces
$\gamma=0$ on ${\cal S}$.  Finally, since the choice of ${\cal S}$ was
arbitrary, if follows that $\chi^i$ and $\gamma$ vanish on all of
${\cal N}$.

\section{Consistency check of symmetry algebra}
\label{sec:consistency}

In this appendix we verify that for vector fields $\xi^a$ satisfying the conditions
(\ref{symmetrydeflinear}) and (\ref{condition1}) of the symmetry algebra, the corresponding metric perturbation
(\ref{metricsym}) satisfies the boundary conditions (\ref{zcondts}) derived in Sec.\ \ref{sec:bcs}.

Taking the Lie derivative of Eq.\ (\ref{calFdef1}) with respect to $\xi^a$ gives
\be
\lie_\xi {\hat \ell}_a \hateq \lie_\xi g_{ab} \ell^b + g_{ab} \lie_\xi \ell^b.
\ee
Making use of Eqs.\ (\ref{restriction}), (\ref{metricsym}), (\ref{betadef}), (\ref{gammadef1}) and (\ref{condition1}) gives
\be
h_{ab} \ell^b \hateq (\gamma - \beta) {\hat \ell}_a \hateq 0,
\label{z1proof}
\ee
which establishes the condition (\ref{z1}).

Next for simplicity and without loss of generality we specialize to a
representative of the boundary structure with $\kappa=0$.
We write the definition (\ref{gammadef1}) in the form, using (\ref{tangent}) and (\ref{calFdef1}),
\be
\xi^b \nabla_b \ell^a + \ell^b \nabla^a \xi_b \hateq \gamma \ell^a,
\ee
and take the Lie derivative with respect to $\ell^a$.  The right hand
side becomes $(\lie_\ell \gamma) \ell^a$, which vanishes by Eqs.\
(\ref{condition1}) and (\ref{sdl1}).  Writing $v^a$ for the expression
on the left hand side, the left hand side becomes $\ell^c \nabla_c v^a
- v^c \nabla_c \ell^a$, and the second term can be written as $\gamma
\ell^c \nabla_c \ell^a \hateq \gamma \kappa \ell^a \hateq 0$.
We thus obtain
\be
0 \hateq \ell^c \nabla_c \xi^b \nabla_b \ell^a + \ell^c \xi^b \nabla_c
\nabla_b \ell^a + \ell^c \ell^b \nabla_c \nabla^a \xi_b.
\ee
The first term can be written using the definition (\ref{betadef}) as
$-\beta \ell^b \nabla_b \ell^a + \xi^c \nabla_c \ell^b \nabla_b \ell^a
\hateq - \xi^c \ell^b \nabla_c \nabla_b \ell^a$, where we have used
(\ref{kappadef}) and $\kappa=0$.  It follows that
\be
0 \hateq - \ell^c \xi^b R_{cbda} \ell^d + \ell^c \ell^b \nabla_c \nabla_a
\xi_b \hateq \ell^c \ell^b \nabla_a \nabla_c \xi_b,
\ee
from which the condition (\ref{z2}) follows.

\section{Choice of reference solution}
\label{app:background}

As explained in Sec.\ \ref{sec:cov-phase}, the dynamics of a theory fix
the symmetry generator charges on phase space only up to an overall ``constant of integration''.
To fix that constant of integration, following Wald and Zoupas \cite{WZ},
we choose a reference solution and demand that the charges vanish on that
solution.  There are two different cases, complete intrinsic
structures, and incomplete intrinsic structures associated with
nontrivial boundaries $\partial {\cal N}$ of the null hypersurface
${\cal N}$, as discussed in Sec.\ \ref{sec:uiss1}

In the first case of complete intrinsic structures, we choose a
one-parameter family of reference solutions $g_{ab}(\varepsilon)$
and demand that the limit $\varepsilon \to 0$ of the charges evaluated on the reference solution
vanish.  (We use a one parameter family rather than a single solution
since our chosen family
of solutions does not have a continuous limit as $\varepsilon \to 0$.)
The reference solution is maximally extended Schwarzschild written in Kruskal coordinates
\be
ds^2 = - 2 e^{2 \mu(s)} dU dV + m^2 \rho(s)^2 d \Omega^2,
\label{Sch}
\ee
where $s = U V/m^2$, $m$ is the mass, and $\mu(s)$ and $\rho(s)$ are functions whose exact forms are unimportant
for what follows.  We also need to specify how this manifold is to be identified with our
given boundary structure $(M, {\cal N}, \mf p)$.  We identify ${\cal N}$ with the horizon $U=0$,
and pick $\mf p$ to be determined by the representative $(\ell^a, {\hat \ell}_a, \kappa)$ where
\begin{subequations}
\label{normalchoice}
\begin{eqnarray}
{\hat \ell}_a &=& (d U)_a, \\
\ell^a &=& - e^{-2 \mu(0)} \left( \frac{\partial}{\partial V} \right)^a, \\
\kappa &=& 0.
\end{eqnarray}
\end{subequations}
We identify the parameter $\varepsilon$ with the mass $m$ and will take the $m \to 0$ limit.

We now show that the charge (\ref{GenHam}) integrated over a fixed
cross section ${\cal S}$ vanishes for the reference solution, in the limit $m
\to 0$, as claimed in Sec.\ \ref{sec:wz}.  The expansion $\theta$ and
Weingarten map ${\cal K}_i^{\ j}$ vanish for this solution with the
choice (\ref{normalchoice}) of normal.   The charge therefore reduces
to
\be
\mc Q_\xi^{\rm loc}({\cal S}) = - \frac{1}{8 \pi} \int_{\cal S} \varepsilon_{ijk}
\beta(\chi^i) \ell^i.
\ee
The only quantity that depends on the metric in this expression is the
volume form $\varepsilon_{ijk}$, which
from Eq.\ (\ref{Sch}) is of the form $\varepsilon_{ijk} = m^2
\varepsilon^0_{ijk}$ where $\varepsilon^0_{ijk}$ is independent of
$m$.  Hence $\mc Q_\xi^{\rm loc}({\cal S})  \to 0$ as $m \to 0$ as required.

Note that this conclusion is unchanged if we replace the reference
solution $g_{ab}(m)$ with $\psi_* g_{ab}(m)$ for any diffeomorphism
$\psi: M \to M$ which preserves the boundary structure $\mf p$.  The
only effect of this change on the argument is to replace
$\varepsilon^0_{ijk}$ with $\varphi_*  \varepsilon^0_{ijk}$, where
$\varphi$ is the restriction of $\psi$ to ${\cal N}$, which does not
affect the conclusion.  Thus the consistency condition\footnote{The
  charges need not
  vanish in the $m \to 0$ limit for the transformed reference solution $\psi(m)_* g_{ab}(m)$
  which allows the diffeomorphism $\psi$ to depend on $m$.  However, there is
  no physical argument for imposing this more stringent requirement.}
discussed by Wald
and Zoupas \cite{WZ} is satisfied.

Turn now to the second case of a nontrivial boundary $\partial {\cal N}$.
If the boundary $\partial {\cal N}$ is a twosphere,
we take the reference solution to be the Schwarzschild solution
(\ref{Sch}), with the hypersurface ${\cal N}$ now being restricted to
$U > 0$, so that the boundary $\partial {\cal N}$ is identified with the bifurcation twosphere of
Schwarzschild.  Apart from this modifications the analysis and conclusions
are unchanged.

The case where the boundary $\partial {\cal N}$ is a single point $\{ {\cal P} \}$
is slightly more complicated.  We choose the reference solution to be the Schwarzschild solution (\ref{Sch})
for $U > U_0(m)$, and a spherically symmetric ingoing Vaidya solution at
earlier advanced times, so that the origin of the event horizon is mapped onto ${\cal P}$.
The reference boundary structure is chosen to satisfy Eqs.\ (\ref{normalchoice}) in the Schwarzschild region,
which determines its definition everywhere.
If we choose the function $U_0(m)$ to go to zero as $m \to 0$, then the charge
(\ref{GenHam}) integrated over a fixed cross section ${\cal S}$ is evaluated entirely in the Schwarzschild region
for sufficiently small $m$, and the rest of the argument follows as before.
Roughly speaking, we are taking the limit of small black holes formed in the distant past
to define the reference solution in this case.

Of course, we could dispense with the reference solutions and simply say that
we are picking the constant of integration to enforce the expression
(\ref{GenHam}) starting from its variation.  The reference solutions clarify
the physical interpretation of that assumption.

\section{Consistency of two expressions for flux of localized charge}
\label{app:consistent1}

In this appendix we show explicitly that the two expressions
(\ref{flux1}) and (\ref{flux2}) for the
flux of the localized charge coincide, as they must from the
general Wald-Zoupas framework reviewed in Sec.\ \ref{sec:cov-phase}.

The expression (\ref{flux1}) was derived from Eq.\ (\ref{iid}).  The
variation in the second term in (\ref{iid}) can be replaced with a
Lie derivative with respect to $\xi$, from Eq.\ (\ref{metricsym}),
giving from the expression (\ref{eq:alpha-gr}) a contribution to
$\Theta_{ijk}$ of
\be
-\frac{1}{8 \pi} \lie_\chi (\theta \varepsilon_{ijk}).
\ee
Using Cartan's formula $\lie_v \df \omega = i_v d {\df \omega} + d (
i_v {\df \omega})$ and the definition (\ref{divergenceop}) of the divergence
operator shows that this contribution matches the second
term in Eq.\ (\ref{flux2}).  Hence, using the expression (\ref{theta11}) for
$\theta_{ijk}$, it remains to show that
\be
\ell^f (\nabla_f h - \nabla_e h_f^{\ e}) = 2 {\hat D}_i ( \chi^j {\cal
  K}_j^{\ \,i} - \beta \ell^i).
\label{step1}
\ee

Inserting the expression (\ref{metricsym}) for the metric perturbation
$h_{ab}$ into the left hand side of Eq.\ (\ref{step1}), commuting
derivatives and making use of the vacuum equation of motion $R_{ab}=0$
gives the expression
\be
\ell^f \nabla_f (\nabla_a \xi^a) - \ell^f \nabla_e \nabla^e \xi_f.
\label{eee}
\ee
It follows from Eqs.\ (\ref{zcondts}) and (\ref{dividentity})
that
\be
\nabla_a \left[ ( \nabla^a \xi^b + \nabla^b \xi^a) \ell_b \right]=0,
\ee
and simplifying by once again commuting derivatives acting on $\xi^a$
and inserting into (\ref{eee}) gives that the left hand side of Eq.\
(\ref{step1}) is
\be
2 \lie_\ell (\nabla_a \xi^a) + (\nabla^a \xi^b + \nabla^b \xi^a )
\nabla_a \ell_b.
\label{step2}
\ee
Note that this expression is independent of the definition of $\ell^a$
off ${\cal N}$, by Eq.\ (\ref{z1}).

We now turn to evaluating the right hand side of Eq.\ (\ref{step1}).
We define the vector
\be
v^a = \xi^a \nabla_a \ell^b - \beta \ell^b,
\label{vadef}
\ee
which satisfies ${\hat \ell}_a v^a =0$, in terms of which the right
hand side can be written as $2 {\hat D}_i v^i$.
We now make use of the relation (\ref{dividentity}) between the three dimensional
and four dimensional divergence operators, and the definition
(\ref{varpidef}), which yields for the right hand side
\be
2 \nabla_a v^a + 2n^b ( v^a \nabla_a {\hat \ell}_b + {\hat \ell}_a
\nabla_b v^a).
\label{D22}
\ee
Here $n^b$ is any null vector field which satisfies $n_a \ell^a =-1$.
Now using the definition (\ref{betadef}) of $\beta$ in Eq.\ (\ref{vadef})
we obtain  $v^a = \ell^c \nabla_c \xi^a$, and
substituting into (\ref{D22}) gives
\be
2 \nabla_a \ell^c \nabla_c \xi^a
+ 2 \ell^c \nabla_a \nabla_c \xi^a
+ 2 n^b \ell^c \nabla_a {\hat \ell}_b \nabla_c \xi^a
+ 2 n^b {\hat \ell}_a \nabla_b \ell^c \nabla_c \xi^a
+ 2 n^b {\hat \ell}_a \ell^c \nabla_b \nabla_c \xi^a.
\label{step3}
\ee

It remains to show that the expressions (\ref{step2}) and
(\ref{step3}) coincide.
Commuting the derivatives in the second term in (\ref{step3}) and using the
vacuum equation of motion $R_{ab}=0$ shows that this term matches the
first term in Eq.\ (\ref{step2}).  The last term in (\ref{step3})
vanishes by Eqs.\ (\ref{zcondts}).  In the third term, the derivative
acting on ${\hat \ell}_a$ is entirely along the surface, since $\ell^c
\ell_a \nabla_c \xi^a=0$ by Eq.\ (\ref{z1}).  Hence we can replace
${\hat \ell}_a$ with $\ell_a$ in this term, and also in the fourth
term.  Next, we have that $\ell^a$ is hypersurface orthogonal on
${\cal N}$, so $\ell_{[a} \nabla_b \ell_{c]}\hateq 0$.
It follows that $\nabla_a \ell_b \hateq \nabla_{(a} \ell_{b)} + w_{[a}
\ell_{b]}$
for some $w_a$ with $w_a n^a=0$.  Substituting this into
the first, third and fourth terms in Eq.\ (\ref{step3}) we find that
the dependence on $w_a$ cancels out, so that $\nabla_a \ell_b$ can be
replaced in these terms with $\nabla_{(a} \ell_{b)}$.
The first term in (\ref{step3}) then matches the second term in (\ref{step2}).
The third and fourth terms can be written as $4 p_a \ell_b \nabla^{(a}
\xi^{b)}$ where $p_a = n^b \nabla_{(a} \ell_{b)}$, which vanishes by
Eq.\ (\ref{z1}).  Thus the expressions (\ref{step2}) and (\ref{step3})
coincide as desired.

\section{Symplectic currents on black holes horizons}
\label{sec:bh}

Our explicit expressions for the symplectic current and charges for
general null surfaces
allow us to establish a number of results about
black hole horizons.

First, in vacuum general relativity, the obstruction (\ref{Hamiltonian
  Condition}) to defining the contribution to a
global symmetry generator charge $\mc Q_\xi$ from an integral over a
future horizon ${\cal H}^+$ vanishes,
\be
\int_{{\cal H}^+_\pm} i_\xi \df \omega =0,
\label{noobst}
\ee
as discussed in Sec.\ \ref{sec:globalcharges}
above, assuming certain fall off conditions on the shear along the
horizon at the future boundary ${\cal H}^+_+$, which we now discuss.
Consider a cross section ${\cal S}$ of the horizon that
approaches ${\cal H}^+_+$.  The integrand in Eq.\
(\ref{noobst}) is given explicitly for a null surface
in Eq.\ (\ref{omegapb}), and scales as a product of a symmetry
generator $\chi^i$, times the expansion or shear of the background, times two
factors of metric perturbation $h_{ij}$.  Denoting an affine
parameter along the horizon by $v$, the symmetry generator scales
$\sim v$ as $v \to \infty$, by Eq.\ (\ref{groupt1}).
If the shear of the background and perturbations scales as
\be
\sigma_{ij} \sim v^{-p}
\label{shearcondt}
\ee
for some $p>1$ as $v \to \infty$, then it follows from Eq.\
(9.2.32) of Wald \cite{Wald-book} that the expansion $\theta$ is
negligible. Also
from Eq.\ (\ref{identity22}) it follows $h_{ij} \sim v^{-(p-1)} +
({\rm const})$, and hence the condition (\ref{noobst}) will be satisfied
at ${\cal H}^+_+$.

Is the condition (\ref{shearcondt}) on the late time decay of the shear physically
realistic?  Consider first linear gravitational perturbations of a Kerr black hole with initial data of compact spatial
support.  For this case Barack showed that the Weyl scalars $\Psi_0$ and $\Psi_4$ decay
along the horizon at late
times like $v^{-7}$ or smaller \cite{Barack:1999st}.
It then follows from Eqs.\
(9.2.32) and (9.2.33) of Wald \cite{Wald-book} that $\sigma_{ij} \sim
v^{-6}$. For more general solutions with incoming radiation at $\scri^-$,
we conjecture that imposing that the News tensor fall off along $\scri^-$
as $\sim v^{-p}$ with $p>1$ in the limit $v \to \infty$ towards $\scri^-_+$
will be sufficient to ensure the fall off condition (\ref{shearcondt}) along the
event horizon, both linearly and nonlinearly.  This conjecture is based on the intuition that
backscattering should serve to decrease rather than increase the
incoming flux at late advanced times $v$.

For eternal black holes with a
bifurcation two-sphere ${\cal H}^+_-$,
the condition (\ref{noobst}) will be satisfied at ${\cal H}^+_-$ from
Eqs.\ (\ref{condt3}) and (\ref{omegapb}).

Second, we show that the contribution to any global symmetry generator
charge $\mc Q_\xi$ from the integral over a future event horizon
${\cal H}^+$ can be expressed in terms of corresponding localized
charges $\mc Q_\xi^{\rm loc}$ evaluated on the components ${\cal
  H}^+_\pm$ of $\partial
{\cal H}^+$, as discussed in Sec.\ \ref{sec:WZ-charge} above.
This requires the vanishing of the correction term $i_\xi \df
\Theta$ in the definition (\ref{localcharge1}) of the localized charge:
\be
\int_{{\cal H}^+_\pm} i_\xi \df \Theta =0.
\label{vvv}
\ee
Using the explicit expression (\ref{Theta11}), an argument
analogous to that given in the last paragraph shows that the
quantity (\ref{vvv}) vanishes at ${\cal H}^+_+$, under the same assumptions on the shear as above.
For eternal black holes with a
bifurcation two-sphere ${\cal H}^+_-$, the corresponding integral (\ref{vvv})
vanishes by the condition (\ref{condt3}).

\section{Alternative definition of field configuration space and associated symmetry algebra}
\label{app:alternative}

In the body of this paper we have presented a specific definition of a field configuration space $\ms F$ for general
relativity in the presence of a null boundary, and derived from that definition a symmetry algebra and various types of charges.
A natural question is whether there is any freedom in the choice of
definition of $\ms F$.  In this appendix, we explore a modification of
the definition of $\ms F$, in which we allow a larger set of metrics.
A key motivation for this exploration is the fact is that
the new metric variations which are now allowed do not
correspond to degeneracies of the symplectic form, and so can be
regarded as physical degrees of freedom.
We will show that our analysis of the symmetry algebra
can be straightforwardly generalized, but that it is not possible to
implement the Wald-Zoupas prescription described in Sec.\ \ref{sec:WZ-charge} to compute localized
charges in this context.  One can obtain expressions for localized
charges but they are not unique.

The starting point for the modified field configuration
space definition is to omit the non-affinity $\kappa$ in the
definition (\ref{eq:gs}) of intrinsic structure $\mf u$. Thus, $\mf u$
consists of an equivalence class of normals $\ell^i$ that are related
by rescalings of the form (\ref{rescale}).  The symmetry group is
modified by replacing the transformation (\ref{groupta}) with an
arbitrary smooth mapping ${\bar u} = {\bar u}(u,\theta^A)$, and the algebra
(\ref{symmetrydeflinear}) is modified by dropping the requirement (\ref{sdl1}).
The definition of the boundary structure $\mf p$ in Sec.\ \ref{sec:cpsnb} is
correspondingly modified by omitting the non-affinity $\kappa$ from the definition
(\ref{frakpdef}), and omitting the requirement (\ref{kappaequiv}) from the definition of
the equivalence relation.  The definition of the field configuration
space $\ms F_{\mf p}$ is
modified by omitting the requirement (\ref{calFdef2}).
The conclusions (\ref{condition1}), (\ref{Hpdef2}) and (\ref{res1})
then continue to hold.  In particular, a key point is that the arguments of Appendix \ref{app:trivial} continue to apply,
and so none of the new symmetry generators $\chi^i$ on the null surface correspond to degeneracies
of the symplectic form.

In the following subsection \ref{sec:bcs}, the conditions (\ref{z2}) and (\ref{Gammaconstraint}) on
variations of the metric are no longer valid.  Also the non-affinity
$\kappa$ is no longer preserved under variation of the metric, its
variation is given by $\delta \kappa = - \Gamma_i \ell^i/2$, from Eqs.\
(\ref{Gammadef}), (\ref{kkk00}) and (\ref{sffp2}).

The computation of charges in Sec.\ \ref{sec:wzcharges} is modified as
follows.
The expression (\ref{noether2}) for the Noether charge is still valid,
as is its variation (\ref{vnoether1}).
In Sec.\ \ref{sec:ham}, the expression (\ref{theta11}) for the presymplectic potential
$\theta_{ijk}$ is valid, but the subsequent expression (\ref{theta22})
acquires the extra term $- \varepsilon_{ijk} \Gamma_a
\ell^a / (16 \pi)$, and there is the corresponding correction
$ \varepsilon_{ijk} \chi^k \Gamma_a \ell^a / (16 \pi)$ to Eq.\ (\ref{H2a}).
In the computation of localized charges, we are unable to find a
presymplectic potential $\df \Theta$ satisfying all the requirements
listed in Sec.\ \ref{sec:WZ-charge}.  Specifically, if we use the
choice (\ref{eq:alpha-gr}) of the 3-form $\df \alpha$, then
the extra term in
$\df \theta$ implies that $\df \Theta$ no longer vanishes on
stationary backgrounds.  One could cancel this extra term by adding a
term proportional to $\kappa \varepsilon_{ijk}$ to $\alpha_{ijk}$, but
this term is not invariant under the rescaling (\ref{rescale}) as it
must be.  The expression $\kappa - \lie_\ell \ln \theta$ is invariant
under rescaling, but from Raychaudhuri's equation in vacuum it is
equivalent to $\theta/2 + \sigma_{AB} \sigma^{AB}/\theta$ which is not
well defined in the limit $\theta \to 0$.
It does not appear to be
possible to find a presymplectic potential $\df \Theta$ satisfying all
the requirements.

Of course, one can drop the requirements related to stationarity, and
choose the same expression
(\ref{eq:alpha-gr}) for the 3-form $\df \alpha$ as before.  Then the argument
of Appendix \ref{app:background} shows that, assuming the localized
charges $\df {\mc Q}_\xi^{\rm loc}$ vanish on the reference solution, the expressions (\ref{anss}) and
(\ref{GenHam}) for the localized charge are still valid.  However,
since we are no longer imposing any assumptions related to
stationarity, the relation (\ref{eq:WZ-flux}) between the flux
$d \df{\mc Q}_{\xi}^{\rm loc}$ and presymplectic potential
$\df{\Theta}$ need not hold, and the flux will not vanish on
stationary backgrounds.  In addition, one could have picked other
expressions for $\df \alpha$, so the expression for the localized
charge is not unique.  It may be possible in this context to find
some other criterion that could be used to determine a unique charge expression.

\bibliographystyle{JHEP}
\bibliography{hamiltonians-null-surfaces}

\providecommand{\href}[2]{#2}\begingroup\raggedright\begin{thebibliography}{10}

\bibitem{Bondi:1962px}
H.~Bondi, M.~G.~J. van~der Burg and A.~W.~K. Metzner, \emph{{Gravitational
  waves in general relativity. 7. Waves from axisymmetric isolated systems}},
  \href{https://doi.org/10.1098/rspa.1962.0161}{\emph{Proc. Roy. Soc. Lond.}
  {\bfseries A269} (1962) 21}.

\bibitem{Sachs:1962wk}
R.~K. Sachs, \emph{{Gravitational waves in general relativity. 8. Waves in
  asymptotically flat space-times}},
  \href{https://doi.org/10.1098/rspa.1962.0206}{\emph{Proc. Roy. Soc. Lond.}
  {\bfseries A270} (1962) 103}.

\bibitem{Sachs:1962zza}
R.~Sachs, \emph{{Asymptotic symmetries in gravitational theory}},
  \href{https://doi.org/10.1103/PhysRev.128.2851}{\emph{Phys. Rev.} {\bfseries
  128} (1962) 2851}.

\bibitem{He:2014cra}
T.~He, P.~Mitra, A.~P. Porfyriadis and A.~Strominger, \emph{{New Symmetries of
  Massless QED}}, \href{https://doi.org/10.1007/JHEP10(2014)112}{\emph{JHEP}
  {\bfseries 10} (2014) 112} [\href{https://arxiv.org/abs/1407.3789}{{\ttfamily
  1407.3789}}].

\bibitem{Kapec:2015ena}
D.~Kapec, M.~Pate and A.~Strominger, \emph{{New Symmetries of QED}},
  \href{https://arxiv.org/abs/1506.02906}{{\ttfamily 1506.02906}}.

\bibitem{WZ}
R.~M. Wald and A.~Zoupas, \emph{{A General definition of `conserved quantities'
  in general relativity and other theories of gravity}},
  \href{https://doi.org/10.1103/PhysRevD.61.084027}{\emph{Phys. Rev.}
  {\bfseries D61} (2000) 084027}
  [\href{https://arxiv.org/abs/gr-qc/9911095}{{\ttfamily gr-qc/9911095}}].

\bibitem{Strominger:2017zoo}
A.~Strominger, \emph{{Lectures on the Infrared Structure of Gravity and Gauge
  Theory}},  \href{https://arxiv.org/abs/1703.05448}{{\ttfamily 1703.05448}}.

\bibitem{Ashtekar:1981bq}
A.~Ashtekar and M.~Streubel, \emph{{Symplectic Geometry of Radiative Modes and
  Conserved Quantities at Null Infinity}},
  \href{https://doi.org/10.1098/rspa.1981.0109}{\emph{Proc. Roy. Soc. Lond.}
  {\bfseries A376} (1981) 585}.

\bibitem{Dray:1984rfa}
T.~Dray and M.~Streubel, \emph{{Angular momentum at null infinity}},
  \href{https://doi.org/10.1088/0264-9381/1/1/005}{\emph{Class. Quant. Grav.}
  {\bfseries 1} (1984) 15}.

\bibitem{Donnay:2016ejv}
L.~Donnay, G.~Giribet, H.~A. González and M.~Pino, \emph{{Extended Symmetries
  at the Black Hole Horizon}},
  \href{https://doi.org/10.1007/JHEP09(2016)100}{\emph{JHEP} {\bfseries 09}
  (2016) 100} [\href{https://arxiv.org/abs/1607.05703}{{\ttfamily
  1607.05703}}].

\bibitem{Donnay:2015abr}
L.~Donnay, G.~Giribet, H.~A. Gonzalez and M.~Pino, \emph{{Supertranslations and
  Superrotations at the Black Hole Horizon}},
  \href{https://doi.org/10.1103/PhysRevLett.116.091101}{\emph{Phys. Rev. Lett.}
  {\bfseries 116} (2016) 091101}
  [\href{https://arxiv.org/abs/1511.08687}{{\ttfamily 1511.08687}}].

\bibitem{Eling:2016xlx}
C.~Eling and Y.~Oz, \emph{{On the Membrane Paradigm and Spontaneous Breaking of
  Horizon BMS Symmetries}},
  \href{https://doi.org/10.1007/JHEP07(2016)065}{\emph{JHEP} {\bfseries 07}
  (2016) 065} [\href{https://arxiv.org/abs/1605.00183}{{\ttfamily
  1605.00183}}].

\bibitem{Cai:2016idg}
R.-G. Cai, S.-M. Ruan and Y.-L. Zhang, \emph{{Horizon supertranslation and
  degenerate black hole solutions}},
  \href{https://doi.org/10.1007/JHEP09(2016)163}{\emph{JHEP} {\bfseries 09}
  (2016) 163} [\href{https://arxiv.org/abs/1609.01056}{{\ttfamily
  1609.01056}}].

\bibitem{Hawking:2015qqa}
S.~W. Hawking, \emph{{The Information Paradox for Black Holes}},
  \href{https://arxiv.org/abs/1509.01147}{{\ttfamily 1509.01147}}.

\bibitem{Hawking:2016sgy}
S.~W. Hawking, M.~J. Perry and A.~Strominger, \emph{{Superrotation Charge and
  Supertranslation Hair on Black Holes}},
  \href{https://doi.org/10.1007/JHEP05(2017)161}{\emph{JHEP} {\bfseries 05}
  (2017) 161} [\href{https://arxiv.org/abs/1611.09175}{{\ttfamily
  1611.09175}}].

\bibitem{Hawking:2016msc}
S.~W. Hawking, M.~J. Perry and A.~Strominger, \emph{{Soft Hair on Black
  Holes}}, \href{https://doi.org/10.1103/PhysRevLett.116.231301}{\emph{Phys.
  Rev. Lett.} {\bfseries 116} (2016) 231301}
  [\href{https://arxiv.org/abs/1601.00921}{{\ttfamily 1601.00921}}].

\bibitem{Carlip:2017xne}
S.~Carlip, \emph{{Black Hole Entropy from Bondi-Metzner-Sachs Symmetry at the
  Horizon}}, \href{https://doi.org/10.1103/PhysRevLett.120.101301}{\emph{Phys.
  Rev. Lett.} {\bfseries 120} (2018) 101301}
  [\href{https://arxiv.org/abs/1702.04439}{{\ttfamily 1702.04439}}].

\bibitem{Blau:2015nee}
M.~Blau and M.~O'Loughlin, \emph{{Horizon Shells and BMS-like Soldering
  Transformations}}, \href{https://doi.org/10.1007/JHEP03(2016)029}{\emph{JHEP}
  {\bfseries 03} (2016) 029}
  [\href{https://arxiv.org/abs/1512.02858}{{\ttfamily 1512.02858}}].

\bibitem{Penna:2017bdn}
R.~F. Penna, \emph{{Near-horizon BMS symmetries as fluid symmetries}},
  \href{https://doi.org/10.1007/JHEP10(2017)049}{\emph{JHEP} {\bfseries 10}
  (2017) 049} [\href{https://arxiv.org/abs/1703.07382}{{\ttfamily
  1703.07382}}].

\bibitem{Grumiller:2018scv}
D.~Grumiller and M.~M. Sheikh-Jabbari, \emph{{Membrane Paradigm from Near
  Horizon Soft Hair}},  \href{https://arxiv.org/abs/1805.11099}{{\ttfamily
  1805.11099}}.

\bibitem{Koga:2001vq}
J.-i. Koga, \emph{{Asymptotic symmetries on Killing horizons}},
  \href{https://doi.org/10.1103/PhysRevD.64.124012}{\emph{Phys. Rev.}
  {\bfseries D64} (2001) 124012}
  [\href{https://arxiv.org/abs/gr-qc/0107096}{{\ttfamily gr-qc/0107096}}].

\bibitem{Mao:2016pwq}
P.~Mao, X.~Wu and H.~Zhang, \emph{{Soft hairs on isolated horizon implanted by
  electromagnetic fields}},
  \href{https://doi.org/10.1088/1361-6382/aa59da}{\emph{Class. Quant. Grav.}
  {\bfseries 34} (2017) 055003}
  [\href{https://arxiv.org/abs/1606.03226}{{\ttfamily 1606.03226}}].

\bibitem{Penna:2015gza}
R.~F. Penna, \emph{{BMS invariance and the membrane paradigm}},
  \href{https://doi.org/10.1007/JHEP03(2016)023}{\emph{JHEP} {\bfseries 03}
  (2016) 023} [\href{https://arxiv.org/abs/1508.06577}{{\ttfamily
  1508.06577}}].

\bibitem{Strominger:2014pwa}
A.~Strominger and A.~Zhiboedov, \emph{{Gravitational Memory, BMS
  Supertranslations and Soft Theorems}},
  \href{https://doi.org/10.1007/JHEP01(2016)086}{\emph{JHEP} {\bfseries 01}
  (2016) 086} [\href{https://arxiv.org/abs/1411.5745}{{\ttfamily 1411.5745}}].

\bibitem{Hollands:2016oma}
S.~Hollands, A.~Ishibashi and R.~M. Wald, \emph{{BMS Supertranslations and
  Memory in Four and Higher Dimensions}},
  \href{https://doi.org/10.1088/1361-6382/aa777a}{\emph{Class. Quant. Grav.}
  {\bfseries 34} (2017) 155005}
  [\href{https://arxiv.org/abs/1612.03290}{{\ttfamily 1612.03290}}].

\bibitem{Bousso:1999xy}
R.~Bousso, \emph{{A Covariant entropy conjecture}},
  \href{https://doi.org/10.1088/1126-6708/1999/07/004}{\emph{JHEP} {\bfseries
  07} (1999) 004} [\href{https://arxiv.org/abs/hep-th/9905177}{{\ttfamily
  hep-th/9905177}}].

\bibitem{Wall:2011hj}
A.~C. Wall, \emph{{A proof of the generalized second law for rapidly changing
  fields and arbitrary horizon slices}},
  \href{https://doi.org/10.1103/PhysRevD.87.069904,
  10.1103/PhysRevD.85.104049}{\emph{Phys. Rev.} {\bfseries D85} (2012) 104049}
  [\href{https://arxiv.org/abs/1105.3445}{{\ttfamily 1105.3445}}], [Erratum:
  Phys. Rev.D87,no.6,069904(2013)].

\bibitem{Casini:2017roe}
H.~Casini, E.~Teste and G.~Torroba, \emph{{Modular Hamiltonians on the null
  plane and the Markov property of the vacuum state}},
  \href{https://doi.org/10.1088/1751-8121/aa7eaa}{\emph{J. Phys.} {\bfseries
  A50} (2017) 364001} [\href{https://arxiv.org/abs/1703.10656}{{\ttfamily
  1703.10656}}].

\bibitem{Brown:2015lvg}
A.~R. Brown, D.~A. Roberts, L.~Susskind, B.~Swingle and Y.~Zhao,
  \emph{{Complexity, action, and black holes}},
  \href{https://doi.org/10.1103/PhysRevD.93.086006}{\emph{Phys. Rev.}
  {\bfseries D93} (2016) 086006}
  [\href{https://arxiv.org/abs/1512.04993}{{\ttfamily 1512.04993}}].

\bibitem{Lehner:2016vdi}
L.~Lehner, R.~C. Myers, E.~Poisson and R.~D. Sorkin, \emph{{Gravitational
  action with null boundaries}},
  \href{https://doi.org/10.1103/PhysRevD.94.084046}{\emph{Phys. Rev.}
  {\bfseries D94} (2016) 084046}
  [\href{https://arxiv.org/abs/1609.00207}{{\ttfamily 1609.00207}}].

\bibitem{Hopfmuller:2016scf}
F.~Hopfmuller and L.~Freidel, \emph{{Gravity Degrees of Freedom on a Null
  Surface}}, \href{https://doi.org/10.1103/PhysRevD.95.104006}{\emph{Phys.
  Rev.} {\bfseries D95} (2017) 104006}
  [\href{https://arxiv.org/abs/1611.03096}{{\ttfamily 1611.03096}}].

\bibitem{Wieland:2017zkf}
W.~Wieland, \emph{{New boundary variables for classical and quantum gravity on
  a null surface}},
  \href{https://doi.org/10.1088/1361-6382/aa8d06}{\emph{Class. Quant. Grav.}
  {\bfseries 34} (2017) 215008}
  [\href{https://arxiv.org/abs/1704.07391}{{\ttfamily 1704.07391}}].

\bibitem{Wieland:2017cmf}
W.~Wieland, \emph{{Fock representation of gravitational boundary modes and the
  discreteness of the area spectrum}},
  \href{https://doi.org/10.1007/s00023-017-0598-6}{\emph{Annales Henri
  Poincare} {\bfseries 18} (2017) 3695}
  [\href{https://arxiv.org/abs/1706.00479}{{\ttfamily 1706.00479}}].

\bibitem{Donnelly:2016auv}
W.~Donnelly and L.~Freidel, \emph{{Local subsystems in gauge theory and
  gravity}}, \href{https://doi.org/10.1007/JHEP09(2016)102}{\emph{JHEP}
  {\bfseries 09} (2016) 102}
  [\href{https://arxiv.org/abs/1601.04744}{{\ttfamily 1601.04744}}].

\bibitem{Speranza:2017gxd}
A.~J. Speranza, \emph{{Local phase space and edge modes for
  diffeomorphism-invariant theories}},
  \href{https://doi.org/10.1007/JHEP02(2018)021}{\emph{JHEP} {\bfseries 02}
  (2018) 021} [\href{https://arxiv.org/abs/1706.05061}{{\ttfamily
  1706.05061}}].

\bibitem{Hopfmuller:2018fni}
F.~Hopfmüller and L.~Freidel, \emph{{Null Conservation Laws for Gravity}},
  \href{https://doi.org/10.1103/PhysRevD.97.124029}{\emph{Phys. Rev.}
  {\bfseries D97} (2018) 124029}
  [\href{https://arxiv.org/abs/1802.06135}{{\ttfamily 1802.06135}}].

\bibitem{Brady:1995na}
P.~R. Brady, S.~Droz, W.~Israel and S.~M. Morsink, \emph{{Covariant double null
  dynamics: (2+2) splitting of the Einstein equations}},
  \href{https://doi.org/10.1088/0264-9381/13/8/015}{\emph{Class. Quant. Grav.}
  {\bfseries 13} (1996) 2211}
  [\href{https://arxiv.org/abs/gr-qc/9510040}{{\ttfamily gr-qc/9510040}}].

\bibitem{Epp:1995uc}
R.~J. Epp, \emph{{The Symplectic structure of general relativity in the double
  null (2+2) formalism}},
  \href{https://arxiv.org/abs/gr-qc/9511060}{{\ttfamily gr-qc/9511060}}.

\bibitem{Reisenberger:2007pq}
M.~P. Reisenberger, \emph{{The Symplectic 2-form and Poisson bracket of null
  canonical gravity}},  \href{https://arxiv.org/abs/gr-qc/0703134}{{\ttfamily
  gr-qc/0703134}}.

\bibitem{Parattu:2015gga}
K.~Parattu, S.~Chakraborty, B.~R. Majhi and T.~Padmanabhan, \emph{{A Boundary
  Term for the Gravitational Action with Null Boundaries}},
  \href{https://doi.org/10.1007/s10714-016-2093-7}{\emph{Gen. Rel. Grav.}
  {\bfseries 48} (2016) 94} [\href{https://arxiv.org/abs/1501.01053}{{\ttfamily
  1501.01053}}].

\bibitem{1987thyg.book..676C}
C.~{Crnkovic} and E.~{Witten}, \emph{{Covariant description of canonical
  formalism in geometrical theories.}},  in \emph{Three hundred years of
  gravitation} (S.~W. Hawking and W.~Israel, eds.), pp.~676--684.
\newblock Cambridge University Press, 1987.

\bibitem{ASHTEKAR1991417}
A.~Ashtekar, L.~Bombelli and O.~Reula, \emph{The covariant phase space of
  asymptotically flat gravitational fields},  in \emph{Mechanics, Analysis and
  Geometry: 200 Years After Lagrange} (M.~Francaviglia, ed.), North-Holland
  Delta Series, pp.~417--450.
\newblock Elsevier, Amsterdam, 1991.

\bibitem{LW}
J.~Lee and R.~M. Wald, \emph{{Local symmetries and constraints}},
  \href{https://doi.org/10.1063/1.528801}{\emph{J. Math. Phys.} {\bfseries 31}
  (1990) 725}.

\bibitem{W-noether-entropy}
R.~M. Wald, \emph{{Black hole entropy is the Noether charge}},
  \href{https://doi.org/10.1103/PhysRevD.48.R3427}{\emph{Phys. Rev.} {\bfseries
  D48} (1993) 3427} [\href{https://arxiv.org/abs/gr-qc/9307038}{{\ttfamily
  gr-qc/9307038}}].

\bibitem{Khavkine:2014kya}
I.~Khavkine, \emph{{Covariant phase space, constraints, gauge and the Peierls
  formula}}, \href{https://doi.org/10.1142/S0217751X14300099}{\emph{Int. J.
  Mod. Phys.} {\bfseries A29} (2014) 1430009}
  [\href{https://arxiv.org/abs/1402.1282}{{\ttfamily 1402.1282}}].

\bibitem{2017NuPhB.924..312G}
M.~{Geiller}, \emph{{Edge modes and corner ambiguities in 3d Chern-Simons
  theory and gravity}},
  \href{https://doi.org/10.1016/j.nuclphysb.2017.09.010}{\emph{Nucl. Phys. B}
  {\bfseries 924} (2017) 312}
  [\href{https://arxiv.org/abs/1703.04748}{{\ttfamily 1703.04748}}].

\bibitem{KP-first-law}
K.~Prabhu, \emph{{The First Law of Black Hole Mechanics for Fields with
  Internal Gauge Freedom}},
  \href{https://doi.org/10.1088/1361-6382/aa536b}{\emph{Class. Quant. Grav.}
  {\bfseries 34} (2017) 035011}
  [\href{https://arxiv.org/abs/1511.00388}{{\ttfamily 1511.00388}}].

\bibitem{IW-euclidean}
V.~Iyer and R.~M. Wald, \emph{{A Comparison of Noether charge and Euclidean
  methods for computing the entropy of stationary black holes}},
  \href{https://doi.org/10.1103/PhysRevD.52.4430}{\emph{Phys. Rev.} {\bfseries
  D52} (1995) 4430} [\href{https://arxiv.org/abs/gr-qc/9503052}{{\ttfamily
  gr-qc/9503052}}].

\bibitem{SeiW}
M.~D. Seifert and R.~M. Wald, \emph{{A General variational principle for
  spherically symmetric perturbations in diffeomorphism covariant theories}},
  \href{https://doi.org/10.1103/PhysRevD.75.084029}{\emph{Phys. Rev.}
  {\bfseries D75} (2007) 084029}
  [\href{https://arxiv.org/abs/gr-qc/0612121}{{\ttfamily gr-qc/0612121}}].

\bibitem{Wald-book}
R.~M. Wald, \emph{{General Relativity}}. The University of Chicago Press, 1984.

\bibitem{IW-noether-entropy}
V.~Iyer and R.~M. Wald, \emph{{Some properties of Noether charge and a proposal
  for dynamical black hole entropy}},
  \href{https://doi.org/10.1103/PhysRevD.50.846}{\emph{Phys. Rev.} {\bfseries
  D50} (1994) 846} [\href{https://arxiv.org/abs/gr-qc/9403028}{{\ttfamily
  gr-qc/9403028}}].

\bibitem{Christodoulou:1993uv}
D.~Christodoulou and S.~Klainerman, \emph{{The global nonlinear stability of
  the Minkowski space}}. Princeton University Press, 1993.

\bibitem{Szabados2009}
L.~B. Szabados, \emph{Quasi-local energy-momentum and angular momentum in
  general relativity}, \href{https://doi.org/10.12942/lrr-2009-4}{\emph{Living
  Reviews in Relativity} {\bfseries 12} (2009) 4}.

\bibitem{Flanagan:2015pxa}
{\'E}.~{\'E}. Flanagan and D.~A. Nichols, \emph{{Conserved charges of the
  extended Bondi-Metzner-Sachs algebra}},
  \href{https://doi.org/10.1103/PhysRevD.95.044002}{\emph{Phys. Rev.}
  {\bfseries D95} (2017) 044002}
  [\href{https://arxiv.org/abs/1510.03386}{{\ttfamily 1510.03386}}].

\bibitem{Jacobson:1993vj}
T.~Jacobson, G.~Kang and R.~C. Myers, \emph{{On black hole entropy}},
  \href{https://doi.org/10.1103/PhysRevD.49.6587}{\emph{Phys. Rev.} {\bfseries
  D49} (1994) 6587} [\href{https://arxiv.org/abs/gr-qc/9312023}{{\ttfamily
  gr-qc/9312023}}].

\bibitem{W-closed}
R.~M. Wald, \emph{On identically closed forms locally constructed from a
  field}, \href{https://doi.org/http://dx.doi.org/10.1063/1.528839}{\emph{J.
  Math. Phys.} {\bfseries 31} (1990) 2378}.

\bibitem{Ashtekar:2001jb}
A.~Ashtekar, C.~Beetle and J.~Lewandowski, \emph{{Geometry of generic isolated
  horizons}}, \href{https://doi.org/10.1088/0264-9381/19/6/311}{\emph{Class.
  Quant. Grav.} {\bfseries 19} (2002) 1195}
  [\href{https://arxiv.org/abs/gr-qc/0111067}{{\ttfamily gr-qc/0111067}}].

\bibitem{Gourgoulhon:2005ng}
E.~Gourgoulhon and J.~L. Jaramillo, \emph{{A 3+1 perspective on null
  hypersurfaces and isolated horizons}},
  \href{https://doi.org/10.1016/j.physrep.2005.10.005}{\emph{Phys. Rept.}
  {\bfseries 423} (2006) 159}
  [\href{https://arxiv.org/abs/gr-qc/0503113}{{\ttfamily gr-qc/0503113}}].

\bibitem{Bardeen:1973gs}
J.~M. Bardeen, B.~Carter and S.~W. Hawking, \emph{{The Four laws of black hole
  mechanics}}, \href{https://doi.org/10.1007/BF01645742}{\emph{Commun. Math.
  Phys.} {\bfseries 31} (1973) 161}.

\bibitem{Ashtekar:2014zsa}
A.~Ashtekar, \emph{{Geometry and Physics of Null Infinity}},  in \emph{{One
  hundred years of general relativity}} (L.~Bieri and S.-T. Yau, eds.), vol.~20
  of \emph{Surveys in Differential Geometry}.
\newblock International Press of Boston, Inc., 2015.
\newblock \href{https://arxiv.org/abs/1409.1800}{{\ttfamily 1409.1800}}.

\bibitem{Hotta:2000gx}
M.~Hotta, K.~Sasaki and T.~Sasaki, \emph{{Diffeomorphism on horizon as an
  asymptotic isometry of Schwarzschild black hole}},
  \href{https://doi.org/10.1088/0264-9381/18/10/301}{\emph{Class. Quant. Grav.}
  {\bfseries 18} (2001) 1823}
  [\href{https://arxiv.org/abs/gr-qc/0011043}{{\ttfamily gr-qc/0011043}}].

\bibitem{Lust:2017gez}
D.~Lust, \emph{{Supertranslations and Holography near the Horizon of
  Schwarzschild Black Holes}},
  \href{https://doi.org/10.1002/prop.201800001}{\emph{Fortsch. Phys.}
  {\bfseries 66} (2018) 1800001}
  [\href{https://arxiv.org/abs/1711.04582}{{\ttfamily 1711.04582}}].

\bibitem{Hou:2017pes}
S.~Hou, \emph{{Asymptotic Symmetries of the Null Infinity and the Isolated
  Horizon}},  \href{https://arxiv.org/abs/1704.05701}{{\ttfamily 1704.05701}}.

\bibitem{Morales:2008}
E.~Morales, \emph{{On a Second Law of Black Hole Mechanics in a Higher
  Derivative Theory of Gravity}}, Ph.D. thesis, University of Gottingen, 2008.

\bibitem{Strominger:2013jfa}
A.~Strominger, \emph{{On BMS Invariance of Gravitational Scattering}},
  \href{https://doi.org/10.1007/JHEP07(2014)152}{\emph{JHEP} {\bfseries 07}
  (2014) 152} [\href{https://arxiv.org/abs/1312.2229}{{\ttfamily 1312.2229}}].

\bibitem{Strominger:2017aeh}
A.~Strominger, \emph{{Black Hole Information Revisited}},
  \href{https://arxiv.org/abs/1706.07143}{{\ttfamily 1706.07143}}.

\bibitem{PhysRevLett.43.181}
A.~Ashtekar and A.~Magnon-Ashtekar, \emph{Energy-momentum in general
  relativity}, \href{https://doi.org/10.1103/PhysRevLett.43.181}{\emph{Phys.
  Rev. Lett.} {\bfseries 43} (1979) 181}.

\bibitem{Campiglia:2015lxa}
M.~Campiglia, \emph{{Null to time-like infinity Green’s functions for
  asymptotic symmetries in Minkowski spacetime}},
  \href{https://doi.org/10.1007/JHEP11(2015)160}{\emph{JHEP} {\bfseries 11}
  (2015) 160} [\href{https://arxiv.org/abs/1509.01408}{{\ttfamily
  1509.01408}}].

\bibitem{Campiglia:2017mua}
M.~Campiglia and R.~Eyheralde, \emph{{Asymptotic $U(1)$ charges at spatial
  infinity}}, \href{https://doi.org/10.1007/JHEP11(2017)168}{\emph{JHEP}
  {\bfseries 11} (2017) 168}
  [\href{https://arxiv.org/abs/1703.07884}{{\ttfamily 1703.07884}}].

\bibitem{Troessaert:2017jcm}
C.~Troessaert, \emph{{The BMS4 algebra at spatial infinity}},
  \href{https://doi.org/10.1088/1361-6382/aaae22}{\emph{Class. Quant. Grav.}
  {\bfseries 35} (2018) 074003}
  [\href{https://arxiv.org/abs/1704.06223}{{\ttfamily 1704.06223}}].

\bibitem{Prabhu2018}
K.~Prabhu, \emph{{Conservation of asymptotic charges from past to future null
  infinity: Maxwell fields}},
  \href{https://doi.org/10.1007/JHEP10(2018)113}{\emph{JHEP} {\bfseries 2018}
  (2018) 113} [\href{https://arxiv.org/abs/1808.07863}{{\ttfamily
  1808.07863}}].

\bibitem{brown1986}
J.~D. Brown and M.~Henneaux, \emph{{Central Charges in the Canonical
  Realization of Asymptotic Symmetries: An Example from Three-Dimensional
  Gravity}}, \href{https://doi.org/10.1007/BF01211590}{\emph{Commun. Math.
  Phys.} {\bfseries 104} (1986) 207}.

\bibitem{Barnich:2001jy}
G.~Barnich and F.~Brandt, \emph{{Covariant theory of asymptotic symmetries,
  conservation laws and central charges}},
  \href{https://doi.org/10.1016/S0550-3213(02)00251-1}{\emph{Nucl. Phys.}
  {\bfseries B633} (2002) 3}
  [\href{https://arxiv.org/abs/hep-th/0111246}{{\ttfamily hep-th/0111246}}].

\bibitem{Barnich:2007bf}
G.~Barnich and G.~Compere, \emph{{Surface charge algebra in gauge theories and
  thermodynamic integrability}},
  \href{https://doi.org/10.1063/1.2889721}{\emph{J. Math. Phys.} {\bfseries 49}
  (2008) 042901} [\href{https://arxiv.org/abs/0708.2378}{{\ttfamily
  0708.2378}}].

\bibitem{Arnold}
V.~I. Arnol'd, \emph{{Mathematical methods of classical mechanics}}, Graduate
  texts in mathematicals. Springer, New York, NY, 1978.

\bibitem{Barnich:2011ct}
G.~Barnich and C.~Troessaert, \emph{{Supertranslations call for
  superrotations}}, \href{https://doi.org/10.22323/1.127.0010}{\emph{PoS}
  {\bfseries CNCFG2010} (2010) 010}
  [\href{https://arxiv.org/abs/1102.4632}{{\ttfamily 1102.4632}}], [Ann. U.
  Craiova Phys.21,S11(2011)].

\bibitem{2002clme.book.....G}
H.~{Goldstein}, C.~{Poole} and J.~{Safko}, \emph{{Classical mechanics}}. 2002.

\bibitem{1986JMP....27..489B}
J.~D. {Brown} and M.~{Henneaux}, \emph{{On the Poisson brackets of
  differentiable generators in classical field theory}},
  \href{https://doi.org/10.1063/1.527249}{\emph{J. Math. Phys.} {\bfseries 27}
  (1986) 489}.

\bibitem{Seraj:2016cym}
A.~Seraj, \emph{{Consderved charges, surface degrees of freedom, and black hole
  entropy}}, Ph.D. thesis, IPM, Tehran, 2016.
\newblock \href{https://arxiv.org/abs/1603.02442}{{\ttfamily 1603.02442}}.

\bibitem{Guo:2002ed}
H.-Y. Guo, C.-G. Huang and X.-n. Wu, \emph{{Noether charge realization of
  diffeomorphism algebra}},
  \href{https://doi.org/10.1103/PhysRevD.67.024031}{\emph{Phys. Rev.}
  {\bfseries D67} (2003) 024031}
  [\href{https://arxiv.org/abs/gr-qc/0208067}{{\ttfamily gr-qc/0208067}}].

\bibitem{Barnich:2011mi}
G.~Barnich and C.~Troessaert, \emph{{BMS charge algebra}},
  \href{https://doi.org/10.1007/JHEP12(2011)105}{\emph{JHEP} {\bfseries 12}
  (2011) 105} [\href{https://arxiv.org/abs/1106.0213}{{\ttfamily 1106.0213}}].

\bibitem{Andy2018talk}
A.~Strominger.
\newblock presentation at Strings 2018,
  \url{https://indico.oist.jp/indico/event/5/picture/106.pdf}.

\bibitem{Haco:2018ske}
S.~Haco, S.~W. Hawking, M.~J. Perry and A.~Strominger, \emph{{Black Hole
  Entropy and Soft Hair}},  \href{https://arxiv.org/abs/1810.01847}{{\ttfamily
  1810.01847}}.

\bibitem{Ashtekar:2004gp}
A.~Ashtekar, J.~Engle, T.~Pawlowski and C.~Van Den~Broeck, \emph{{Multipole
  moments of isolated horizons}},
  \href{https://doi.org/10.1088/0264-9381/21/11/003}{\emph{Class. Quant. Grav.}
  {\bfseries 21} (2004) 2549}
  [\href{https://arxiv.org/abs/gr-qc/0401114}{{\ttfamily gr-qc/0401114}}].

\bibitem{CentralExt}
V.~Chandrasekaran, {\'E}.~{\'E}. Flanagan and K.~Prabhu In preparation.

\bibitem{Kapec:2015vwa}
D.~Kapec, V.~Lysov, S.~Pasterski and A.~Strominger, \emph{{Higher-Dimensional
  Supertranslations and Weinberg's Soft Graviton Theorem}},
  \href{https://arxiv.org/abs/1502.07644}{{\ttfamily 1502.07644}}.

\bibitem{Pate:2017fgt}
M.~Pate, A.-M. Raclariu and A.~Strominger, \emph{{Gravitational Memory in
  Higher Dimensions}},
  \href{https://doi.org/10.1007/JHEP06(2018)138}{\emph{JHEP} {\bfseries 06}
  (2018) 138} [\href{https://arxiv.org/abs/1712.01204}{{\ttfamily
  1712.01204}}].

\bibitem{Hollands:2003ie}
S.~Hollands and A.~Ishibashi, \emph{{Asymptotic flatness and Bondi energy in
  higher dimensional gravity}},
  \href{https://doi.org/10.1063/1.1829152}{\emph{J. Math. Phys.} {\bfseries 46}
  (2005) 022503} [\href{https://arxiv.org/abs/gr-qc/0304054}{{\ttfamily
  gr-qc/0304054}}].

\bibitem{Barnich:2009se}
G.~Barnich and C.~Troessaert, \emph{{Symmetries of asymptotically flat 4
  dimensional spacetimes at null infinity revisited}},
  \href{https://doi.org/10.1103/PhysRevLett.105.111103}{\emph{Phys. Rev. Lett.}
  {\bfseries 105} (2010) 111103}
  [\href{https://arxiv.org/abs/0909.2617}{{\ttfamily 0909.2617}}].

\bibitem{Campiglia:2015yka}
M.~Campiglia and A.~Laddha, \emph{{New symmetries for the Gravitational
  S-matrix}}, \href{https://doi.org/10.1007/JHEP04(2015)076}{\emph{JHEP}
  {\bfseries 04} (2015) 076}
  [\href{https://arxiv.org/abs/1502.02318}{{\ttfamily 1502.02318}}].

\bibitem{Compere:2018ylh}
G.~Compère, A.~Fiorucci and R.~Ruzziconi, \emph{{Superboost transitions,
  refraction memory and super-Lorentz charge algebra}},
  \href{https://arxiv.org/abs/1810.00377}{{\ttfamily 1810.00377}}.

\bibitem{1993CQGra..10..773H}
S.~A. {Hayward}, \emph{{The general solution to the Einstein equations on a
  null surface}},
  \href{https://doi.org/10.1088/0264-9381/10/4/012}{\emph{Class. Quant. Grav.}
  {\bfseries 10} (1993) 773}.

\bibitem{Barack:1999st}
L.~Barack, \emph{{Late time decay of scalar, electromagnetic, and gravitational
  perturbations outside rotating black holes}},
  \href{https://doi.org/10.1103/PhysRevD.61.024026}{\emph{Phys. Rev.}
  {\bfseries D61} (2000) 024026}
  [\href{https://arxiv.org/abs/gr-qc/9908005}{{\ttfamily gr-qc/9908005}}].

\end{thebibliography}\endgroup

\end{document}